\documentclass[aps,pra,twocolumn,superscriptaddress,amsmath,amssymb,floatfix]{revtex4-1}
\usepackage{color}
\usepackage{graphicx}
\usepackage{bm}
\usepackage{textcomp}
\usepackage[normalem]{ulem}
\newcommand{\njp}{New J. Phys.}
\newcommand{\jms}{J. Mol. Spectrosc.}
\newcommand{\epjd}{Eur. Phys. J. D}

\begin{document}

%\preprint{}

% \title{Optical production of long-range tetrameric complex in ultracold regime}
\title{Photoassociation of ultracold long-range polyatomic molecules}


\author{Marko Gacesa}
\affiliation{Bay Area Environmental Research Institute, Moffett Field, CA 94035-0001, USA}
\affiliation{Space Science Division, NASA Ames Research Center, MS 245-3, Moffett Field, CA 94035, USA}
\affiliation{Department of Physics, Khalifa University, Abu Dhabi 127788, United Arab Emirates}
\email{marko.gacesa@ku.ac.ae}

\author{Jason N. Byrd}
\affiliation{ENSCO, Inc., 4849 North Wickham Road, Melbourne, FL 32940, USA}

\author{Jonathan Smucker}
\affiliation{Department of Physics, University of Connecticut, Storrs, CT 06269-3046, USA}

\author{John A. Montgomery, Jr.}
\affiliation{Department of Physics, University of Connecticut, Storrs, CT 06269-3046, USA}

\author{Robin C\^ot\'e}
\affiliation{Department of Physics, University of Connecticut, Storrs, CT 06269-3046, USA}
\affiliation{Physics Department, University of Massachusetts Boston, Boston, Massachusetts 02125, USA}
\email{robin.cote@uconn.edu}

\date{\today}

\begin{abstract}
We explore the feasibility of optically forming long-range tetratomic and larger polyatomic molecules in their ground electronic state from ultracold pairs of polar molecules aligned by external fields.
Depending on the relative orientation of the interacting diatomic molecules, we find that a tetratomic can be formed either as a weakly bound complex in a very extended halo state or as a pure long-range molecule composed of collinear or nearly-collinear diatomic molecules. The latter is a novel type of tetratomic molecule comprised of two diatomic molecules bound at long intermolecular range and predicted to be stable in cold and ultracold regimes.
Our numerical studies were conducted for ultracold KRb and RbCs, resulting in production of (KRb)$_2$ and (RbCs)$_2$ complexes, respectively. Based on universal properties of long-range interactions between polar molecules, we identify triatomic and tetratomic linear polar molecules with favorable ratio of dipole and quadrupole moments for which the apporach could be generalized to form  larger polyatomic molecules.

\end{abstract}

\pacs{}
%\keywords{}

\maketitle

\section{\label{intro}Introduction}

Ultracold polar molecules have been proposed as an ideal model system to explore novel physical phenomena at the intersection of molecular physics with few- and many-body quantum physics. For example, they could be used to engineer and study lattice spin models in strongly interacting many-body hamiltonians\cite{2013Natur.501..521Y,2013MolPh.111.1908G,2013PhRvL.111z0401E,2014Natur.511..198R,2016PhRvL.116m5301D,2020arXiv200111792K}, supersolidity\cite{2011PhRvA..83e3611C}, unconventional superfluid phases and quantum magnetism\cite{2013PhRvL.110g5301H,2015famr.book....3W}, to name a few. Moreover, the tunability of interactions combined with the extraordinary degree of control available in current experiments\cite{chin2010feshbach} make such systems suitable for engineering quantum simulators\cite{micheli2006toolbox,2008RvMP...80..885B}, as well as quantum entanglement and information processing \cite{demille2002quantum,krems2009cold,kuznetsova2011atom,doi:10.1002/9781118742631.ch14,micheli2006toolbox,yelin2006schemes,2010PhRvA..81c0301K}.
Unlike atomic gases, where inter-particle interactions are isotropic and short-range, gases of ultracold polar molecules exhibit much richer dynamics and macroscopic properties due to the long-range anisotropic electric dipole-dipole interactions between their constituents. While dipole-dipole interactions can be realized using atomic magnetic dipoles\cite{PhysRevLett.95.150406} or between excited Rydberg atoms \cite{2002PhRvL..88m3004B,2005JPhB...38S.295S}, they will be much weaker (ground state atoms) or be short-lived (Rydberg atoms) than those between permanent electric dipoles of polar molecules, and allow less tunability via external electromagnetic fields\cite{jaksch2000fast,lukin2001dipole,cote2005tutorial,stanojevic2006long,2014PhRvL.113s5302H}.

Producing ultracold polar molecules was achieved for only a handful of diatomic species and up to this day remains challenging.
Fermionic $^{40}$K$^{87}$Rb\cite{Ni231,2015Sci...350..659M} were the first heteronuclear molecules to be produced in deeply bound molecular states, with NaK \cite{park2015ultracold,seesselberg2018modeling}, RbCs\cite{PhysRevLett.113.205301,PhysRevLett.113.255301,2017PhRvL.118g3201R}, NaRb\cite{2016PhRvL.116t5303G}, and LiNa\cite{heo2012formation,son2020collisional} being added to the list more recently.
In these experiments, pairs of atoms in an ultracold atomic gas were coherently associated into loosely bound molecules by a magnetic field sweep across a Fano-Feshbach resonance, and stimulated Raman adiabatic passage (STIRAP)\cite{1990JChPh..92.5363G} was used to transfer the population into the molecular ground state.
Alternative approaches to produce ultracold molecules is the photoassociation (PA) of ultracold atomic pairs\cite{PhysRevLett.58.2420,2006RvMP...78..483J} or Feshbach Optimized PA (FOPA) \cite{2008PhRvL.101e3201P,pellegrini2009probing,FOPA-2015-85Rb2,FOPA-2020-KCs}.

Thus far, those techniques were limited to diatomic molecules, even though the theory suggests that they could be extended to larger molecules \cite{dulieu2009formation,balakrishnan2016perspective}. 
In practice, in addition to a significantly increased complexity of the experiments, one of the greatest obstacles to production of larger molecules is the rapid loss of cold molecules from the trap, caused by chemical reactions (\textit{e.g.}, KRb+KRb $\rightarrow$ K$_2$+Rb$_2$ is exothermic) or not well-understood ``complex nature of scattering processes'' found to occur even for endothermic intermolecular reactions \cite{2017PhRvL.118g3201R}. 
In order to explain the loss due to bi-molecular exothermic reactions, and how to prevent it, detailed studies of ultracold reactions between KRb molecules including energetics, were conducted \cite{2010PhRvA..82a0502B,2012JChPh.137k4305B}. Byrd \textit{et al.} \cite{2012JChPh.136a4306B} have carried out calculations of structure, energetics, and reactions of alkali metal tetramer molecules and analyzed conditions for their controllable binding in a trap with attractive dipoles \cite{2012PhRvL.109h3003B,2012PhRvA..86c2711B}.
These studies can be interpreted as initial investigations of electronic structure of ground-state tetramers composed of two alkali-metal polar molecules.

In addition, the knowledge of electronic potential energy surfaces (PESs) for ground and excited electronic states over a wide range of internuclear separations is critical for experiments' design and selection of suitable atomic and molecular systems.
Thus, preliminary theoretical studies have been focused on understanding long-range atom-diatom \cite{PhysRevA.82.042711,2015PhRvL.115g3201P} and diatom-diatom interactions \cite{2012PhRvA..86c2711B,2012PhRvA..86c2711B,2012PhRvL.109h3003B,2012JChPh.136a4306B,2015JChPh.142u4303V,2016JPhB...49a4004L,2017arXiv170309174Q} for experimentally accessible species including Cs-Cs$_2$, KRb-KRb, and other alkali metal diatomic molecules. Notably, Perez-Rios \textit{et al.} \cite{2015PhRvL.115g3201P} computed the rate coefficient for the PA of Cs$_3$ and found that it would be comparable to the formation rates observed for diatomic cases, and, more recently, Schnabel et al. \cite{PhysRevA.103.022820}  theoretically investigated the prospects for photoassociation (PA) of Rb$_3$.

In this paper, we explore the possibility that ultracold long-range tetratomic and larger polyatomic molecular complexes can be formed by photoassociation of pairs of ultracold polar molecules.
The properties of the long-range complexes can be predicted from the ratio of dipole and quadrupole electric moments of the pairs, while their orientation determines if such long-range states exist. The concept is robust and extends beyond known diatomic pairs to larger polyatomic complexes.

This article is organized as follows. In Section \ref{s2} we outline the working concepts used in the photoassociation of pairs of polar molecules. In Section \ref{s3} we describe the theoretical methods employed on two benchmark ultracold molecules, KRb and RbCs, both of which have been experimentally produced. In Section \ref{s4}, the numerical results and their implications for the production of larger polyatomic molecules are presented. We summarize the main findings of this study and conclude in Section \ref{s5}.

\section{\label{s2}Concept: Photoassociation of diatomic into tetratomic molecules}

\subsection{Background}

The photoassociation (PA) of a pair of ultracold atoms was initially proposed by Thorsheim \textit{et al.} \cite{PhysRevLett.58.2420} in the context of molecular spectroscopy. 
The PA is particularly effective at ultralow temperatures, where two atoms colliding with extremely low relative kinetic energy ($k_B T < 1$ mK $\simeq$ 21 MHz) can efficiently absorb a photon from a laser field tuned to a quasi-resonant free-bound electronic dipolar transition and form a molecule in an excited electronic state, typically in a weakly bound ro-vibrational level close to the dissociation energy. 
The PA has found numerous applications in low-temperature atomic, molecular, and optical physics, as described in a number of excellent review articles\cite{1995ARPC...46..423L,1999JMoSp.195..194S,2006RvMP...78..483J,krems2009cold,dulieu2009formation,balakrishnan2016perspective,cote2010forming}.

The PA can be used to form ultracold molecules in deeply bound ro-vibrational levels of the electronic ground state. A demonstrated approach consists of two steps: the PA of an ultracold atomic pair into a molecule in an excited electronic state, followed by the relaxation into the ground state via spontaneous emission\cite{1995ARPC...46..423L,PhysRevLett.80.4402,2002PhRvL..89f3001V,2004EPJD...28..351V}.
The main difficulty with the PA formation can be traced to the Franck-Condon principle, according to which the PA is almost always most effective at large internuclear separations where the spontaneous emission has very low probability to populate deeply bound levels. Early proposals to remedy this obstacle suggested to use re-pumping lasers to reach excited molecular levels that overlap better with deeply molecular ground state levels \cite{1997CPL...279...50C,1999JMoSp.195..236C}.
% If the electronic structure of the two states is favorable, the spontaneous emission will populate deeply bound ro-vibrational levels of the ground state and produce stable ultracold molecules\cite{1995ARPC...46..423L,PhysRevLett.80.4402,2002PhRvL..89f3001V,2004EPJD...28..351V}.
Another approach to enhancing photoassociative production efficiency of stable ultracold molecules can be achieved via Feshbach resonances \cite{2008PhRvL.101e3201P,2014PhRvA..89e2712H,2014JMoSp.300..124G,2020JChPh.152q4307H,sun2020formation}, short laser pulses \cite{2006PhRvA..73c3408K,2009NJPh...11e5011G,2014PhRvA..89e2712H,2015PhRvL.115q3003C,2017PhRvA..96d3417W}, or both \cite{2013PhRvA..88f3418G,2015PhRvA..92c2709H}.

\subsection{Photoassociation in specific geometries}

\begin{figure}[t]
 \centering
 \includegraphics[clip,width=1.05\linewidth]{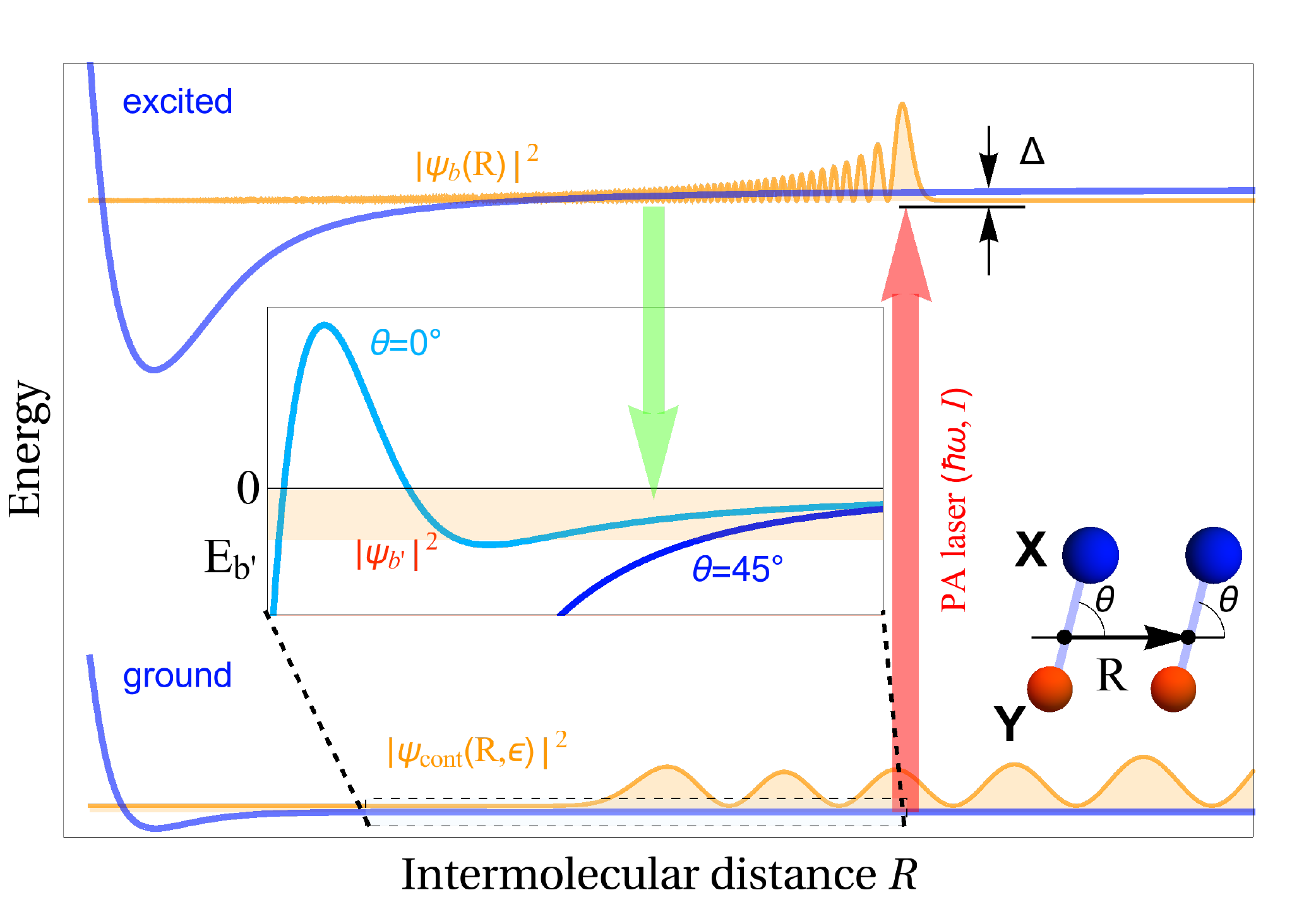}
 \caption{(Color online) Schematic representation of photoassociative formation of a ground-state molecule (XY)$_2$ from two aligned ultracold polar dimers XY whose relative orientation is given by the angle $\theta$. The excited tetramer in the ro-vib. state $|b\rangle$ (wave function $\psi_b$), detuned by $\Delta$ from its dissociation energy, is created by absorbing a photon of energy $\hbar \omega$ from the laser field of intensity $I$. The excited state spontaneously decays into the ground state, ro-vib. level $|b'\rangle$ with binding energy $E_{b'}$ (wave function $\psi_{b'}$). 
 \textit{Inset}: The long-range region of the ground state. Long-range potential well (shown for $\theta$=0\textdegree) capable of supporting several bound states exists for $\theta<\theta_c$, where $\theta_c \approx 20$\textdegree \; for alkali metal pairs.}
 \label{fig1}
\end{figure}

In the two-step PA formation of tetratomic molecules, and larger polyatomic molecules, the relative orientation of the colliding pair will determine the Franck-Condon factors and whether deeply bound molecules in the ground state can be formed in significant numbers. The basic assumption is that ultracold polar molecules aligned by external fields will approach each other very slowly with a specific relative angle. The otherwise very complex potential energy surface can be replaced at large separation between two interacting polar molecules by a simpler set of curves parametrized by the separation $R$ between their center-of-mass, and the angle $\theta$ between their orientation and the intermolecular axis (see Fig.~\ref{fig1}). Since PA into the attractive excited electronic state occurs at very large separation, the massive molecules do not accelerate significantly towards each other before spontaneous decay takes place red, with the molecular partners therefore still well separated and at large distances.
While one of the partners remains in its initial state and aligned by the external field, the other absorbed a photon and gained angular momentum leading to its rotation and misalignment of the orientation between molecular partners.
However, since the Franck-Condon principle dictates that the transition will be most probable where the wave functions overlap best, decay into geometries with long-range wells in the vicinity of the location of the ground-excited molecular pair will be favored.
As we explain in the next section, this decay at large separation will lead to the formation of polyatomic molecules in their electronic ground state, albeit in high excited ro-vibrational levels.
These assumptions allow a drastic simplification of the full problem, by reducing the full multidimensional problem to a much simpler one that can described by the intermolecular separation $R$ and the alignment angle $\theta$.

%While exploring the interaction of two ultracold KRb molecules in a trap,
Byrd \textit{et al.} \cite{2012PhRvL.109h3003B} found that strong quadrupole interactions between the molecules create a repulsive barrier in potential energy curves for parallel and collinear alignment of the diatomic molecules.
The barrier, found to be several Kelvin high, gives rise to outer wells (for collinear alignments) deep enough to support bound states between two KRb molecules (see Fig. \ref{fig1}).
Byrd \textit{et al.} \cite{2012PhRvL.109h3003B} also determined that the first two terms in the long-range potential expansion determine the existence and height of the barrier for KRb-KRb pairs as well as for several other alkali metal dimers. For a more general geometry, other terms of the long-range expansion have to be included\cite{mulder1979anisotropy,van1980ab,2011JChPh.135x4307B}.
In case of KRb-KRb, the barrier was found to survive deviations up to 20\textdegree \; from the collinear alignment, as well as similar deviations from a purely planar geometry.

The existence of the barrier in the ground state is serendipitous for the PA formation of the complex: the long-range potential well in the ground state will, in general, have a significantly higher probability to be populated by the spontaneous emission from an electronic excited state accessible to the PA. Again, this can be understood from the Franck-Condon principle\cite{krems2009cold}.
In Fig. \ref{fig1}, we depict a sketch for the PA formation of the (XY)$_2$ complex from aligned two polar diatomics XY, with relative orientation described by the angle $\theta$. Here, the spontaneous emission will populate deeply bound ro-vibrational levels in the long-range well for nearly-aligned dimers ($\theta<20$), while only weakly bound and very extended levels (that would not be stable against collisons and other dissociation processes in the trap) would be populated in the absence of the barrier.

\subsection{Approximations and Simplifications}
\label{sec:approximations}

In principle, treating the problem exactly would require the calculation of multi-dimensional potential energy surfaces for both ground and electronically excited states for all possible orientations of the approaching molecules, as well as the corresponding transition dipole moments. In addition, the molecules need to be aligned using external fields, such as electric fields, in order to take advantage of the long-range angle-dependent barrier. The coupling of this external field with the internal degrees of freedom also adds to the level of complexity to be dealt with.

To render this problem tractable, a first assumption is made that the individual molecules can be aligned via external fields. In that case, each pair of molecules is always in the same plane and the angle $\theta$ between the molecular axis and the line joining the molecules' center of mass is the same for both molecules. Relaxing this condition would modify the long-range interaction \cite{2012PhRvL.109h3003B} and complicate the analysis. However, we assume a strong enough field so that the alignment is possible, and long-range barriers and wells persist even if the alignment is not perfect.

Second, in order to further simplify the treatment, we take advantage of the conditions offered by ultracold temperatures, and the relatively low densities encountered in experimental settings. The low density allows us to consider only binary event, {\it i.e.} the possibility of having a third molecule in the vicinity of a pair of molecules can be neglected. The ultralow temperatures imply very slow moving particles, which is amplified by the considerable mass of the polar molecules considered, anywhere between roughly 50-150 $u$ or more (atomic mass unit: $u=1$ g/mol = 1822.8885 $m_e$). For example, assuming $m\sim 100$ $u$, $E\sim k_B T \sim \frac{1}{2}mv^2$ for temperatures $T\sim 1$ nK - 1 $\mu$K  lead to velocities $v\sim 0.4 - 12.9$ mm/s. Neglecting the small momentum kick of the photon absorbed during the PA process, a pair of molecule will travel a very short distance while excited. 
In fact,  the lifetime of the molecular electronic excited states being in the range of a few nsec (say 10), and during that time, the molecules cover roughly $4.08\times 10^{-12} - 1.29\times 10^{-10}$ m, {\it i.e.} a fraction to a few Bohr radii 0.08 - 2.4 $a_0$ (1 $a_0=5.29\times 10^{-11}$ m) . During that time and distance, the molecules do not have significant opportunity to exert a torque on each other and their alignment should remain intact.

In addition, as we will describe in the following sections, the PA excitation takes place in the 
vicinity of the classical outer turning point of the excited curve, which is selected to be at large 
separation by the laser detuning $\Delta$. At those large distances of several tens of Bohr radii, each ``compact" molecule is well separated from the colliding partner, and does not change its orientation significantly. Since both molecules are originally aligned in their electronic ground state, we thus can assume that they maintain their alignment during the PA process.

However, before the second step of the scheme takes place, {\it i.e.} the de-excitation into the long-range ground state intermolecular well,  we need to account for the molecular rotation induced by the PA process in the first step.  Indeed, once a molecule in the pair of approaching molecules is excited, it will start rotating, and during the time it is excited (roughly its lifetime of tens of nsec or so), it will execute many full rotations (classically speaking), and hence the ground and excited molecules will become misaligned. 

We recall the Franck-Condon principle, which states that the transition will be most probable where the wave
function overlap best. In fact, the Einstein $A$-coefficient for spontaneous emission of a photon of energy 
$\hbar\omega$ is proportional to $\omega^3 |\langle \psi_g|D|\psi_e\rangle|^2$,
{\it i.e.} the overlap between the ground ($\psi_g$) and excited ($\psi_e$) wave functions will dictate the most probable decay 
(assuming the transition dipole moment $D\sim {\rm const.}$ in the appropriate range of separation). Because
the ultracold molecules move a very short distance before decaying to the ground state (a few $a_0$ as described
above), the de-excitation will take place while the molecules are still at large separations. For example,
assuming that the initial PA takes place at the outer turning point of the barrier (roughly 125 $a_0$ for KRb or 45 $a_0$ for RbCs: see next sections), during the roughly 10 nsec in the existed state, the pair should still be at large separations after de-excitation.

So, while the ground state molecule would remain oriented in the laboratory frame ({\it e.g.}, by the electric field), 
we can thus assume $\theta_1=0$ for convenience. The excited molecule will execute many rotation before decaying, and 
the transition itself being vertical, the decay will preserve the angles when it takes place. Hence the decay will be most 
probable in tetramer ground state curves that provide the largest overlap within the few bohr radii of the barrier of the initial PA step. This should happen when there is a turning point nearby ({\it i.e.} for $\theta_1=0$, and any $\theta_2$ and $\phi$ leading to a barrier nearby). For any other orientation of the pair of molecules, the tetramer surface will be repulsive or deeply attractive in that region, leading to small wave function probability density and overlap in that range, and thus small $A$-coefficients.

Therefore, the most probable decay will take place into long-range wells, with a probability given by an average $A$-coefficient over all angles. A more explicit discussion is given in the next sections.

\section{\label{s3}Methods}

\subsection{Electronic structure calculation}

\begin{figure*}[t]
 \centering
 \includegraphics[clip,width=0.9\linewidth]{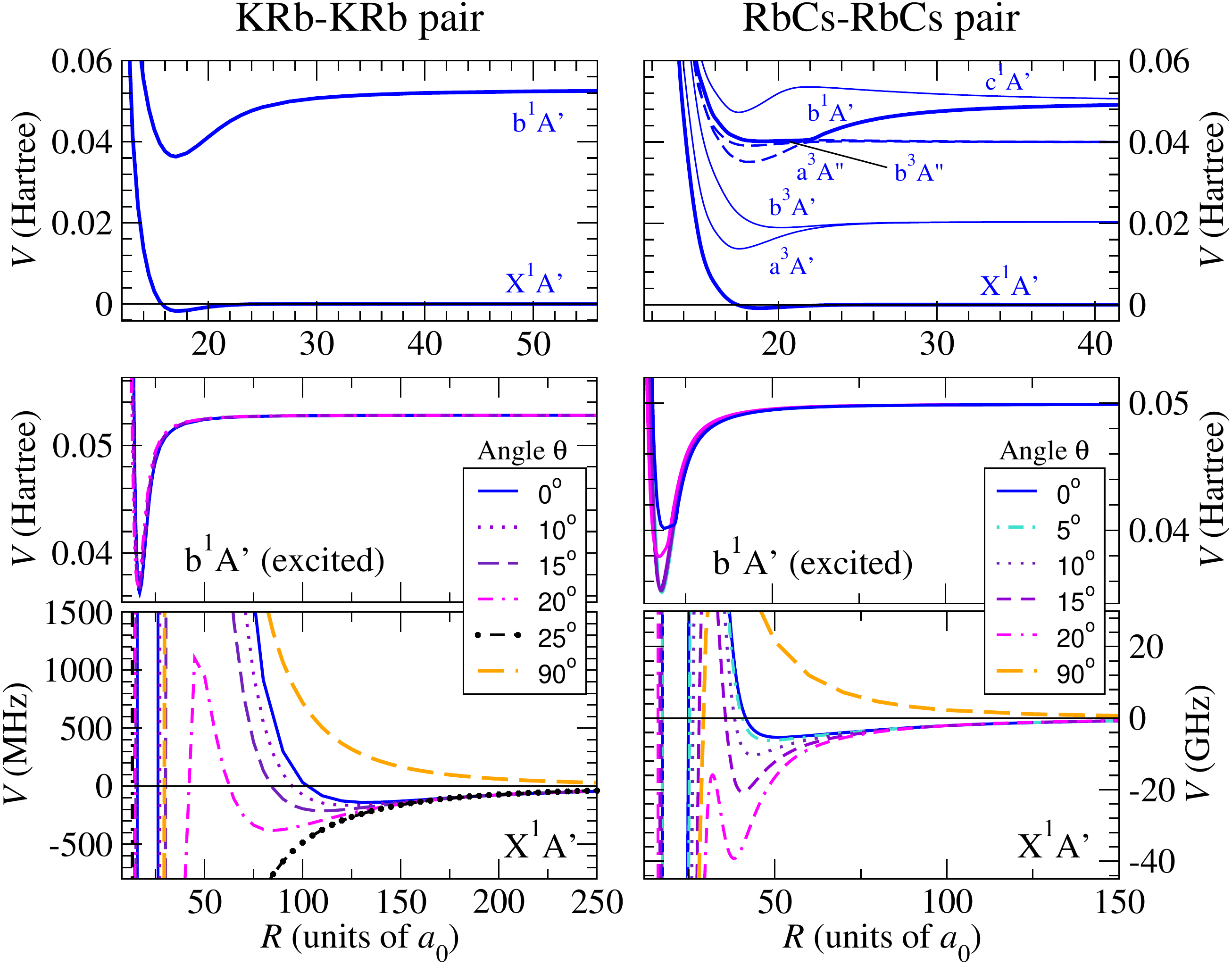}
 \caption{(Color online) \textit{Top:} Calculated potential energy curves for (KRb)$_2$ (left panels) and (RbCs)$_2$ (right panels)
 for $\theta=0^\mathrm{\circ}$.
 \textit{Middle}: Zoom on the excited electronic state b$^1\mathrm{A}'$ for selected orientations $\theta$ (legend).
 \textit{Bottom}: Same as above for the ground state. The potential barrier and long-range well are present for both complexes and diminish with increasing angle $\theta$.}
 \label{fig2}
\end{figure*}

To keep our study relevant to ongoing experiments, we carried out detailed numerical analyses of long-range tetratomic molecules' formation from ultracold pairs of KRb and RbCs molecules. KRb was the first heteronuclear molecule to be cooled to ultracold temperatures\cite{Ni231} and remains one of the most studied system in the context of novel physical regimes and phenomena. 
Although the exothermal reaction KRb+KRb$\rightarrow$ K$_2$ + Rb$_2$ occurs \cite{2010PhRvA..82a0502B}, restricted geometries allowed to trap and study those molecules \cite{de2011controlling,2015Sci...350..659M,2017Sci...357.1002B}. Recently, a degenerate Fermi gas of KRb molecules was produced\cite{De_Marco853}. 
On the other hand, RbCs is interesting because it is an experimentally accessible non-reactive bosonic molecule.
Moreover, RbCs  is the most extended and least polar of bi-alkali molecules \cite{2005JChPh.122t4302A}: its long-range attractive dipole-dipole interaction is less significant and the repulsive exchange interaction is more significant than for other bi-alkali molecules, allowing us to analyze a practical limiting case.

% RbCs+RbCs reaction: 
% 1. barrierless starting (from both RbCs+RbCs and Rb2+Cs2 sides) \cite{PhysRevA.78.022705}; likely all alkali dimers are. However, the reaction is prevented by spins at ultracold temperatures. 

We constructed \textit{ab initio} electronic potential energy curves (PECs) $V_Y(R,\theta)$ for (KRb)$_2$ and (RbCs)$_2$ complexes in the planar geometry, where $R$ and $\theta$ are the distance between the center-of-mass of two diatomics and their relative orientation angle, respectively (see Fig. \ref{fig1}). 
\begin{figure}[b!]
 \centering
 \includegraphics[clip,width=0.9\linewidth]{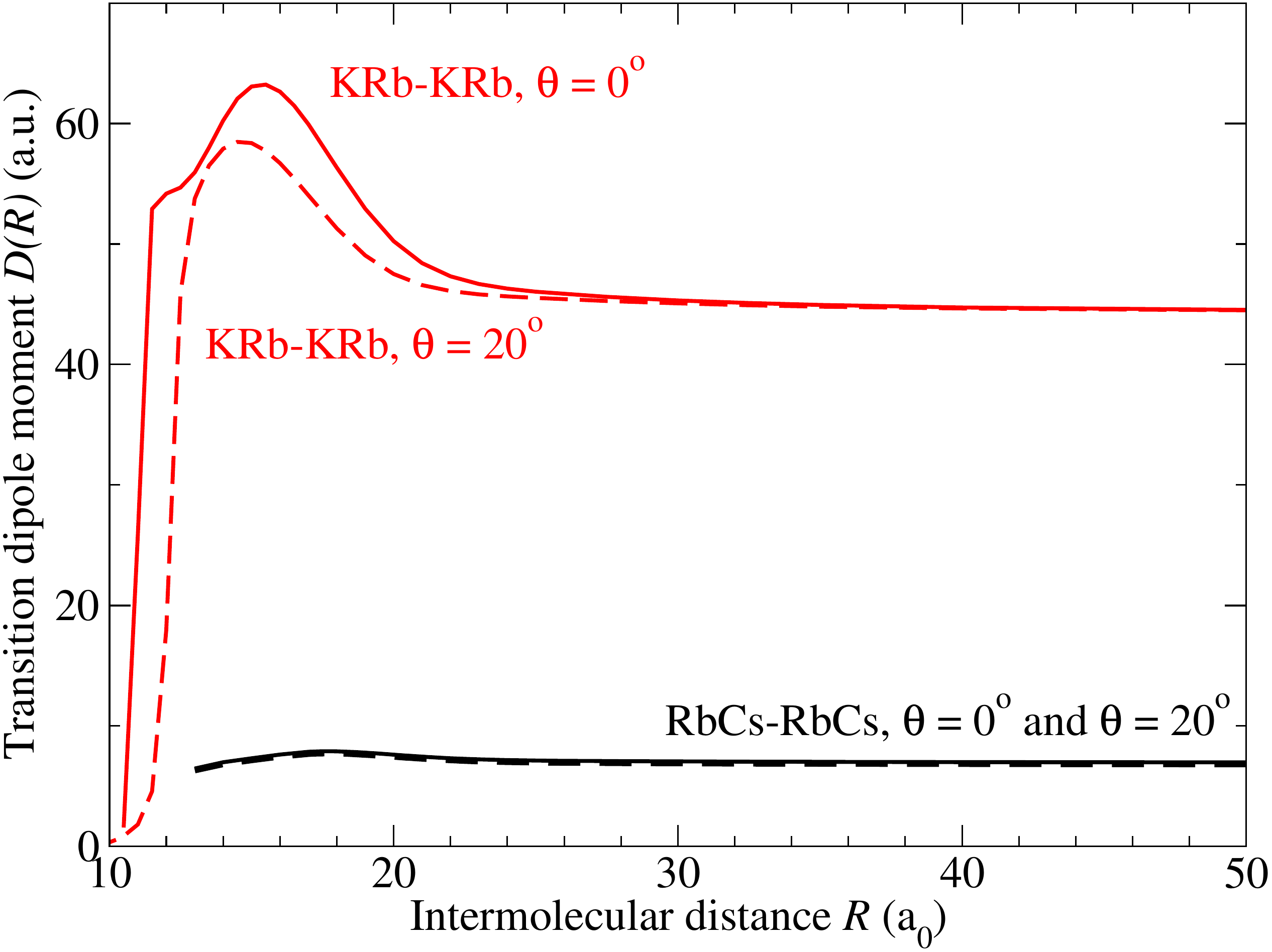}
 \caption{(Color online) Transition dipole moments for X$^{1}A'$-b$^{1}A'$ transitions in (KRb)$_2$ and (RbCs)$_2$ (solid curves for $\theta = 0^\mathrm{o}$, dashed curves for $\theta =20^\mathrm{o}$) . In case of (RbCs)$_2$, differences between relative orientations given by $\theta=0^\mathrm{o} \dots 20^\mathrm{o}$ are less than 2\%.}
 \label{fig2b}
\end{figure}
For (KRb)$_2$, the ground (X$^1A'$) and two lowest excited singlet PECs (b$^1A'$, and c$^1A'$) were constructed. For (RbCs)$_2$, we constructed the ground (X$^1A'$) and lowest six excited PECs (a$^3A'$, b$^3A'$, a$^3A''$, b$^3A''$, b$^1A'$, c$^1A'$) in order to correctly account for the spin-orbit coupling. The KRb and RbCs bond lengths were set to 2.2014 a.u. and 4.4271 a.u., respectively.
The \textit{ab initio} points were calculated for 40 values of $R$ and 7 angles, $\theta$=\{0\textdegree,5\textdegree,10\textdegree,15\textdegree,18\textdegree,20\textdegree,45\textdegree,90\textdegree\}, at CCSD(T) level of theory \cite{purvis1982,raghavachari1989,bartlett2007} with the Karlsruhe def2-TZVPP basis set \cite{weigend2005}. The inner valence electrons of Rb and Cs were replaced with Stuttgart small-core relativistic ECP-28 and ECP-46 effective core potentials (ECPs), respectively \cite{leininger1996}. The excited state energies were computed using EE-EOM-CCSD (no frozen core) theory\cite{stanton1993,bartlett2007} with the same basis and ECPs as for the ground state. All calculations were performed in MOLPRO 2010.1\cite{molpro10_short}.
In Fig. \ref{fig2} (top panels) we show the \textit{ab initio} PECs for ground and excited states for both molecules.
The long-range van der Waals tails of the interaction potentials were constructed from the theory presented in Ref. \cite{2012PhRvA..86c2711B} (Eqs. (9-13), pp. 4). The first- and second-order interaction energies (electrostatic, dispersion, and induction) were evaluated up to 8th order\cite{codelongrange}. 
Final PECs were constructed separately for each angle $\theta$ by fitting a spline to the \textit{ab initio} points and ensuring a smooth transition to the long-range form. Ground state potentials were joined before the barrier, typically at internuclear distances between 20 and 30 Bohr radii, to ensure the correct long-range form. The most important characteristic values of the potential energy curves are given in Table \ref{t_pec}.
\begin{table}[t] %[H] add [H] placement to break table across pages
\caption{\label{t_pec}Characteristic points of potential energy curves for (KRb)$_2$ and (RbCs)$_2$ for selected orientation angles $\theta$. Inner well depth ($V^{\mathrm{in}}_\mathrm{min}$), equilibrium distance $r_\mathrm{eq}$, outer well depth ($V^\mathrm{out}_\mathrm{min}$), barrier height ($h_\mathrm{b}$), and the inner turning point of the outer well ($R_Q$) are given. All values are in atomic units, and $[n]$ stand for $\times 10^n$. }
\begin{ruledtabular}
\begin{tabular}{c|ccccc}
% \multicolumn{3}{c|}{(KRb)$_2$}                   &       \\
(KRb)$_2$ & $V^{\mathrm{in}}_\mathrm{min}$ & $r_\mathrm{eq}$ & $h_\mathrm{b}$  &  $V^\mathrm{out}_\mathrm{min}$  & $R_Q$      \\
\hline
$\theta=0^\circ$  & -1.74 [-3] & 17.21 & 1.95 [-5]       &  -2.13 [-8]          & 102.07     \\
$\theta=20^\circ$ & -2.37 [-3] & 16.11 & 1.76 [-7]       &  -5.68 [-8]          & 62.48     \\
$\theta=25^\circ$ & -2.37 [-3] & 16.11 &               &                    &           \\
$\theta=90^\circ$ & -2.63 [-3] & 10.68 &               &                    &           \\
\hline
(RbCs)$_2$        &      &          &       &                    &            \\
\hline
$\theta=0^\circ$  & -8.45 [-4] & 18.81 &  3.41 [-5]      &  -8.16 [-8]          & 41.59     \\
$\theta=10^\circ$ & -9.65 [-4] & 18.55 &  2.03 [-5]      &  -1.57 [-8]          & 38.32     \\
$\theta=20^\circ$ & -1.32 [-3] & 17.74 & -2.41 [-6]      &  -5.94 [-6]          & 31.87     \\
$\theta=90^\circ$ & -2.96 [-3] & 11.07 &               &                    &           \\
% Lines of table here ending with \\
\end{tabular}
\end{ruledtabular}
\end{table}

Similarly, transition dipole moments for different orientations were constructed from \textit{ab initio} values and smoothly interpolated to the asymptotic values (Fig. \ref{fig2b}).

\subsection{Photoassociation}

Using the PECs computed above, we estimate the production rates of (KRb)$_2$ and (RbCs)$_2$ in the ground electronic state. We assume a two-step process: the first step is using PA to form tetratomic molecules in the first excited singlet electronic state, b$^1A'$, which is followed by the second step, the spontaneous relaxation that transfers the population to the ground electronic state (Fig. \ref{fig1}).
For both molecules, the PA step consists of a single-photon excitation from a pair of diatomics in their ro-vibrational ground state of the ground electronic X${}^1\Sigma^+$ state, red-detuned by $\Delta$ from a pair in which one of the diatomic is in its ground ro-vibrational level of its excited $2^1\Sigma^+$ electronic state.
In case of RbCs molecules, as the two diatomics approach each other, the b${}^1A'$ undergoes a crossing with the a${}^3A''$ and b${}^3A''$ states, which for RbCs corresponds to the single excitation to the $1^3\Pi$ state. Consequently, for RbCs, it is necessary to perform a $3\times3$ spin-orbit calculation mixing the b${}^1A'$, a${}^3A''$ and b${}^3A''$ states. This was performed by computing the $3\times3$ spin-orbit matrix using a three-state multi-configuration self consistent field (MCSCF) \cite{werner1985,knowles1985} calculation at every \textit{ab initio} point. As depicted in Fig.~\ref{fig2}, the spin-orbit coupling impacts the b${}^1A'$ electronic state for separation $R$ smaller than 24 Bohr radii, much smaller than the long-range barrier considered to increase the formation rate of tetratomic molecules, and hence will not affect our results (discussed in the following sections).
% Note that the spin-orbit interaction is only important for the intermolecular distances smaller than 24 \AA), where the diagonal elements of the SO matrix were replaced with the unperturbed EOM-CCSD energies.

%(RbCs)$_2$ molecules in the ro-vibrational state $v$ of the excited electronic state can be expressed as $K^\mathrm{PA}_v = \langle v_\mathrm{rel} \sigma^{\mathrm{PA}}_v \rangle$, where $v_\mathrm{rel}$ is the relative velocity of the interacting pair of RbCs dimers, and $\sigma^{\mathrm{PA}}_v$ is the PA cross section. 
The PA rate coefficient can be expressed as\cite{PhysRevLett.58.2420} 
% $K^\mathrm{PA}_b = \langle v_\mathrm{rel} \sigma^{\mathrm{PA}}_b \rangle$, 
\begin{equation}
K^\mathrm{PA}_b = \langle v_\mathrm{rel} \sigma^{\mathrm{PA}}_b \rangle \; ,
\end{equation}
where $v_\mathrm{rel}$ is the relative velocity of the colliding dimers, and $\sigma^{\mathrm{PA}}_b$ is the PA cross section for forming the complex in the ro-vibrational state $|b\rangle = |v,\ell \rangle$ of the selected excited electronic state. Here, $v$ and $\ell$ are the vibrational and total orbital quantum number, respectively, and the brackets $\langle \cdots \rangle$ indicate averaging over a Maxwellian distribution of initial velocities at the mean gas temperature $T$.
At ultracold temperatures ($s$-wave regime, $\ell=0$) and assuming a low PA laser intensity $I$, the optimal PA rate coefficient $K_{b}^{\mathrm{PA}}(T,I)$ is given by\cite{1994PhRvL..73.1352N,2006JPhB...39S.965J,2016PhRvA..94a3407G,juarros2006one}:
\begin{equation}
 K_{b}^{\mathrm{PA}}(T,I) = \frac{8 \pi^3}{h^2 c} \frac{I}{Q_T} e^{-1/2} |\langle b | D(R) | \varepsilon, \ell=0 \rangle |^2 ,
 \label{eq:PA3}
\end{equation}
% $\mu$(Rb$_2$Cs$_2$) = 198528.2712429 amu
where $Q_T = (2 \pi \mu k_B T / h^2)^{3/2}$ is the translational partition function, $\mu$ is the reduced mass of the pair, $\varepsilon$ is the asymptotic relative collision energy, and $D(R)$ is the transition dipole moment between the initial continuum state $|\varepsilon, \ell=0 \rangle$ of the colliding pair and the target bound state $|b \rangle = | v, \ell=1 \rangle$ of the complex in an electronic excited state. Here, $k_B$ and $h$ are Boltzmann and Planck constants, and $c$ is the speed of light in vacuum.

\subsection{Formation rate of ground state molecules}

As described in Section~\ref{sec:approximations}, we assume that the most probable decay takes place in the same geometry as the initial PA step. We account for other final geometries by a coefficient $\eta$ discussed at the end of this section.

The rate coefficient $K_{b,b'}(T,I)$, for the formation of tetratomic molecules in the ro-vibrational state $|b' \rangle = |v',\ell' \rangle$ of their electronic ground state by spontaneous emission from the state $|b \rangle$ of an excited electronic state, is given by\cite{2006JPhB...39S.965J,2016PhRvA..94a3407G}
\begin{equation}
  K_{b,b'}(T,I) = K_{b}^{\mathrm{PA}}(T,I) r_{b,b'}^{(\alpha)} \;, 
  \label{eq:K}
\end{equation}
where $r_{b,b'}^{(\alpha)}$ is the branching ratio for the spontaneous radiative emission for the branch $\alpha = \{R, Q, P\}$. The branching ratio can be expressed in terms of Einstein $A$ coefficients weighted by H\"onl-London factors \cite{Herzberg}:
\begin{equation}
 r_{b,b'}^{(\alpha)} = {A_{b,b'}^{(\alpha)}} / \left[{\sum_{b'} A_{b,b'}^{(\alpha)} + \int A_{b}^{(\alpha)} (\varepsilon') \mathrm{d} \varepsilon'} \right] \, .
 \label{eq:r}
\end{equation}
The Einstein $A$ coefficients for bound-bound and bound-free transitions, $A_{b, b'}^{(\alpha)}$ and $A_{b}^{(\alpha)}(\mathrm{\varepsilon})$, respectively, are
\begin{eqnarray}
  A_{b,b'}^{(\alpha)} & = & \frac{4 e^2 [\omega_{b,b'}^{(\alpha)}]^3}{3 \hbar c^3}  W_{J_{b'}}^{(\alpha)} |\langle b' | D(R) | b \rangle|^2 \nonumber \\
  A_{b}^{(\alpha)}(\mathrm{\varepsilon}) & = & \frac{4 e^2 \omega_{b}^{(\alpha)}(\varepsilon)^3}{3 \hbar c^3} W_{J_{b'}}^{(\alpha)} |\langle \varepsilon \ell | D(R) | b \rangle|^2 ,
 \label{eq:Abf}
\end{eqnarray}
where $\hbar \omega_{b,b'}^{(\alpha)} = |E_{b}-E_{b'}|$ and $\hbar \omega_{b}^{(\alpha)}(\varepsilon) = |E_{b}-E_{\varepsilon \ell}|$, are the frequencies for the bound-bound  and bound-free transition, respectively, while $D(R)$ is the dipole transition moment between the two electronic states and $W_{J_{b'}}^{(\alpha)}$ are the H\"onl-London factors for the branch $\alpha$\cite{Herzberg}. 
The largest contribution to the PA rate coefficients comes from small values of $\varepsilon$, validating the use of the dipole transition moment in free-bound transitions in ultracold regime.

In evaluating the above expressions, we included all bound-bound transitions and bound-free transitions for energies up to $\varepsilon/k_B=1$ mK, as in Refs. \cite{2006JPhB...39S.965J,2016PhRvA..94a3407G}. We assumed that all optical transitions are $Q$-branch transitions for $\ell = \ell' = 0$. This approximation has minimal impact on the production rate coefficients because $|v \rangle \approx |v,\ell \rangle$ for both $v$ and $v'$ states. Thus, we simplify the notation by dropping the index $\alpha$, \textit{i.e.}, $A_{b,b'} \equiv A_{b,b'}^{(\alpha)}$.

The comparison between different pair alignments and different molecules is simplified if the rate coefficients are expressed in terms of binding energies $E_{b}$ and $E_{b'}$ of the ro-vibrational levels $b$ and $b'$. In places where the energy notation is used, we replaced the indices $b$ and $b'$ in the rate coefficients with $E_{b}$ and $E_{b'}$, respectively. Moreover, the binding energy $E_{b}$ can expressed in terms of the PA laser detuning $\Delta$, selected such that $E_{b}=h\Delta$ (see Fig. \ref{fig1}). In this notation, the PA rate coefficient and formation rate coefficient are expressed as $K_{\Delta}^{\mathrm{PA}}(T,I)$ and $K_{\Delta,E_{b'}}(T,I)$, and the Einstein $A$ coefficients as $A_{\Delta,E_{b'}}$ and $A_{\Delta}(\varepsilon)$, respectively.

The matrix elements in Eqs. (\ref{eq:PA3}) and (\ref{eq:Abf}) were evaluated numerically by diagonalizing the radial Schr\"odinger equation for electronic motion in $\theta$-dependent interaction potentials (Fig. \ref{fig2}) and using the corresponding transition dipole moments (Fig. \ref{fig2b}). The wave functions and bound state energies were calculated numerically using mapped Fourier grid method (MFGR)\cite{1999JChPh.110.9865K}, to simultaneously obtain bound and quasi-discretized continuum spectrum.
The MFGR calculation was performed independently for different relative orientations  $\theta$ of the pairs (no couplings between different potential curves), assuming a variable grid step size determined by the total box size ($R_\mathrm{max}=5000$ $a_0$), and the mapping potential numerically evaluated from the local momentum. The accuracy of the wave functions in the highly oscillatory short-range region was ensured by requiring at least 20 points per a single period of the wave function.
The continuum wave functions were found to be in excellent agreement with a calculation performed using the renormalized Numerov method\cite{ren_numerov} for the energies greater than 500 nK.

Finally, we account for the change of geometry due to the rotation of the excited molecule arising from the photon absorption.
Omitting the index $\alpha$, we can rewrite the branching ratio $r^{(\alpha )}_{b,b'}$ in Eq.(\ref{eq:r}) as 
$r_{b,b'} = A_{b,b'} \tau_b $ where $\tau_b$ is simply the lifetime of the state $b$ due to
decay. In principle, this lifetime includes all possible final geometries. As before, we can use the detuning $\Delta$ to
label the bound level $b$, and write $r_{\Delta,b'} = A_{\Delta,b'} \tau_\Delta $. We now account for the rotation
of the excited molecule using an averaged $A$-coefficient $\bar{A}_{\Delta, b'}$, so that we can introduce the averaged branching ratio $\bar{r}_{\Delta,b'} = \bar{A}_{\Delta,b'} \tau_\Delta$.

We define a coefficient $\eta$ as follows
\begin{equation}
   \bar{A}_{\Delta, b'} =  \frac{1}{4\pi} \int d\Omega A_{\Delta, b'} (\Omega) \equiv \eta A_{\Delta, b'} (0) ,
   \label{eq:eta_def}
\end{equation}
where the solid angle $\Omega$ stands for all possible orientations, and $A_{\Delta, b'} (0)$ is the Einstein $A$-coefficient for the fully aligned case with $\theta_1=\theta_2=\phi=0$. As we will discuss in the next section on results, $A_{\Delta, b'} (\Omega)$ is sizable only when a long-range barrier together with a long-range well exist. All other configurations lead to very poor Franck-Condon overlaps, and much smaller $A$-coefficients. We can therefore rewrite $\bar{r}_{\Delta,b'} = \eta r_{\Delta,b'}  (0)$, where $r_{\Delta,b'}  (0)$ is the ratio corresponding to fully aligned case with $A_{\Delta, b'} (0)$, so that $K_{b,b'}(T,I)$ in Eq.(\ref{eq:K}) becomes
\begin{eqnarray}
    \bar{K}_{\Delta,E_{b'}}(T,I) &= & \eta K_{\Delta}^{\mathrm{PA}}(T,I) r_{\Delta,b'}  (0) \;, \\
    & =& \eta K_{\Delta,E_{b'}}(T,I,0)\; ,
  \label{eq:K_eta}
\end{eqnarray}
where $ K_{\Delta,E_{b'}}(T,I,0)$ is the rate coefficient for the fully aligned case.
Thus $\bar{K}_{\Delta,E_{b'}}(T,I)$ is the averaged rate coefficient for the formation of ultracold bound tetratomic (or polyatomic) molecules.

\section{\label{s4}Results and discussion} %%%

\subsection{(KRb)$_2$ and (RbCs)$_2$ tetratomic molecules}

\subsubsection{Photoassociation rates and transition probablities}

\begin{figure}[t]
 \centering
 \includegraphics[clip,width=1.0\linewidth]{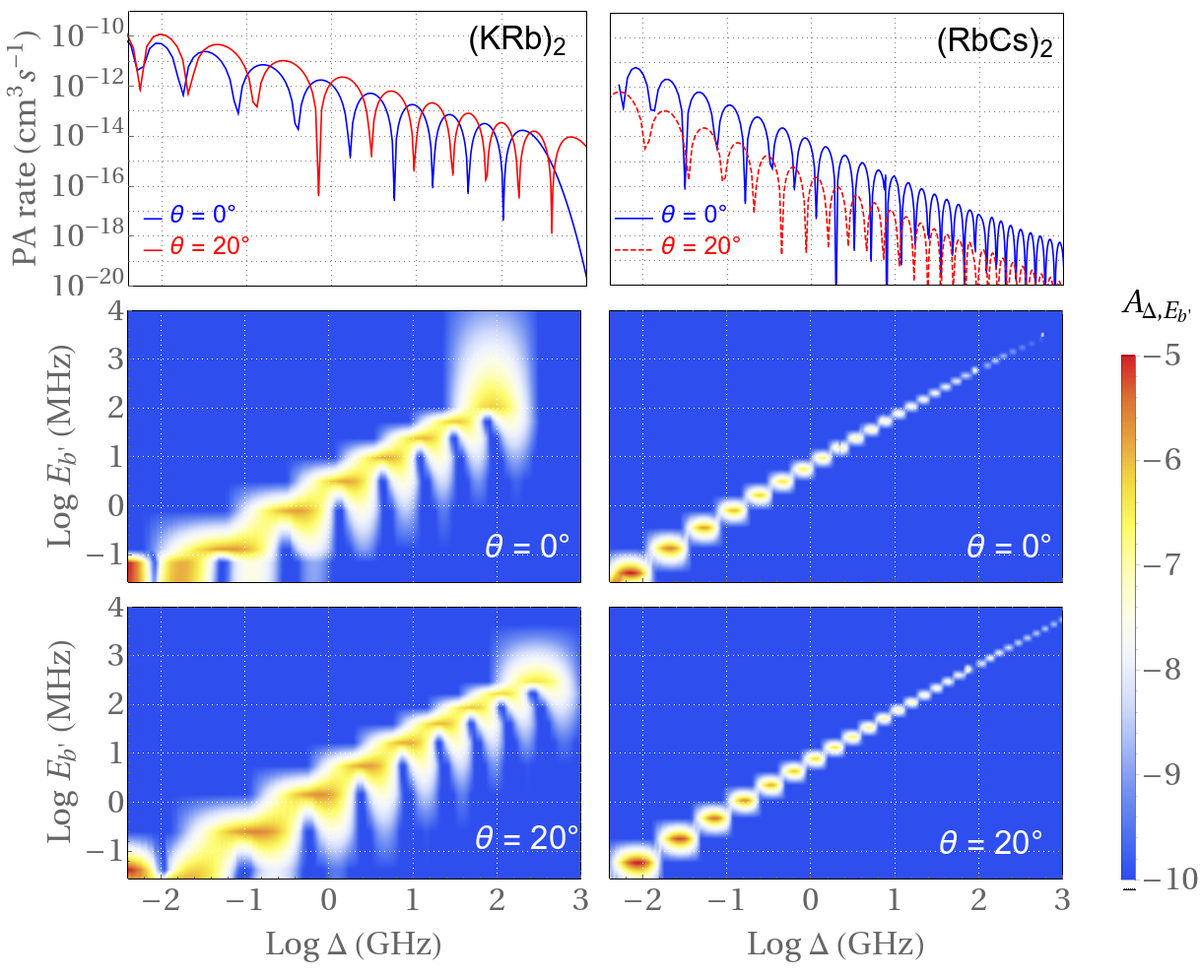}
 \caption{(Color online) \textit{Top panels}: PA rate coefficients $K_{\Delta}^{\mathrm{PA}}(T_0,I_0)$ for the production of (KRb)$_2$ (left) and (RbCs)$_2$ (right) complexes, evaluated for $\theta=0$\textdegree \, and 20\textdegree. \newline
 \textit{Middle \& bottom panels}: Einstein $A$ coefficients, $A_{\Delta,E_{b'}}$, for the spontaneous emission (bound-bound) from the excited $b^1\mathrm{A}'$ state to the ground state, shown in energy-notation. Both quantities are expressed as base ten logarithms. Note that binding energies $E_b=h\Delta$ and $E_{b'}=h\nu{_b'}$ given in terms of frequency units.
 }
 \label{fig3}
\end{figure}

Using Eqs. (\ref{eq:PA3}) and (\ref{eq:Abf}), we calculated the PA rates and Einstein $A$ coefficients for angles $\theta=0\dots45$\textdegree \; for both molecules. The most interesting relative alignments that characterize two different physical regimes based on the barrier height are obtained for $\theta=0^\circ$ (highest barrier) and $\theta=\theta_c \approx 20^\circ$, where $\theta_c$ is the angle for which the barrier dips below the asymptotic threshold (see Fig. \ref{fig2}).

For the two relative orientations, in Fig. \ref{fig3} (top panels) we show the PA rate coefficients $K_{\Delta}^{\mathrm{PA}}(T_0,I_0)$, calculated assuming the PA laser intensity $I_0=1$ kW/cm$^2$ and average gas temperature $T_0=1$ $\mu$K. %and Einstein $A$ coefficients for the b$^1\mathrm{A}'-\mathrm{X}^1\mathrm{A}'$ electronic transition.
%The contributions from the bound-free transitions (Eq. (\ref{eq:Abf}) are a second-order contribution and can be neglected.
For (KRb)$_2$ (Fig. \ref{fig3}, top left), the PA rates vary between about 10$^{-10}$ cm$^3$/s and $10^{-14}$ cm$^3$/s for the detuning $\Delta$ set to about 10 MHz and $10^2$ to $10^3$ GHz, respectively. For larger detunings, the PA rate rapidly drops to negligible values. The relative alignment of the dimers does not strongly affect the magnitude of the PA rates even though the oscillation period is extended for larger angles $\theta$ as the binding energies of the target states are shifted. Notably, for $\theta=20^\circ$, the PA rates are as high as $10^{-14}$ cm$^3$/s for the largest detunings considered, $\Delta>2.7\times10^3$ GHz, suggesting that deeper ro-vibrational levels in the excited state could be populated for $\theta \approx \theta_c$, in contrast to $\theta=0^\circ$.
% A similar conclusion was reached for the PA of atom-molecule pairs into trimers\cite{}. 
For (RbCs)$_2$ (Fig. \ref{fig3}, top right), we computed the PA rates to be at least an order of magnitude smaller, varying between about 10$^{-11}$  and $10^{-18}$ cm$^3$/s, as well as to decrease more rapidly with the detuning. Moreover, the PA rates are about an order of magnitude larger for the collinear alignment, $\theta=0^\circ$, than for $\theta=\theta_c$, suggesting that the potential barrier has a more significant role in this system.

The results suggest that the long-range excited tetratomic molecules, with intermolecular distances $R>50$ Bohr, could be photoassociated regardless of the orientation $\theta$.
As expected, the PA rate coefficients are the largest for the formation of weakly bound ($\Delta<10$ GHz) excited complexes. The cutoff value of the detuning $\Delta$, beyond which the PA rates are negligible, increases with the relative orientation $\theta$, even though smaller angles yield larger overall PA rates. 

% The production rates of ground state tetramers depend on the PA rates and the Einstein $A$ coefficients (Eqs. (\ref{eq:K}-\ref{eq:Abf})). 
In order to determine the production rates of ground state tetramers, we calculated the Einstein $A$ coefficients, $A_{b,b'}^{(\alpha)}$, for all pairs of ro-vibrational levels $|b\rangle$ and $|b'\rangle$ in the ground and excited electronic state, for all considered relative orientations $\theta$.
Again, the results for two relative alignments $\theta=0^\circ$ and $\theta \approx \theta_c=20^\circ$ are given in Fig. \ref{fig3} (middle and bottom panels).
As expected, the transition probability is the largest on the diagonal, where the wave functions of the pair of ro-vibrational states involved in the transition have similar nodal structure and extent, leading to large Franck-Condon overlaps.

In the case of (KRb)$_2$ (Fig. \ref{fig3}, left column), for the collinear alignment ($\theta=0^\circ$), the effect of the long-range barrier in the ground state manifests as an extended ``diffuse crest'' on top of the last oscillation on the diagonal, around $\Delta=10^2$ GHz, where the Einstein $A$ coefficients remain significant for the transitions to the ground-state ro-vibrational levels with binding energies as large as $E_b' = 10^4$ MHz. The enhanced transition probabilities are due to the increased Franck-Condon overlap between the excited state and the outer turning point of the long-range well in the ground state.
For $\theta=20^\circ$, the Einstein $A$ coefficients remain significant for even larger detunings, up to $\Delta=10^3$ GHz, corresponding to the ro-vibrational levels as deep as $E_b' = 10^3$ MHz. The difference is due to the fact that the outer potential well becomes deeper and more extended as the barrier height is reduced (see Fig. \ref{fig2}, bottom panels). Finally, for $\Delta<10$ GHz, the Einstein $A$ coefficients for the two alignments become very similar.

\begin{figure}[t]
 \centering
 \includegraphics[clip,width=1.05\linewidth]{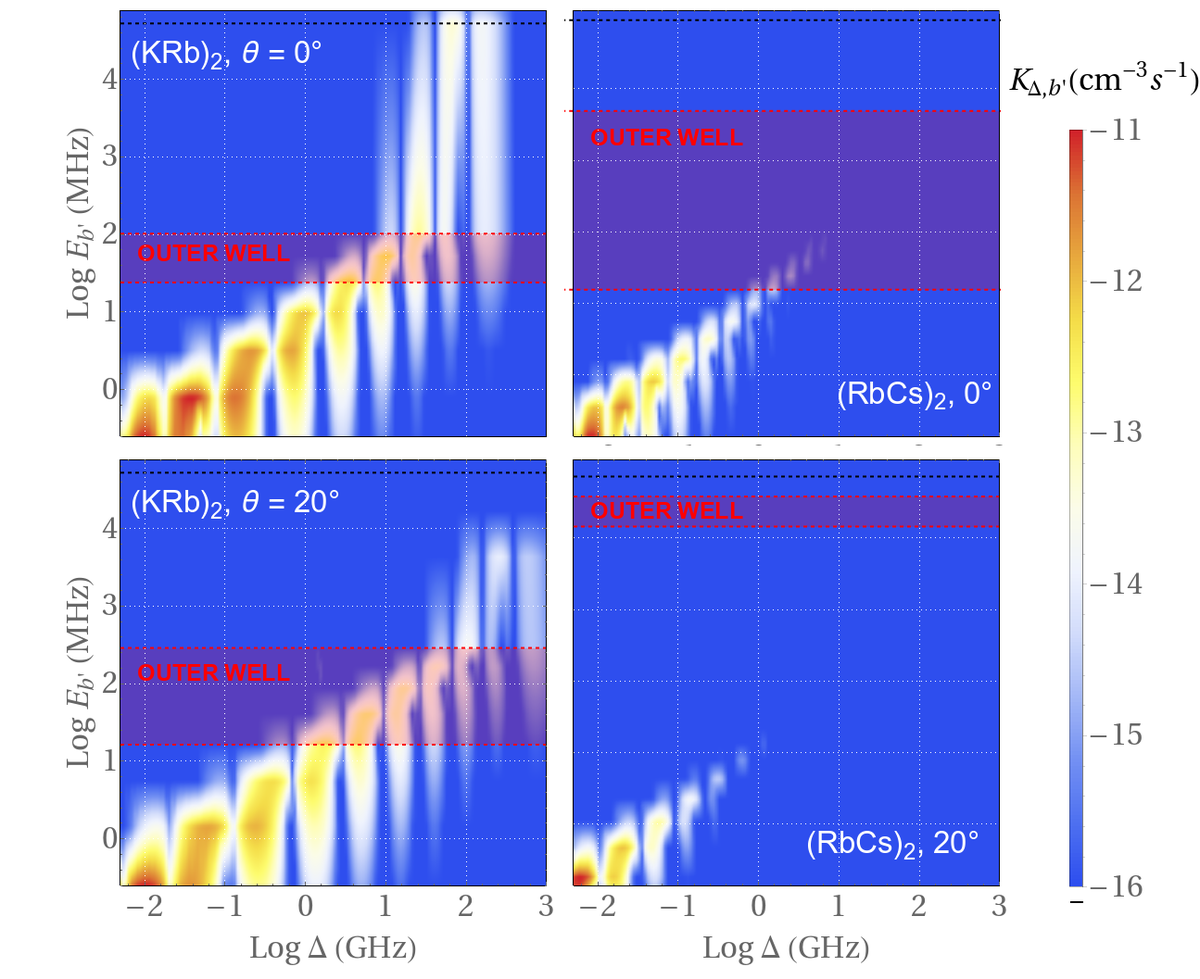}
 \caption{(Color online) Production rate of ground state molecules as a function of PA laser detuning $\Delta$ and binding energy $E_{b'}$ shown for $\theta=0^\circ$ (left) and $\theta=20^\circ$ (right). All axes are in log scale.
 }
 \label{fig4}
\end{figure}

\subsubsection{Ground-state production rate coefficients}

The rate coefficients $K_{\Delta,E_{b'}}(T_0,I_0)$, computed using Eq. (\ref{eq:K}) for the production of (KRb)$_2$ and (RbCs)$_2$ tetratomic molecules in the electronic ground state, are shown in Fig. \ref{fig4} for the PA laser detuning $\Delta$ and the binding energy $E_{b'}$.
The figure illustrates the main result of the study and warrants a more detailed discussion.
As above, for both molecules, we analyzed the two relative alignments of the diatomics, given by $\theta=0^\circ$ (top panels) and $\theta=20^\circ$ (bottom panels). Note that the critical angle $\theta_c$, for which the potential barrier dips below the kinetic interaction energy of the pair, is $\theta_c=20.6^\circ$ for (KRb)$_2$ vs $\theta_c=19^\circ$ for (RbCs)$_2$.

The common feature in all cases is the diagonal distribution of distinct vertically-elongated lobes, where $K_{\Delta,E_{b'}} > 10^{-16}$ cm$^{-3}$s$^{-1}$. 
The lobes correspond to the bound ro-vibrational levels in the electronic ground state of the particular case close to the dissociation limit that have the binding energy $E_{b'}$ between 0.5 and $10^{5}$ MHz. The binding energies corresponding to the bound ro-vibrational states in the outer well are marked as the purple shaded area; the energies below it correspond to the weakly bound near-dissociation levels present regardless of the barrier, and the energies above it to the deeply bound levels in the inner well.

\begin{figure}[t]
 \centering
 \includegraphics[clip,width=1.0\linewidth]{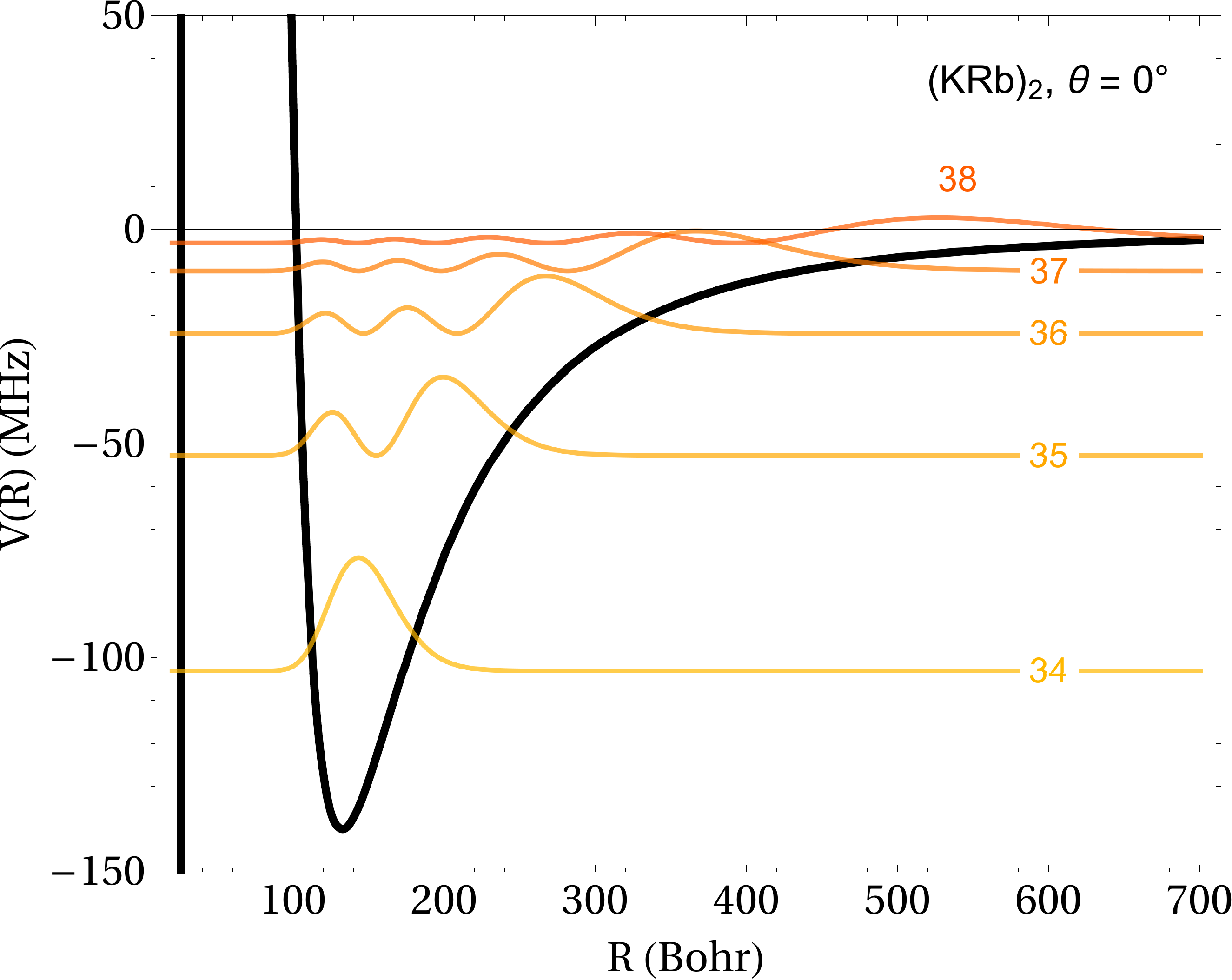}
\caption{(Color online) Probability density of the five lowest vibrational levels ($v'=34 \dots 38$, indicated on the wave functions) in the outer potential well of (KRb)$_2$ molecule at collinear alignment $(\theta=0^\circ)$. The colors correspond to the production rates given in Fig. \ref{fig4}.
 }
 \label{fig_wfs_krb-0}
\end{figure}

For (KRb)$_2$ with $\theta=0^\circ$ (Fig. \ref{fig4}, top left), we count a total of nine lobes, four of which correspond to the weakly-bound levels, three at least partly overlap with the outer well (and likely correspond to the outer well bound levels), and two correspond to the deeply bound states in the inner potential well. 
For $\theta=20^\circ$ (Fig. \ref{fig4}, bottom left), the outer well is deeper and more extended and supports up to seven bound levels. However, no inner well levels can be populated with a significant probability. 
In Fig. \ref{fig_wfs_krb-0}, we illustrate the probability density $|\psi_{v'}(R)|^2$ of the ro-vibrational wave functions $\psi_{v'}(R)$ in the outer well ($v'=34\dots38$) for collinear (KRb)$_2$ $(\theta=0^\circ)$.

In the case of (RbCs)$_2$, the situation is somewhat less favorable for molecule formation: only the four highest-energy bound levels in the outer well are accessible for $\theta=0^\circ$ (Fig. \ref{fig4}, top right) and none for $\theta=20^\circ$ (Fig. \ref{fig4}, bottom right). This is the case because the outer well in (RbCs)$_2$ is much deeper (see Fig. \ref{fig2}), and bound levels more localized but more difficult to populate, according to Franck-Condon principle.
For both molecules, in case of collinear diatomics ($\theta=0^\circ$), the production rates are about hundred times greater than for $\theta=20^\circ$. The dependence on the orientation is the most pronounced for small binding energies, $E_{b'}<50$ MHz in (KRb)$_2$ as well as for $10<E_{b'}<100$ MHz in (RbCs)$_2$, where the rates are negligible for $\theta=20^\circ$.
Once the tetratomic molecules are formed in the ground electronic state, they will spontaneously relax to the absolute ground state via a radiative cascade, providing they are not destroyed by collisions or lost from the trap through other mechanisms. 

In order to estimate the experimental feasibility of the proposed approach, we can approximate the number of tetratomic complexes formed per second in a trapped ultracold molecular gas as 
\begin{equation}
 N = \eta_g\eta n^2 V K_{\Delta,E_{b'}}(T_0,I_0) \; ,
\end{equation}
where $n$ the gas density, $V$ is the total volume of the gas illuminated by the PA laser of intensity $I_0$, $\eta$ is the fraction decaying into the right geometry as defined in Eq,(\ref{eq:eta_def}), 
and $\eta_g$ is the fraction of initially correctly aligned pairs at the mean gas temperature $T_0$ (before the PA process).
Based on the existing experiments with KRb and RbCs, we take the molecular gas density and the PA volume to be $n = 10^{12}$ cm$^{-3}$, and $V=1$ mm$^3$, respectively. We also assume that the polar molecules are aligned before undergoing the PA, and set 
$\eta_g =1$.

\begin{figure}[t]
 \centering
 \includegraphics[clip,width=1.0\linewidth]{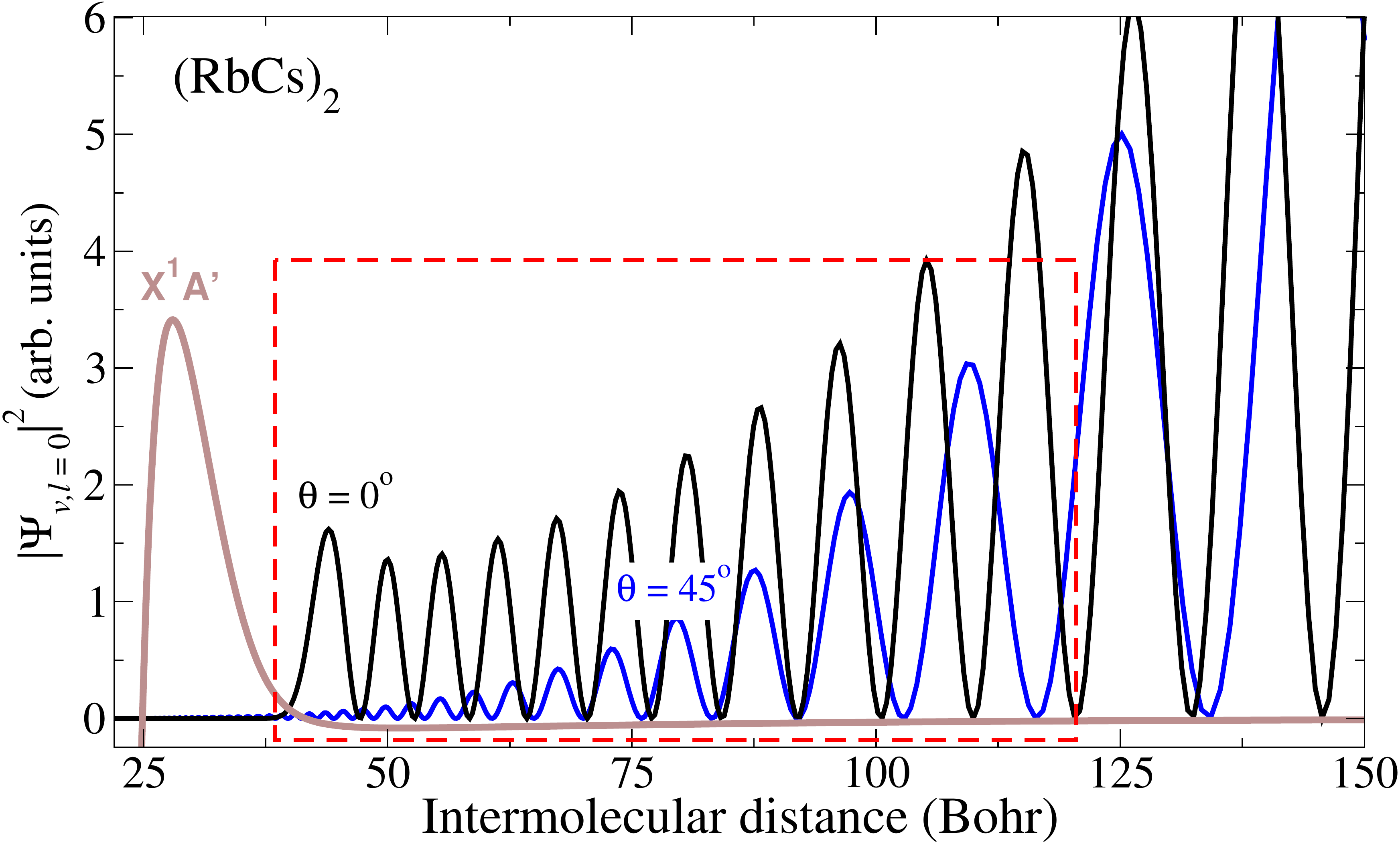}
\caption{(Color online) Comparison of the probability density of the ground state wave function $|\psi_g|^2$ for different orientations ($\theta_1=\theta_2=\theta$ and $\phi=0$ corresponding to geometries with ($\theta = 0$) and without ($\theta = 45^{o}$) long-range barrier/wells for RbCs+RbCs. The dashed box approximately indicates the outer well and the region where the transitions take place. The $|\psi_g|^2$ in that region will be about 5 to 100 times larger for the orientations for which the long-range barrier is present.
 }
 \label{fig:overlap_w/o_barrier}
\end{figure}

Before describing the numbers of molecules produced, we first discuss the value of $\eta$. 
From Eq.(\ref{eq:eta_def}), we can write
\begin{equation}
   \eta =  \frac{1}{4\pi} \frac{ \int d\Omega A_{\Delta, b'} (\Omega)}{A_{\Delta, b'} (0)} .
\end{equation}
As shown in Fig.~\ref{fig:overlap_w/o_barrier}, the probability density of the ground state wave function $|\psi_g|^2$
in the vicinity of the outer well (roughly 40-70 $a_0$ for RbCs+RbCs) is 50-100 times larger in the presence of a long-range barrier/well than without. Hence, the $A$-coefficient will be 50-100 times larger than for cases without long-range barriers 
and wells. Furthermore, the $A$-coefficient does not change significantly as long as the long-range barrier and an outer well exist, and so in these cases we can assume $A_{\Delta, b'} (\Omega) \approx A_{\Delta, b'}(0)$. Consequently, using the long-range expansion (see Eq.(\ref{eq:Vlongrange}) in the next section), we can estimate $\eta$ by integrating over all geometries for which a long-range barrier and an outer well exist:
\begin{equation}
   \eta \approx \frac{1}{4\pi} \int d\phi d(\cos \theta_1) d(\cos \theta_2) f(\theta_1,\theta_2,\phi) ,
\end{equation}
where $f=1$ if a long-range barrier and well exist, and $f=0$ otherwise. 
For the molecules considered, the function $f$ is non-zero in a region roughly defined by $\phi \in [0,20^{o}]$, and $\theta_1+\theta_2 < 50^{o}$. This is illustrated for KRb+KRb in Fig. 2 of Ref. \cite{2012PhRvL.109h3003B}, where the barrier height is given as a function of the angles $\theta_1$, $\theta_2$, and $\phi$.

% \begin{equation}
%   \eta = \frac{1}{4} \int_{0}^{\pi} \int_{0}^{\pi} \sin \theta_1 \sin \theta_2 f(\theta_1,\theta_2) d\theta_1 d\theta_2 \; ,
% \end{equation}
% where $f(\theta_1,\theta_2) = 1$ for $\theta_1+\theta_2 < 50^\circ$ and zero otherwise. The expression evaluates to $\eta = 0.0057$, implying $\eta \approx 0.6\%$. 

% \textcolor{blue}{Comment - the integral is evaluated analytically to:}
% \begin{equation}
%  \eta = \frac{1}{144} \left[36 - 5\pi \cos \left(\frac{2\pi}{9}\right) 
%       - 36\sin\left(\frac{2\pi}{9} \right) \right]
% \end{equation}
% \textcolor{blue}{End comment}

Using the above parameter ranges, the integral evaluates to $\eta=0.0057$, indicating that about 0.6\% of tetramers will decay to a bound ro-vibrational state in the outer well. 
If we assume a conservative estimate of the production rates, where $\eta=0.005$ and $K_{\Delta,E_{b'}}(T_0,I_0) \approx 10^{-14}$ cm$^{-3}$, a total of $N=5 \times 10^4$ molecules per second would be produced. These numbers compare favorably with estimates of molecular production rates in ongoing experiments with ultracold molecules. 

We note that if the molecules are in a cloud (and not a restrictive trapping geometry), although perfectly aligned by an external field ({\it i.e.} $\theta_1=\theta_2=\theta$ with $\phi=0$), they can approach each other from any direction, leading to all possible values $\theta$. Since only $\theta<20^{o}$ leads to long-range wells, averaging over all $\theta$ gives $\eta_g \approx 0.04$, for which we get $N=2000$ molecules per second.

A similar order-of-magnitude analysis for (RbCs)$_2$ yields $N \approx 10^{4}$ molecules per second (with $\eta_g=1$) for four ro-vibrational states in the outer well of the complex if the pairs are aligned, and $N \approx 400$ s$^{-1}$ in the absence of pair alignment ($\eta_g = 0.04$).

Finally, in the case of KRb, our calculation for $\theta=0^\circ$ suggests that deeply bound states of a tetratomic in a short-range well could be produced if the PA laser is red-detuned by more than 60 GHz from the dissociation limit of the excited ground state. While the computed production rates, $K_{\Delta,E_{b'}}(T_0,I_0) < 10^{-14}$ cm$^{-3}$ would be on the low side, the produced tetratomics would be bound by more than 40 GHz, making them stable against collisional destruction in the trap.

\subsection{Other systems}

The results obtained for KRb and RbCs can be generalized to other tetratomics composed of two dipolar molecules, as well as to larger polyatomic complexes, if their electronic structure  supports a long-range potential well. As we discuss below, this can be accomplished by aligning polar molecules with appropriate properties with preferentially collinear orientations ($\theta<\theta_c$) . The existence of a barrier can be inferred from the long-range intermolecular potential expansion
\begin{equation}
  V(R,\theta_1,\theta_2,\phi) \xrightarrow{R \rightarrow \infty} - \sum_{n}\frac{W_n(\theta_1,\theta_2,\phi)}{R^n} \; ,
  \label{eq:Vlongrange}
\end{equation}
where the angles $\theta_1$,$\theta_2$, and $\phi$ define the relative orientation of the pair in a general case, and the functions $W_n$ contain electrostatic and/or dispersion contributions \cite{mulder1979anisotropy,2011JChPh.135x4307B}.

For two aligned identical molecules in a plane, \textit{i.e.}, $\theta_1 = \theta_2 \equiv \theta$ and $\phi=0$, Eq. (\ref{eq:Vlongrange}) simplifies to\cite{2012PhRvL.109h3003B}
\begin{equation}
  V(R,\theta) \simeq -\frac{W_3}{R^3} - \frac{W_5}{R^5} \; .
  \label{eq:Vlongrange2}
\end{equation}
For collinear molecules ($\theta=0^\circ$), the terms $W_3$ and $W_5$ become $W_3=2 \mathcal{D}^2$ and $W_5=-6\mathcal{Q}^2 + 4\mathcal{D}\mathcal{Q} \simeq -6\mathcal{Q}^2$, 
where $\mathcal{D}$, $\mathcal{Q}$, and $\mathcal{O}$ are the electrostatic dipole, quadrupole, and octupole moments, respectively. 

Consequently, if the ratio $|\mathcal{D}/ \mathcal{Q} | \ll 1$, a long-range barrier and potential well can be present and sufficiently deep (on the order of several Kelvin for the studied alkali metals dimers) to support bound states. This occurs when the long-range attractive $R^{-3}$ dipole-dipole interaction is overtaken by the shorter-range repulsive $R^{-5}$ quadrupole-quadrupole interaction at shorter intermolecular distances where the chemical bonding does not yet dominate.

By setting $V=0$ in Eq. (\ref{eq:Vlongrange2}) and solving for the intermolecular distance, we obtain $R_\mathcal{Q} \simeq \sqrt{3\mathcal{Q}^2/\mathcal{D}^2}$, the distance where the $R^{-5}$ quadrupole repulsive interaction overtakes $R^{-3}$ dipolar attraction. 
Similarly, we can estimate the intermolecular distance where the short-range attraction becomes dominant as $R_\mathrm{sr} = -W_6 / W_5$, where $W_6$ is the collinear orientation term\cite{2012PhRvL.109h3003B}. For the collinear molecules ($\theta=0$), the $W_6$ term simplifies to $W_6 = C_{6,0} + 4\left( C_{6,1} + C_{6,2} \right) \simeq C_{6,0}$, to the leading term (the $C_{n,i}$ coefficients include dispersion and induction contributions \cite{2011JChPh.135x4307B,2012PhRvA..86c2711B,2012PhRvL.109h3003B}). 
If $R_\mathcal{Q}$ is outside the region where the bonds are strongly perturbed (i.e., $R_\mathrm{sr} \ll R_\mathcal{Q}$) and higher terms in Eq. (\ref{eq:Vlongrange}) can be neglected, we expect the barrier to be present and the long-range bond between the two polar molecules to be possible.

The $C_{6,0}$ term can be approximated using \cite{2010JChPh.132x4305Z}
\begin{equation}
    C_{6,0} \simeq \frac{3}{2} \frac{\alpha_1 \alpha_2 \Delta E_1 \Delta E_2}{(\Delta E_1 + \Delta E_2)} \;,
    \label{eq:C60-general}
\end{equation}
where $\alpha_i$ and $\Delta E_i$ are the static dipole polarizability and the excitation energy to the first electronic excited state of molecule $i$, respectively. For identical molecules, Eq.(\ref{eq:C60-general}) 
reduces to
\begin{equation}
    C_{6,0} \simeq \frac{3}{4} \alpha^2 \Delta E \; .
    \label{eq:C60-general2}
\end{equation}

\begin{table} %[H] add [H] placement to break table across pages
\caption{\label{t1}\textit{ab initio} values of the electric dipole ($\mathcal{D}$), quadrupole ($\mathcal{Q}$), and octupole ($\mathcal{O}$) electrostatic moments (defined from the center-of-mass), long-range expansion coefficient $C_{6,0}$, and outer turning point $R_\mathcal{Q}$ for sample molecules likely to have an outer well in the ground electronic state. All values are in atomic units. The values for diatomics were adopted from Ref.\cite{2012PhRvA..86c2711B}.}
\begin{ruledtabular}
\begin{tabular}{c|cccc|c}
Molecule  & \multicolumn{4}{c|}{\textit{Ab initio} calculation}   &       \\ 
Diatomic & $\mathcal{D}$  & $\mathcal{Q}$ & $\mathcal{O}$ & $C_{6,0}$ & $R_\mathcal{Q}$ \\
\hline
LiNa  & 0.20    &  10.07   &  -47.33  &  3289    &    95  \\
KRb   & 0.23    &  16.99   &   -3.16  & 13540    &   126   \\
KCs   & 0.75    &  13.00   & -105.70  & 17260    &    38  \\
NaK   & 1.12    &  10.56   &  -26.54  &  7777    &    19  \\
RbCs  & 0.55    &  14.19   &   -5.39  & 19210    &    60  \\
\hline
Triatomic & & & & & \\
\hline
LiONa    & 0.800  & 19.948  &  -69.046  & 238   &  43.2  \\
NaOK     & 0.597  & 24.088  &  -73.389  & 465   &  69.9  \\
KORb     & 0.005  & 23.527  & -115.538  & 571   &  8150  \\
RbOCs    & 0.612  & 19.063  &   89.463  & 732   &  53.9  \\
\hline
Tetraatomic & & & & & \\
\hline
HCCCl    & 0.193  &  5.214  &   28.284  & 315   &  46.8  \\
HCCF     & 0.254  &  3.082  &   23.789  & 113   &  21.0  \\
HCCBr    & 0.128  &  6.127  &   38.711  & 424   &  82.9  \\
HCCI     & 0.004  &  7.071  &   51.056  & 608   &  3062  \\
% Lines of table here ending with \\
\end{tabular}
\end{ruledtabular}
\end{table}

\begin{figure}[t]
 \centering
 \includegraphics[clip,width=1.0\linewidth]{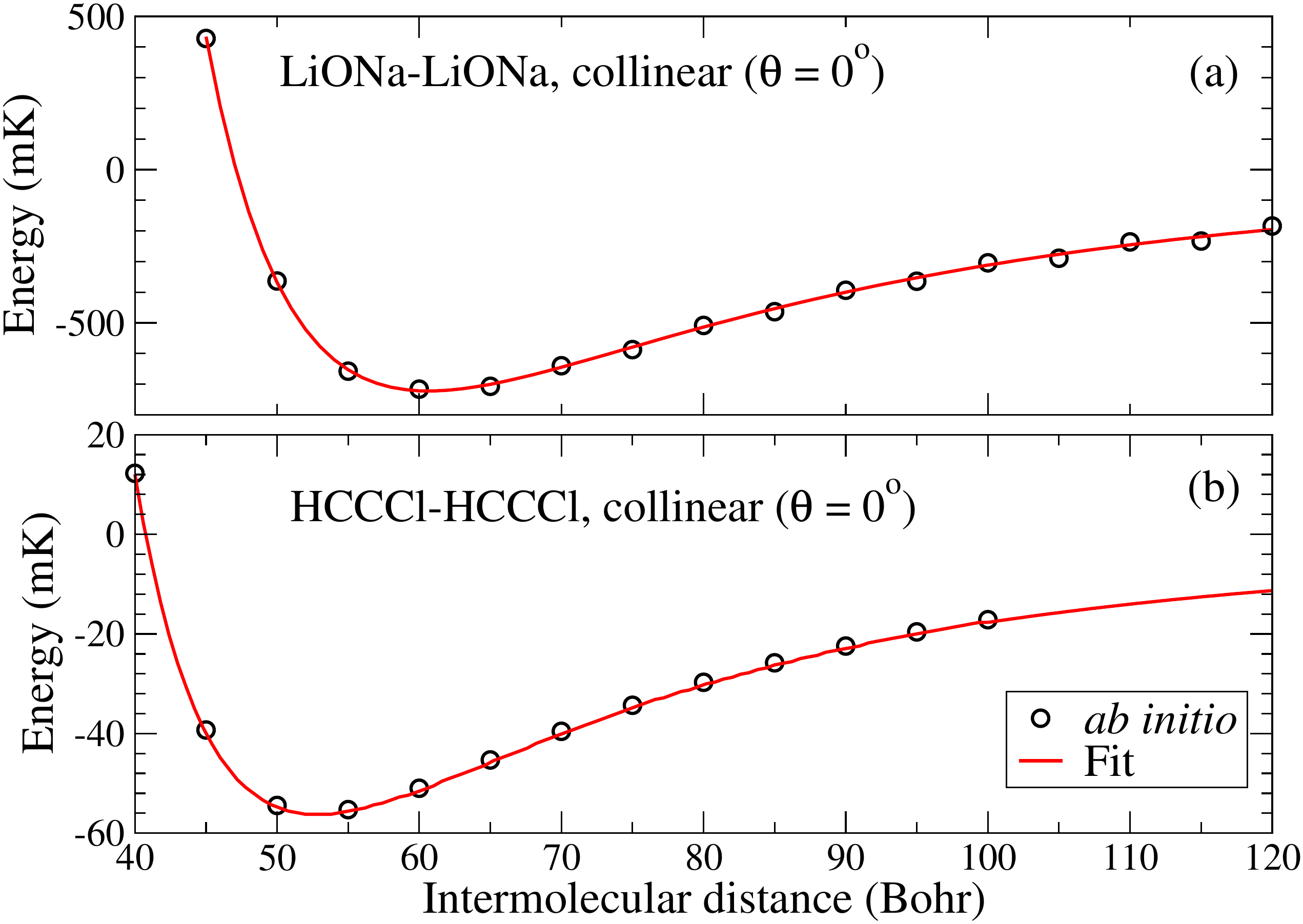}
\caption{(Color online) Fits of long-range intermolecular potential expansion to the \textit{ab initio} points for (LiONa)$_2$ (a) and (HCCCl)$_2$ (b) molecules. Outer well region is shown. The fitting form $V(R) = - (2 \mathcal{D}^2) / (R^3) + (6 \mathcal{Q}^2 - 4 \mathcal{D}\mathcal{O}) / R^5 - W / R^6$ was used, where the parameter $W$ contains $W_6$ and higher-order terms. Fitting coefficients are: $\mathcal{D}=0.792175$, $\mathcal{Q}= 19.8437$, $\mathcal{O}=-68.143$, and $W=-9743.84$ for (LiONa)$_2$ and $\mathcal{D}=0.184$, $\mathcal{Q}= 4.9879$, $\mathcal{O}=30.6274$, and $W=581.217$ for (HCCCl)$_2$. 
$R_\mathcal{Q}$ computed using the fitted values ($R_\mathcal{Q}^\mathrm{fit}=43.38$ (LiONa) and $R_\mathcal{Q}^\mathrm{fit}=46.95$ (HCCCl)) are in agreement with the \textit{ab initio} results (given in Table \ref{t1}).
All quantities are in atomic units.
 }
 \label{fig:LIoNa+HCCCl}
\end{figure}

In Table \ref{t1}, we summarize the moments as well as the $C_{6,0}$ terms obtained using \textit{ab initio} methods \cite{2012PhRvL.109h3003B} or  the above approximate expression. We list the bi-alkali molecules with appropriate ${\cal D/Q}$ ratios \cite{2012PhRvL.109h3003B}, as well as families of triatomic linear polar XOY molecules (with X and Y being different alkali atoms), and of tetratomic linear polar molecules HCCZ (with Z being a halogen atom). For each of them, we also include the value of $R_\mathcal{Q}$. These parameters were computed at the CCSD(T)/def2-QZVPPD level of theory, and as in \cite{2012PhRvL.109h3003B}, we used the center-of-mass to compute the multipole terms. The \textit{ab initio} $C_6$ coefficients were constructed using the London equation\cite{2010JChPh.132x4305Z} with the excitation energies and polarizabilities calculated at the EOM-CCSD/def2-QZVPPD and MP2/def2-QZVPPD level of theory, respectively.

For all the molecules given in Table \ref{t1}, $R_\mathcal{Q}$ is located at large enough separation to allow the existence of a long-range well separated from the shorter range surface by a barrier. We illustrate this in two specific cases, the triatomic LiONa and tetratomic HCCCl molecules. Fig.~\ref{fig:LIoNa+HCCCl} shows the long-range well for those two cases in the collinear geometry ($\theta=0$).  The long-range wells are located at large separation, and fairly shallow (roughly $10^{-6}$ and $10^{-7}$ hartree, respectively). Note that long-range expansions fit the \textit{ab initio} points very well for both molecules. 
% We constructed fits for other listed molecules and found very good agreement.

By orienting polar molecules with the appropriate ${\cal D/Q}$ ratio, it should be possible to form larger polyatomic molecules, as long as one can keep them in the ultracold regime. It is worth noting that more flexibility can be achieved if mixed molecular species are interacting. In fact, assuming two different aligned linear polar molecules, some terms in the long-range expansion (\ref{eq:Vlongrange}), that would otherwise cancel out for identical molecules, remain. For example, a $R^{-4}$ term will appear, to give (for $\theta =0$)
\[
  V(R) \simeq -\frac{2 {\cal D}_1 {\cal D}_2}{R^3} + \frac{3({\cal D}_1{\cal Q}_2 - {\cal D}_2{\cal Q}_1)}{R^4} + \frac{6 {\cal Q}_1{\cal Q}_1}{R^5} - \frac{W_6}{R^6} \; .
\]
The sign of the $R^{-4}$ term depends on the relative value of ${\cal D}_i$ and ${\cal Q}_i$ of each molecule; if ${\cal D}_1{\cal Q}_2 > {\cal D}_2{\cal Q}_1$ a longer range barrier could exist, while the reverse might reduce the size of the barrier or even prevent it. Naturally, mixing ultracold molecules in the same trap is currently experimentally challenging, but its realization would provide added flexibility in forming more complex molecules.

\section{\label{s5}Summary and Conclusions}

In this study, we theoretically investigated photoassociative production of cold tetratomic molecules, and potentially larger polyatomic molecules, from pairs of ultracold polar molecules. 
Our working hypothesis was that the {\it standard} PA of ultracold atom pairs into diatomic molecules could be extended to form a polyatomic complex from two ultracold polar molecules. We demonstrated the concept quantitatively using two ultracold polar molecules that have been cooled and trapped successfully, namely KRb and RbCs, to form (KRb)$_2$ and (RbCs)$_2$ tetratomic molecules. 
We computed realistic electronic potentials for specific geometries obtained by aligning these molecules using external fields. Such alignments and restricted geometries, beside being used for our proof-of-concept, have been realized in laboratory settings \cite{2015Sci...350..659M,2017NatPh..13...13M}. These electronic potential energy curves were computed from first principles for the first several excited states of (KRb)$_2$ and (RbCs)$_2$ (including spin-orbit coupling), and various relative orientations ($\theta$ between $0^\circ$ and $45^\circ$).
The PECs are shown for angles $0^\circ \leqslant \theta \leqslant 20^\circ$ in Fig. \ref{fig2}, while the transition dipole moments between the ground state and the first excited singlet state are given in Fig. \ref{fig2b}. The \textit{ab initio} points are available on request.

We found that the PA rate coefficients for the formation of excited (KRb)$_2$ and (RbCs)$_2$ tetratomics are comparable to the rates reported for the PA of atomic pairs, as expected, based on our knowledge of long-range potential interactions and general properties of the PA. We also confirmed that the radiative decay into the electronic ground state of the tetratomic molecule is highly impacted by the orientation of the diatomic molecules, as dictated by the Franck-Condon principle. In fact, since the
PA takes place at a large separation $R$ where the excited electronic energy surface is not steep, the pair of diatomics does not have sufficient time to significantly accelerate (due also to their large mass) before spontaneously decaying. 
This allows one to consider the decay as a vertical transition down to the ground state. Since the Franck-Condon overlap is enhanced by the existence of a long-range barrier and well for the appropriate geometry, the polyatomic ground state molecules will be mostly produced in the collinear geometry (with $\theta \approx 0$), with the production rate basically proportional to the rate of fully aligned pairs with $\theta=0$.
%
% {\color{blue} \sout{allow to consider the orientation of the pairs to remain essentially the same as in their original approach. Hence the vertical transition down to the ground state is approximated by using the same curve as the original approach; the existence of a long-range well therefore impacts significantly the Franck-Condon overlap and the corresponding polyatomic production rate.}}

Specifically, if the diatomics approach each other in a collinear geometry ($\theta=0$) or near-collinear ($\theta < \theta_c \approx 20^\circ$), a long-range outer potential well in the ground electronic state may be present for molecules whose dipole-to-quadrupole electrostatic moments ratio is small, \textit{i.e.}, $|\mathcal{D}/\mathcal{Q}| \ll 1$.
The spontaneous emission rates, from the excited state into bound ro-vibrational states of the outer well, were found to be significant both for (KRb)$_2$ and (RbCs)$_2$. The resulting ground-state molecular complex (\textit{i.e.}, (RbCs)$_2$ tetratomic molecules) formed in this way consists of a pair of polar molecules held together at long range by electrostatic interactions. 

We computed the binding energy of the complexes to be in the range of 30 MHz to 300 MHz (or about 1.4 mK to 14 mK) for (KRb)$_2$, and 20 MHz to 4 GHz (about 1 mK to 190 mK) in the case of (RbCs)$_2$. 
The long-range tetratomics would be considered strongly bound at conditions found in ultracold traps and stable enough against decay through rotation of the pair (estimated to be in milliseconds) to allow the experimental verification (\textit{e.g.}, through a resonantly enhanced multi-photon ionization or a similar technique). 
The predicted long-range complex is a novel type of tetratomic molecule, or, more generally, a long-range complex composed of two nearly-collinear polar molecules.

We generalized the conclusions based on our quantitative study conducted on RbCs and KRb to other systems. In fact, we identified other diatomic, triatomic, and tetratomic linear polar molecules with the appropriate properties, {\it i.e.} $|\mathcal{D}/\mathcal{Q}| \ll 1$. Table \ref{t1} lists families of such molecules, namely XY, XOY, and HCCZ, where X and Y are alkali atoms, and Z a halogen atom. All those molecules exhibit the right behavior, with the corresponding $R_{\cal Q}$ at large distances. We confirmed this result by using {\it ab initio} computation and obtained long-range potential wells for the collinear case of LiONa and HCCCl, in good agreement with the estimated value of $R_{\cal Q}$ (see Fig.~\ref{fig:LIoNa+HCCCl}).
These preliminary calculations suggest that these systems have long-range wells capable of supporting bound states.

Finally, the production of polyatomic molecules using the long-range well will result in the formation of polyatomic molecules in highly excited ro-vibrational levels of the molecular complex. The existence of long-lived states in those long-range complexes is also made possible by the ultracold temperatures that prevent their break up due to collisions. This type of states is usually not accessible in more standard settings, and could allow spectroscopic study of regimes relevant to roaming reactions \cite{bowman2014roaming}.

\begin{acknowledgments}
This work was partially supported by the MURI U.S. Army Research Office grant number W911NF-14-1-0378 (MG, JS), and by the National Science Foundation grant number PHY-2034284 (RC).
\end{acknowledgments}

\bibliography{refs_PA4_v4}

\begin{thebibliography}{101}%
\makeatletter
\providecommand \@ifxundefined [1]{%
 \@ifx{#1\undefined}
}%
\providecommand \@ifnum [1]{%
 \ifnum #1\expandafter \@firstoftwo
 \else \expandafter \@secondoftwo
 \fi
}%
\providecommand \@ifx [1]{%
 \ifx #1\expandafter \@firstoftwo
 \else \expandafter \@secondoftwo
 \fi
}%
\providecommand \natexlab [1]{#1}%
\providecommand \enquote  [1]{``#1''}%
\providecommand \bibnamefont  [1]{#1}%
\providecommand \bibfnamefont [1]{#1}%
\providecommand \citenamefont [1]{#1}%
\providecommand \href@noop [0]{\@secondoftwo}%
\providecommand \href [0]{\begingroup \@sanitize@url \@href}%
\providecommand \@href[1]{\@@startlink{#1}\@@href}%
\providecommand \@@href[1]{\endgroup#1\@@endlink}%
\providecommand \@sanitize@url [0]{\catcode `\\12\catcode `\$12\catcode
  `\&12\catcode `\#12\catcode `\^12\catcode `\_12\catcode `\%12\relax}%
\providecommand \@@startlink[1]{}%
\providecommand \@@endlink[0]{}%
\providecommand \url  [0]{\begingroup\@sanitize@url \@url }%
\providecommand \@url [1]{\endgroup\@href {#1}{\urlprefix }}%
\providecommand \urlprefix  [0]{URL }%
\providecommand \Eprint [0]{\href }%
\providecommand \doibase [0]{http://dx.doi.org/}%
\providecommand \selectlanguage [0]{\@gobble}%
\providecommand \bibinfo  [0]{\@secondoftwo}%
\providecommand \bibfield  [0]{\@secondoftwo}%
\providecommand \translation [1]{[#1]}%
\providecommand \BibitemOpen [0]{}%
\providecommand \bibitemStop [0]{}%
\providecommand \bibitemNoStop [0]{.\EOS\space}%
\providecommand \EOS [0]{\spacefactor3000\relax}%
\providecommand \BibitemShut  [1]{\csname bibitem#1\endcsname}%
\let\auto@bib@innerbib\@empty
%</preamble>
\bibitem [{\citenamefont {{Yan}}\ \emph {et~al.}(2013)\citenamefont {{Yan}},
  \citenamefont {{Moses}}, \citenamefont {{Gadway}}, \citenamefont {{Covey}},
  \citenamefont {{Hazzard}}, \citenamefont {{Rey}}, \citenamefont {{Jin}},\
  and\ \citenamefont {{Ye}}}]{2013Natur.501..521Y}%
  \BibitemOpen
  \bibfield  {author} {\bibinfo {author} {\bibfnamefont {B.}~\bibnamefont
  {{Yan}}}, \bibinfo {author} {\bibfnamefont {S.~A.}\ \bibnamefont {{Moses}}},
  \bibinfo {author} {\bibfnamefont {B.}~\bibnamefont {{Gadway}}}, \bibinfo
  {author} {\bibfnamefont {J.~P.}\ \bibnamefont {{Covey}}}, \bibinfo {author}
  {\bibfnamefont {K.~R.~A.}\ \bibnamefont {{Hazzard}}}, \bibinfo {author}
  {\bibfnamefont {A.~M.}\ \bibnamefont {{Rey}}}, \bibinfo {author}
  {\bibfnamefont {D.~S.}\ \bibnamefont {{Jin}}}, \ and\ \bibinfo {author}
  {\bibfnamefont {J.}~\bibnamefont {{Ye}}},\ }\href {\doibase
  10.1038/nature12483} {\bibfield  {journal} {\bibinfo  {journal} {\nat}\
  }\textbf {\bibinfo {volume} {501}},\ \bibinfo {pages} {521} (\bibinfo {year}
  {2013})}\BibitemShut {NoStop}%
\bibitem [{\citenamefont {{Gorshkov}}\ \emph {et~al.}(2013)\citenamefont
  {{Gorshkov}}, \citenamefont {{Hazzard}},\ and\ \citenamefont
  {{Rey}}}]{2013MolPh.111.1908G}%
  \BibitemOpen
  \bibfield  {author} {\bibinfo {author} {\bibfnamefont {A.~V.}\ \bibnamefont
  {{Gorshkov}}}, \bibinfo {author} {\bibfnamefont {K.~R.~A.}\ \bibnamefont
  {{Hazzard}}}, \ and\ \bibinfo {author} {\bibfnamefont {A.~M.}\ \bibnamefont
  {{Rey}}},\ }\href {\doibase 10.1080/00268976.2013.800604} {\bibfield
  {journal} {\bibinfo  {journal} {Molecular Physics}\ }\textbf {\bibinfo
  {volume} {111}},\ \bibinfo {pages} {1908} (\bibinfo {year} {2013})},\ \Eprint
  {http://arxiv.org/abs/1301.5636} {arXiv:1301.5636 [cond-mat.quant-gas]}
  \BibitemShut {NoStop}%
\bibitem [{\citenamefont {{Eisert}}\ \emph {et~al.}(2013)\citenamefont
  {{Eisert}}, \citenamefont {{van den Worm}}, \citenamefont {{Manmana}},\ and\
  \citenamefont {{Kastner}}}]{2013PhRvL.111z0401E}%
  \BibitemOpen
  \bibfield  {author} {\bibinfo {author} {\bibfnamefont {J.}~\bibnamefont
  {{Eisert}}}, \bibinfo {author} {\bibfnamefont {M.}~\bibnamefont {{van den
  Worm}}}, \bibinfo {author} {\bibfnamefont {S.~R.}\ \bibnamefont {{Manmana}}},
  \ and\ \bibinfo {author} {\bibfnamefont {M.}~\bibnamefont {{Kastner}}},\
  }\href {\doibase 10.1103/PhysRevLett.111.260401} {\bibfield  {journal}
  {\bibinfo  {journal} {\prl}\ }\textbf {\bibinfo {volume} {111}},\ \bibinfo
  {eid} {260401} (\bibinfo {year} {2013})},\ \Eprint
  {http://arxiv.org/abs/1309.2308} {arXiv:1309.2308 [quant-ph]} \BibitemShut
  {NoStop}%
\bibitem [{\citenamefont {{Richerme}}\ \emph {et~al.}(2014)\citenamefont
  {{Richerme}}, \citenamefont {{Gong}}, \citenamefont {{Lee}}, \citenamefont
  {{Senko}}, \citenamefont {{Smith}}, \citenamefont {{Foss-Feig}},
  \citenamefont {{Michalakis}}, \citenamefont {{Gorshkov}},\ and\ \citenamefont
  {{Monroe}}}]{2014Natur.511..198R}%
  \BibitemOpen
  \bibfield  {author} {\bibinfo {author} {\bibfnamefont {P.}~\bibnamefont
  {{Richerme}}}, \bibinfo {author} {\bibfnamefont {Z.-X.}\ \bibnamefont
  {{Gong}}}, \bibinfo {author} {\bibfnamefont {A.}~\bibnamefont {{Lee}}},
  \bibinfo {author} {\bibfnamefont {C.}~\bibnamefont {{Senko}}}, \bibinfo
  {author} {\bibfnamefont {J.}~\bibnamefont {{Smith}}}, \bibinfo {author}
  {\bibfnamefont {M.}~\bibnamefont {{Foss-Feig}}}, \bibinfo {author}
  {\bibfnamefont {S.}~\bibnamefont {{Michalakis}}}, \bibinfo {author}
  {\bibfnamefont {A.~V.}\ \bibnamefont {{Gorshkov}}}, \ and\ \bibinfo {author}
  {\bibfnamefont {C.}~\bibnamefont {{Monroe}}},\ }\href {\doibase
  10.1038/nature13450} {\bibfield  {journal} {\bibinfo  {journal} {\nat}\
  }\textbf {\bibinfo {volume} {511}},\ \bibinfo {pages} {198} (\bibinfo {year}
  {2014})}\BibitemShut {NoStop}%
\bibitem [{\citenamefont {{Do{\c c}aj}}\ \emph {et~al.}(2016)\citenamefont
  {{Do{\c c}aj}}, \citenamefont {{Wall}}, \citenamefont {{Mukherjee}},\ and\
  \citenamefont {{Hazzard}}}]{2016PhRvL.116m5301D}%
  \BibitemOpen
  \bibfield  {author} {\bibinfo {author} {\bibfnamefont {A.}~\bibnamefont
  {{Do{\c c}aj}}}, \bibinfo {author} {\bibfnamefont {M.~L.}\ \bibnamefont
  {{Wall}}}, \bibinfo {author} {\bibfnamefont {R.}~\bibnamefont {{Mukherjee}}},
  \ and\ \bibinfo {author} {\bibfnamefont {K.~R.~A.}\ \bibnamefont
  {{Hazzard}}},\ }\href {\doibase 10.1103/PhysRevLett.116.135301} {\bibfield
  {journal} {\bibinfo  {journal} {\prl}\ }\textbf {\bibinfo {volume} {116}},\
  \bibinfo {eid} {135301} (\bibinfo {year} {2016})},\ \Eprint
  {http://arxiv.org/abs/1512.06177} {arXiv:1512.06177 [cond-mat.quant-gas]}
  \BibitemShut {NoStop}%
\bibitem [{\citenamefont {{Kruckenhauser}}\ \emph {et~al.}(2020)\citenamefont
  {{Kruckenhauser}}, \citenamefont {{Sieberer}}, \citenamefont {{Tobias}},
  \citenamefont {{Matsuda}}, \citenamefont {{De Marco}}, \citenamefont {{Li}},
  \citenamefont {{Valtolina}}, \citenamefont {{Rey}}, \citenamefont {{Ye}},
  \citenamefont {{Baranov}},\ and\ \citenamefont
  {{Zoller}}}]{2020arXiv200111792K}%
  \BibitemOpen
  \bibfield  {author} {\bibinfo {author} {\bibfnamefont {A.}~\bibnamefont
  {{Kruckenhauser}}}, \bibinfo {author} {\bibfnamefont {L.~M.}\ \bibnamefont
  {{Sieberer}}}, \bibinfo {author} {\bibfnamefont {W.~G.}\ \bibnamefont
  {{Tobias}}}, \bibinfo {author} {\bibfnamefont {K.}~\bibnamefont {{Matsuda}}},
  \bibinfo {author} {\bibfnamefont {L.}~\bibnamefont {{De Marco}}}, \bibinfo
  {author} {\bibfnamefont {J.-R.}\ \bibnamefont {{Li}}}, \bibinfo {author}
  {\bibfnamefont {G.}~\bibnamefont {{Valtolina}}}, \bibinfo {author}
  {\bibfnamefont {A.~M.}\ \bibnamefont {{Rey}}}, \bibinfo {author}
  {\bibfnamefont {J.}~\bibnamefont {{Ye}}}, \bibinfo {author} {\bibfnamefont
  {M.~A.}\ \bibnamefont {{Baranov}}}, \ and\ \bibinfo {author} {\bibfnamefont
  {P.}~\bibnamefont {{Zoller}}},\ }\href@noop {} {\bibfield  {journal}
  {\bibinfo  {journal} {arXiv e-prints}\ ,\ \bibinfo {eid} {arXiv:2001.11792}}
  (\bibinfo {year} {2020})},\ \Eprint {http://arxiv.org/abs/2001.11792}
  {arXiv:2001.11792 [cond-mat.quant-gas]} \BibitemShut {NoStop}%
\bibitem [{\citenamefont {{Capogrosso-Sansone}}(2011)}]{2011PhRvA..83e3611C}%
  \BibitemOpen
  \bibfield  {author} {\bibinfo {author} {\bibfnamefont {B.}~\bibnamefont
  {{Capogrosso-Sansone}}},\ }\href {\doibase 10.1103/PhysRevA.83.053611}
  {\bibfield  {journal} {\bibinfo  {journal} {\pra}\ }\textbf {\bibinfo
  {volume} {83}},\ \bibinfo {eid} {053611} (\bibinfo {year} {2011})},\ \Eprint
  {http://arxiv.org/abs/1009.6213} {arXiv:1009.6213 [cond-mat.stat-mech]}
  \BibitemShut {NoStop}%
\bibitem [{\citenamefont {{Hazzard}}\ \emph {et~al.}(2013)\citenamefont
  {{Hazzard}}, \citenamefont {{Manmana}}, \citenamefont {{Foss-Feig}},\ and\
  \citenamefont {{Rey}}}]{2013PhRvL.110g5301H}%
  \BibitemOpen
  \bibfield  {author} {\bibinfo {author} {\bibfnamefont {K.~R.~A.}\
  \bibnamefont {{Hazzard}}}, \bibinfo {author} {\bibfnamefont {S.~R.}\
  \bibnamefont {{Manmana}}}, \bibinfo {author} {\bibfnamefont {M.}~\bibnamefont
  {{Foss-Feig}}}, \ and\ \bibinfo {author} {\bibfnamefont {A.~M.}\ \bibnamefont
  {{Rey}}},\ }\href {\doibase 10.1103/PhysRevLett.110.075301} {\bibfield
  {journal} {\bibinfo  {journal} {\prl}\ }\textbf {\bibinfo {volume} {110}},\
  \bibinfo {eid} {075301} (\bibinfo {year} {2013})},\ \Eprint
  {http://arxiv.org/abs/1209.4076} {arXiv:1209.4076 [cond-mat.quant-gas]}
  \BibitemShut {NoStop}%
\bibitem [{\citenamefont {{Wall}}\ \emph {et~al.}(2015)\citenamefont {{Wall}},
  \citenamefont {{Hazzard}},\ and\ \citenamefont
  {{Rey}}}]{2015famr.book....3W}%
  \BibitemOpen
  \bibfield  {author} {\bibinfo {author} {\bibfnamefont {M.~L.}\ \bibnamefont
  {{Wall}}}, \bibinfo {author} {\bibfnamefont {K.~R.~A.}\ \bibnamefont
  {{Hazzard}}}, \ and\ \bibinfo {author} {\bibfnamefont {A.~M.}\ \bibnamefont
  {{Rey}}},\ }\enquote {\bibinfo {title} {{Quantum Magnetism with Ultracold
  Molecules}},}\ in\ \href {\doibase 10.1142/9789814678704_0001} {\emph
  {\bibinfo {booktitle} {From Atomic to Mesoscale: The Role of Quantum
  Coherence in Systems of Various Complexities}}},\ \bibinfo {editor} {edited
  by\ \bibinfo {editor} {\bibfnamefont {S.}~\bibnamefont {{Malinovskaya}},
  \bibfnamefont {A.}}\ and\ \bibinfo {editor} {\bibnamefont {{et al.}}}}\
  (\bibinfo  {publisher} {World Scientific Publishing Co},\ \bibinfo {year}
  {2015})\ pp.\ \bibinfo {pages} {3--37}\BibitemShut {NoStop}%
\bibitem [{\citenamefont {Chin}\ \emph {et~al.}(2010)\citenamefont {Chin},
  \citenamefont {Grimm}, \citenamefont {Julienne},\ and\ \citenamefont
  {Tiesinga}}]{chin2010feshbach}%
  \BibitemOpen
  \bibfield  {author} {\bibinfo {author} {\bibfnamefont {C.}~\bibnamefont
  {Chin}}, \bibinfo {author} {\bibfnamefont {R.}~\bibnamefont {Grimm}},
  \bibinfo {author} {\bibfnamefont {P.}~\bibnamefont {Julienne}}, \ and\
  \bibinfo {author} {\bibfnamefont {E.}~\bibnamefont {Tiesinga}},\ }\href@noop
  {} {\bibfield  {journal} {\bibinfo  {journal} {Rev. Mod. Phys.}\ }\textbf
  {\bibinfo {volume} {82}},\ \bibinfo {pages} {1225} (\bibinfo {year}
  {2010})}\BibitemShut {NoStop}%
\bibitem [{\citenamefont {Micheli}\ \emph {et~al.}(2006)\citenamefont
  {Micheli}, \citenamefont {Brennen},\ and\ \citenamefont
  {Zoller}}]{micheli2006toolbox}%
  \BibitemOpen
  \bibfield  {author} {\bibinfo {author} {\bibfnamefont {A.}~\bibnamefont
  {Micheli}}, \bibinfo {author} {\bibfnamefont {G.}~\bibnamefont {Brennen}}, \
  and\ \bibinfo {author} {\bibfnamefont {P.}~\bibnamefont {Zoller}},\
  }\href@noop {} {\bibfield  {journal} {\bibinfo  {journal} {Nat. Phys.}\
  }\textbf {\bibinfo {volume} {2}},\ \bibinfo {pages} {341} (\bibinfo {year}
  {2006})}\BibitemShut {NoStop}%
\bibitem [{\citenamefont {{Bloch}}\ \emph {et~al.}(2008)\citenamefont
  {{Bloch}}, \citenamefont {{Dalibard}},\ and\ \citenamefont
  {{Zwerger}}}]{2008RvMP...80..885B}%
  \BibitemOpen
  \bibfield  {author} {\bibinfo {author} {\bibfnamefont {I.}~\bibnamefont
  {{Bloch}}}, \bibinfo {author} {\bibfnamefont {J.}~\bibnamefont {{Dalibard}}},
  \ and\ \bibinfo {author} {\bibfnamefont {W.}~\bibnamefont {{Zwerger}}},\
  }\href {\doibase 10.1103/RevModPhys.80.885} {\bibfield  {journal} {\bibinfo
  {journal} {Rev. Mod. Phys.}\ }\textbf {\bibinfo {volume} {80}},\ \bibinfo
  {pages} {885} (\bibinfo {year} {2008})},\ \Eprint
  {http://arxiv.org/abs/0704.3011} {arXiv:0704.3011} \BibitemShut {NoStop}%
\bibitem [{\citenamefont {DeMille}(2002)}]{demille2002quantum}%
  \BibitemOpen
  \bibfield  {author} {\bibinfo {author} {\bibfnamefont {D.}~\bibnamefont
  {DeMille}},\ }\href@noop {} {\bibfield  {journal} {\bibinfo  {journal}
  {\prl}\ }\textbf {\bibinfo {volume} {88}},\ \bibinfo {pages} {067901}
  (\bibinfo {year} {2002})}\BibitemShut {NoStop}%
\bibitem [{\citenamefont {Krems}\ \emph {et~al.}(2009)\citenamefont {Krems},
  \citenamefont {Friedrich},\ and\ \citenamefont {Stwalley}}]{krems2009cold}%
  \BibitemOpen
  \bibfield  {author} {\bibinfo {author} {\bibfnamefont {R.}~\bibnamefont
  {Krems}}, \bibinfo {author} {\bibfnamefont {B.}~\bibnamefont {Friedrich}}, \
  and\ \bibinfo {author} {\bibfnamefont {W.~C.}\ \bibnamefont {Stwalley}},\
  }\href@noop {} {\emph {\bibinfo {title} {Cold molecules: theory, experiment,
  applications}}}\ (\bibinfo  {publisher} {CRC press},\ \bibinfo {year}
  {2009})\BibitemShut {NoStop}%
\bibitem [{\citenamefont {Kuznetsova}\ \emph {et~al.}(2011)\citenamefont
  {Kuznetsova}, \citenamefont {Yelin},\ and\ \citenamefont
  {C{\^o}t{\'e}}}]{kuznetsova2011atom}%
  \BibitemOpen
  \bibfield  {author} {\bibinfo {author} {\bibfnamefont {E.}~\bibnamefont
  {Kuznetsova}}, \bibinfo {author} {\bibfnamefont {S.~F.}\ \bibnamefont
  {Yelin}}, \ and\ \bibinfo {author} {\bibfnamefont {R.}~\bibnamefont
  {C{\^o}t{\'e}}},\ }\href@noop {} {\bibfield  {journal} {\bibinfo  {journal}
  {Quantum Information Processing}\ }\textbf {\bibinfo {volume} {10}},\
  \bibinfo {pages} {821} (\bibinfo {year} {2011})}\BibitemShut {NoStop}%
\bibitem [{\citenamefont
  {{C{\^o}t{\'e}}}(2014)}]{doi:10.1002/9781118742631.ch14}%
  \BibitemOpen
  \bibfield  {author} {\bibinfo {author} {\bibfnamefont {R.}~\bibnamefont
  {{C{\^o}t{\'e}}}},\ }\enquote {\bibinfo {title} {Ultracold molecules: Their
  formation and application to quantum computing},}\ in\ \href {\doibase
  10.1002/9781118742631.ch14} {\emph {\bibinfo {booktitle} {Quantum Information
  and Computation for Chemistry}}}\ (\bibinfo  {publisher} {John Wiley \& Sons,
  Ltd},\ \bibinfo {year} {2014})\ pp.\ \bibinfo {pages} {403--448}\BibitemShut
  {NoStop}%
\bibitem [{\citenamefont {Yelin}\ \emph {et~al.}(2006)\citenamefont {Yelin},
  \citenamefont {Kirby},\ and\ \citenamefont
  {C{\^o}t{\'e}}}]{yelin2006schemes}%
  \BibitemOpen
  \bibfield  {author} {\bibinfo {author} {\bibfnamefont {S.~F.}\ \bibnamefont
  {Yelin}}, \bibinfo {author} {\bibfnamefont {K.}~\bibnamefont {Kirby}}, \ and\
  \bibinfo {author} {\bibfnamefont {R.}~\bibnamefont {C{\^o}t{\'e}}},\
  }\href@noop {} {\bibfield  {journal} {\bibinfo  {journal} {\pra}\ }\textbf
  {\bibinfo {volume} {74}},\ \bibinfo {pages} {050301(R)} (\bibinfo {year}
  {2006})}\BibitemShut {NoStop}%
\bibitem [{\citenamefont {{Kuznetsova}}\ \emph {et~al.}(2010)\citenamefont
  {{Kuznetsova}}, \citenamefont {{Gacesa}}, \citenamefont {{Yelin}},\ and\
  \citenamefont {{C{\^o}t{\'e}}}}]{2010PhRvA..81c0301K}%
  \BibitemOpen
  \bibfield  {author} {\bibinfo {author} {\bibfnamefont {E.}~\bibnamefont
  {{Kuznetsova}}}, \bibinfo {author} {\bibfnamefont {M.}~\bibnamefont
  {{Gacesa}}}, \bibinfo {author} {\bibfnamefont {S.~F.}\ \bibnamefont
  {{Yelin}}}, \ and\ \bibinfo {author} {\bibfnamefont {R.}~\bibnamefont
  {{C{\^o}t{\'e}}}},\ }\href {\doibase 10.1103/PhysRevA.81.030301} {\bibfield
  {journal} {\bibinfo  {journal} {\pra}\ }\textbf {\bibinfo {volume} {81}},\
  \bibinfo {eid} {030301} (\bibinfo {year} {2010})},\ \Eprint
  {http://arxiv.org/abs/0908.4558} {arXiv:0908.4558 [quant-ph]} \BibitemShut
  {NoStop}%
\bibitem [{\citenamefont {Stuhler}\ \emph {et~al.}(2005)\citenamefont
  {Stuhler}, \citenamefont {Griesmaier}, \citenamefont {Koch}, \citenamefont
  {Fattori}, \citenamefont {Pfau}, \citenamefont {Giovanazzi}, \citenamefont
  {Pedri},\ and\ \citenamefont {Santos}}]{PhysRevLett.95.150406}%
  \BibitemOpen
  \bibfield  {author} {\bibinfo {author} {\bibfnamefont {J.}~\bibnamefont
  {Stuhler}}, \bibinfo {author} {\bibfnamefont {A.}~\bibnamefont {Griesmaier}},
  \bibinfo {author} {\bibfnamefont {T.}~\bibnamefont {Koch}}, \bibinfo {author}
  {\bibfnamefont {M.}~\bibnamefont {Fattori}}, \bibinfo {author} {\bibfnamefont
  {T.}~\bibnamefont {Pfau}}, \bibinfo {author} {\bibfnamefont {S.}~\bibnamefont
  {Giovanazzi}}, \bibinfo {author} {\bibfnamefont {P.}~\bibnamefont {Pedri}}, \
  and\ \bibinfo {author} {\bibfnamefont {L.}~\bibnamefont {Santos}},\ }\href
  {\doibase 10.1103/PhysRevLett.95.150406} {\bibfield  {journal} {\bibinfo
  {journal} {\prl}\ }\textbf {\bibinfo {volume} {95}},\ \bibinfo {pages}
  {150406} (\bibinfo {year} {2005})}\BibitemShut {NoStop}%
\bibitem [{\citenamefont {{Boisseau}}\ \emph {et~al.}(2002)\citenamefont
  {{Boisseau}}, \citenamefont {{Simbotin}},\ and\ \citenamefont
  {{C{\^o}t{\'e}}}}]{2002PhRvL..88m3004B}%
  \BibitemOpen
  \bibfield  {author} {\bibinfo {author} {\bibfnamefont {C.}~\bibnamefont
  {{Boisseau}}}, \bibinfo {author} {\bibfnamefont {I.}~\bibnamefont
  {{Simbotin}}}, \ and\ \bibinfo {author} {\bibfnamefont {R.}~\bibnamefont
  {{C{\^o}t{\'e}}}},\ }\href {\doibase 10.1103/PhysRevLett.88.133004}
  {\bibfield  {journal} {\bibinfo  {journal} {\prl}\ }\textbf {\bibinfo
  {volume} {88}},\ \bibinfo {eid} {133004} (\bibinfo {year} {2002})},\ \Eprint
  {http://arxiv.org/abs/physics/0201022} {arXiv:physics/0201022
  [physics.atom-ph]} \BibitemShut {NoStop}%
\bibitem [{\citenamefont {{Singer}}\ \emph {et~al.}(2005)\citenamefont
  {{Singer}}, \citenamefont {{Stanojevic}}, \citenamefont {{Weidem{\"u}ller}},\
  and\ \citenamefont {{C{\^o}t{\'e}}}}]{2005JPhB...38S.295S}%
  \BibitemOpen
  \bibfield  {author} {\bibinfo {author} {\bibfnamefont {K.}~\bibnamefont
  {{Singer}}}, \bibinfo {author} {\bibfnamefont {J.}~\bibnamefont
  {{Stanojevic}}}, \bibinfo {author} {\bibfnamefont {M.}~\bibnamefont
  {{Weidem{\"u}ller}}}, \ and\ \bibinfo {author} {\bibfnamefont
  {R.}~\bibnamefont {{C{\^o}t{\'e}}}},\ }\href {\doibase
  10.1088/0953-4075/38/2/021} {\bibfield  {journal} {\bibinfo  {journal} {J.
  Phys. B}\ }\textbf {\bibinfo {volume} {38}},\ \bibinfo {pages} {S295}
  (\bibinfo {year} {2005})}\BibitemShut {NoStop}%
\bibitem [{\citenamefont {Jaksch}\ \emph {et~al.}(2000)\citenamefont {Jaksch},
  \citenamefont {Cirac}, \citenamefont {Zoller}, \citenamefont {Rolston},
  \citenamefont {C{\^o}t{\'e}},\ and\ \citenamefont {Lukin}}]{jaksch2000fast}%
  \BibitemOpen
  \bibfield  {author} {\bibinfo {author} {\bibfnamefont {D.}~\bibnamefont
  {Jaksch}}, \bibinfo {author} {\bibfnamefont {J.}~\bibnamefont {Cirac}},
  \bibinfo {author} {\bibfnamefont {P.}~\bibnamefont {Zoller}}, \bibinfo
  {author} {\bibfnamefont {S.}~\bibnamefont {Rolston}}, \bibinfo {author}
  {\bibfnamefont {R.}~\bibnamefont {C{\^o}t{\'e}}}, \ and\ \bibinfo {author}
  {\bibfnamefont {M.}~\bibnamefont {Lukin}},\ }\href@noop {} {\bibfield
  {journal} {\bibinfo  {journal} {\prl}\ }\textbf {\bibinfo {volume} {85}},\
  \bibinfo {pages} {2208} (\bibinfo {year} {2000})}\BibitemShut {NoStop}%
\bibitem [{\citenamefont {Lukin}\ \emph {et~al.}(2001)\citenamefont {Lukin},
  \citenamefont {Fleischhauer}, \citenamefont {Cote}, \citenamefont {Duan},
  \citenamefont {Jaksch}, \citenamefont {Cirac},\ and\ \citenamefont
  {Zoller}}]{lukin2001dipole}%
  \BibitemOpen
  \bibfield  {author} {\bibinfo {author} {\bibfnamefont {M.~D.}\ \bibnamefont
  {Lukin}}, \bibinfo {author} {\bibfnamefont {M.}~\bibnamefont {Fleischhauer}},
  \bibinfo {author} {\bibfnamefont {R.}~\bibnamefont {Cote}}, \bibinfo {author}
  {\bibfnamefont {L.}~\bibnamefont {Duan}}, \bibinfo {author} {\bibfnamefont
  {D.}~\bibnamefont {Jaksch}}, \bibinfo {author} {\bibfnamefont {J.~I.}\
  \bibnamefont {Cirac}}, \ and\ \bibinfo {author} {\bibfnamefont
  {P.}~\bibnamefont {Zoller}},\ }\href@noop {} {\bibfield  {journal} {\bibinfo
  {journal} {\prl}\ }\textbf {\bibinfo {volume} {87}},\ \bibinfo {pages}
  {037901} (\bibinfo {year} {2001})}\BibitemShut {NoStop}%
\bibitem [{\citenamefont {Cote}(2005)}]{cote2005tutorial}%
  \BibitemOpen
  \bibfield  {author} {\bibinfo {author} {\bibfnamefont {R.}~\bibnamefont
  {Cote}},\ }in\ \href@noop {} {\emph {\bibinfo {booktitle} {Active and Passive
  Optical Components for WDM Communications V}}},\ Vol.\ \bibinfo {volume}
  {6014}\ (\bibinfo {organization} {International Society for Optics and
  Photonics},\ \bibinfo {year} {2005})\ p.\ \bibinfo {pages}
  {601415}\BibitemShut {NoStop}%
\bibitem [{\citenamefont {Stanojevic}\ \emph {et~al.}(2006)\citenamefont
  {Stanojevic}, \citenamefont {C{\^o}t{\'e}}, \citenamefont {Tong},
  \citenamefont {Farooqi}, \citenamefont {Eyler},\ and\ \citenamefont
  {Gould}}]{stanojevic2006long}%
  \BibitemOpen
  \bibfield  {author} {\bibinfo {author} {\bibfnamefont {J.}~\bibnamefont
  {Stanojevic}}, \bibinfo {author} {\bibfnamefont {R.}~\bibnamefont
  {C{\^o}t{\'e}}}, \bibinfo {author} {\bibfnamefont {D.}~\bibnamefont {Tong}},
  \bibinfo {author} {\bibfnamefont {S.}~\bibnamefont {Farooqi}}, \bibinfo
  {author} {\bibfnamefont {E.}~\bibnamefont {Eyler}}, \ and\ \bibinfo {author}
  {\bibfnamefont {P.}~\bibnamefont {Gould}},\ }\href@noop {} {\bibfield
  {journal} {\bibinfo  {journal} {\epjd}\ }\textbf {\bibinfo {volume} {40}},\
  \bibinfo {pages} {3} (\bibinfo {year} {2006})}\BibitemShut {NoStop}%
\bibitem [{\citenamefont {{Hazzard}}\ \emph {et~al.}(2014)\citenamefont
  {{Hazzard}}, \citenamefont {{Gadway}}, \citenamefont {{Foss-Feig}},
  \citenamefont {{Yan}}, \citenamefont {{Moses}}, \citenamefont {{Covey}},
  \citenamefont {{Yao}}, \citenamefont {{Lukin}}, \citenamefont {{Ye}},
  \citenamefont {{Jin}},\ and\ \citenamefont {{Rey}}}]{2014PhRvL.113s5302H}%
  \BibitemOpen
  \bibfield  {author} {\bibinfo {author} {\bibfnamefont {K.~R.~A.}\
  \bibnamefont {{Hazzard}}}, \bibinfo {author} {\bibfnamefont {B.}~\bibnamefont
  {{Gadway}}}, \bibinfo {author} {\bibfnamefont {M.}~\bibnamefont
  {{Foss-Feig}}}, \bibinfo {author} {\bibfnamefont {B.}~\bibnamefont {{Yan}}},
  \bibinfo {author} {\bibfnamefont {S.~A.}\ \bibnamefont {{Moses}}}, \bibinfo
  {author} {\bibfnamefont {J.~P.}\ \bibnamefont {{Covey}}}, \bibinfo {author}
  {\bibfnamefont {N.~Y.}\ \bibnamefont {{Yao}}}, \bibinfo {author}
  {\bibfnamefont {M.~D.}\ \bibnamefont {{Lukin}}}, \bibinfo {author}
  {\bibfnamefont {J.}~\bibnamefont {{Ye}}}, \bibinfo {author} {\bibfnamefont
  {D.~S.}\ \bibnamefont {{Jin}}}, \ and\ \bibinfo {author} {\bibfnamefont
  {A.~M.}\ \bibnamefont {{Rey}}},\ }\href {\doibase
  10.1103/PhysRevLett.113.195302} {\bibfield  {journal} {\bibinfo  {journal}
  {\prl}\ }\textbf {\bibinfo {volume} {113}},\ \bibinfo {eid} {195302}
  (\bibinfo {year} {2014})},\ \Eprint {http://arxiv.org/abs/1402.2354}
  {arXiv:1402.2354 [cond-mat.quant-gas]} \BibitemShut {NoStop}%
\bibitem [{\citenamefont {Ni}\ \emph {et~al.}(2008)\citenamefont {Ni},
  \citenamefont {Ospelkaus}, \citenamefont {de~Miranda}, \citenamefont
  {Pe{\textquoteright}er}, \citenamefont {Neyenhuis}, \citenamefont {Zirbel},
  \citenamefont {Kotochigova}, \citenamefont {Julienne}, \citenamefont {Jin},\
  and\ \citenamefont {Ye}}]{Ni231}%
  \BibitemOpen
  \bibfield  {author} {\bibinfo {author} {\bibfnamefont {K.-K.}\ \bibnamefont
  {Ni}}, \bibinfo {author} {\bibfnamefont {S.}~\bibnamefont {Ospelkaus}},
  \bibinfo {author} {\bibfnamefont {M.~H.~G.}\ \bibnamefont {de~Miranda}},
  \bibinfo {author} {\bibfnamefont {A.}~\bibnamefont {Pe{\textquoteright}er}},
  \bibinfo {author} {\bibfnamefont {B.}~\bibnamefont {Neyenhuis}}, \bibinfo
  {author} {\bibfnamefont {J.~J.}\ \bibnamefont {Zirbel}}, \bibinfo {author}
  {\bibfnamefont {S.}~\bibnamefont {Kotochigova}}, \bibinfo {author}
  {\bibfnamefont {P.~S.}\ \bibnamefont {Julienne}}, \bibinfo {author}
  {\bibfnamefont {D.~S.}\ \bibnamefont {Jin}}, \ and\ \bibinfo {author}
  {\bibfnamefont {J.}~\bibnamefont {Ye}},\ }\href {\doibase
  10.1126/science.1163861} {\bibfield  {journal} {\bibinfo  {journal}
  {Science}\ }\textbf {\bibinfo {volume} {322}},\ \bibinfo {pages} {231}
  (\bibinfo {year} {2008})}\BibitemShut {NoStop}%
\bibitem [{\citenamefont {{Moses}}\ \emph {et~al.}(2015)\citenamefont
  {{Moses}}, \citenamefont {{Covey}}, \citenamefont {{Miecnikowski}},
  \citenamefont {{Yan}}, \citenamefont {{Gadway}}, \citenamefont {{Ye}},\ and\
  \citenamefont {{Jin}}}]{2015Sci...350..659M}%
  \BibitemOpen
  \bibfield  {author} {\bibinfo {author} {\bibfnamefont {S.~A.}\ \bibnamefont
  {{Moses}}}, \bibinfo {author} {\bibfnamefont {J.~P.}\ \bibnamefont
  {{Covey}}}, \bibinfo {author} {\bibfnamefont {M.~T.}\ \bibnamefont
  {{Miecnikowski}}}, \bibinfo {author} {\bibfnamefont {B.}~\bibnamefont
  {{Yan}}}, \bibinfo {author} {\bibfnamefont {B.}~\bibnamefont {{Gadway}}},
  \bibinfo {author} {\bibfnamefont {J.}~\bibnamefont {{Ye}}}, \ and\ \bibinfo
  {author} {\bibfnamefont {D.~S.}\ \bibnamefont {{Jin}}},\ }\href {\doibase
  10.1126/science.aac6400} {\bibfield  {journal} {\bibinfo  {journal}
  {Science}\ }\textbf {\bibinfo {volume} {350}},\ \bibinfo {pages} {659}
  (\bibinfo {year} {2015})},\ \Eprint {http://arxiv.org/abs/1507.02377}
  {arXiv:1507.02377 [cond-mat.quant-gas]} \BibitemShut {NoStop}%
\bibitem [{\citenamefont {Park}\ \emph {et~al.}(2015)\citenamefont {Park},
  \citenamefont {Will},\ and\ \citenamefont {Zwierlein}}]{park2015ultracold}%
  \BibitemOpen
  \bibfield  {author} {\bibinfo {author} {\bibfnamefont {J.~W.}\ \bibnamefont
  {Park}}, \bibinfo {author} {\bibfnamefont {S.~A.}\ \bibnamefont {Will}}, \
  and\ \bibinfo {author} {\bibfnamefont {M.~W.}\ \bibnamefont {Zwierlein}},\
  }\href@noop {} {\bibfield  {journal} {\bibinfo  {journal} {\prl}\ }\textbf
  {\bibinfo {volume} {114}},\ \bibinfo {pages} {205302} (\bibinfo {year}
  {2015})}\BibitemShut {NoStop}%
\bibitem [{\citenamefont {See{\ss}elberg}\ \emph {et~al.}(2018)\citenamefont
  {See{\ss}elberg}, \citenamefont {Buchheim}, \citenamefont {Lu}, \citenamefont
  {Schneider}, \citenamefont {Luo}, \citenamefont {Tiemann}, \citenamefont
  {Bloch},\ and\ \citenamefont {Gohle}}]{seesselberg2018modeling}%
  \BibitemOpen
  \bibfield  {author} {\bibinfo {author} {\bibfnamefont {F.}~\bibnamefont
  {See{\ss}elberg}}, \bibinfo {author} {\bibfnamefont {N.}~\bibnamefont
  {Buchheim}}, \bibinfo {author} {\bibfnamefont {Z.-K.}\ \bibnamefont {Lu}},
  \bibinfo {author} {\bibfnamefont {T.}~\bibnamefont {Schneider}}, \bibinfo
  {author} {\bibfnamefont {X.-Y.}\ \bibnamefont {Luo}}, \bibinfo {author}
  {\bibfnamefont {E.}~\bibnamefont {Tiemann}}, \bibinfo {author} {\bibfnamefont
  {I.}~\bibnamefont {Bloch}}, \ and\ \bibinfo {author} {\bibfnamefont
  {C.}~\bibnamefont {Gohle}},\ }\href@noop {} {\bibfield  {journal} {\bibinfo
  {journal} {\pra}\ }\textbf {\bibinfo {volume} {97}},\ \bibinfo {pages}
  {013405} (\bibinfo {year} {2018})}\BibitemShut {NoStop}%
\bibitem [{\citenamefont {Takekoshi}\ \emph {et~al.}(2014)\citenamefont
  {Takekoshi}, \citenamefont {Reichs\"ollner}, \citenamefont {Schindewolf},
  \citenamefont {Hutson}, \citenamefont {Le~Sueur}, \citenamefont {Dulieu},
  \citenamefont {Ferlaino}, \citenamefont {Grimm},\ and\ \citenamefont
  {N\"agerl}}]{PhysRevLett.113.205301}%
  \BibitemOpen
  \bibfield  {author} {\bibinfo {author} {\bibfnamefont {T.}~\bibnamefont
  {Takekoshi}}, \bibinfo {author} {\bibfnamefont {L.}~\bibnamefont
  {Reichs\"ollner}}, \bibinfo {author} {\bibfnamefont {A.}~\bibnamefont
  {Schindewolf}}, \bibinfo {author} {\bibfnamefont {J.~M.}\ \bibnamefont
  {Hutson}}, \bibinfo {author} {\bibfnamefont {C.~R.}\ \bibnamefont
  {Le~Sueur}}, \bibinfo {author} {\bibfnamefont {O.}~\bibnamefont {Dulieu}},
  \bibinfo {author} {\bibfnamefont {F.}~\bibnamefont {Ferlaino}}, \bibinfo
  {author} {\bibfnamefont {R.}~\bibnamefont {Grimm}}, \ and\ \bibinfo {author}
  {\bibfnamefont {H.-C.}\ \bibnamefont {N\"agerl}},\ }\href {\doibase
  10.1103/PhysRevLett.113.205301} {\bibfield  {journal} {\bibinfo  {journal}
  {\prl}\ }\textbf {\bibinfo {volume} {113}},\ \bibinfo {pages} {205301}
  (\bibinfo {year} {2014})}\BibitemShut {NoStop}%
\bibitem [{\citenamefont {Molony}\ \emph {et~al.}(2014)\citenamefont {Molony},
  \citenamefont {Gregory}, \citenamefont {Ji}, \citenamefont {Lu},
  \citenamefont {K\"oppinger}, \citenamefont {Le~Sueur}, \citenamefont
  {Blackley}, \citenamefont {Hutson},\ and\ \citenamefont
  {Cornish}}]{PhysRevLett.113.255301}%
  \BibitemOpen
  \bibfield  {author} {\bibinfo {author} {\bibfnamefont {P.~K.}\ \bibnamefont
  {Molony}}, \bibinfo {author} {\bibfnamefont {P.~D.}\ \bibnamefont {Gregory}},
  \bibinfo {author} {\bibfnamefont {Z.}~\bibnamefont {Ji}}, \bibinfo {author}
  {\bibfnamefont {B.}~\bibnamefont {Lu}}, \bibinfo {author} {\bibfnamefont
  {M.~P.}\ \bibnamefont {K\"oppinger}}, \bibinfo {author} {\bibfnamefont
  {C.~R.}\ \bibnamefont {Le~Sueur}}, \bibinfo {author} {\bibfnamefont {C.~L.}\
  \bibnamefont {Blackley}}, \bibinfo {author} {\bibfnamefont {J.~M.}\
  \bibnamefont {Hutson}}, \ and\ \bibinfo {author} {\bibfnamefont {S.~L.}\
  \bibnamefont {Cornish}},\ }\href {\doibase 10.1103/PhysRevLett.113.255301}
  {\bibfield  {journal} {\bibinfo  {journal} {\prl}\ }\textbf {\bibinfo
  {volume} {113}},\ \bibinfo {pages} {255301} (\bibinfo {year}
  {2014})}\BibitemShut {NoStop}%
\bibitem [{\citenamefont {{Reichs{\"o}llner}}\ \emph
  {et~al.}(2017)\citenamefont {{Reichs{\"o}llner}}, \citenamefont
  {{Schindewolf}}, \citenamefont {{Takekoshi}}, \citenamefont {{Grimm}},\ and\
  \citenamefont {{N{\"a}gerl}}}]{2017PhRvL.118g3201R}%
  \BibitemOpen
  \bibfield  {author} {\bibinfo {author} {\bibfnamefont {L.}~\bibnamefont
  {{Reichs{\"o}llner}}}, \bibinfo {author} {\bibfnamefont {A.}~\bibnamefont
  {{Schindewolf}}}, \bibinfo {author} {\bibfnamefont {T.}~\bibnamefont
  {{Takekoshi}}}, \bibinfo {author} {\bibfnamefont {R.}~\bibnamefont
  {{Grimm}}}, \ and\ \bibinfo {author} {\bibfnamefont {H.-C.}\ \bibnamefont
  {{N{\"a}gerl}}},\ }\href {\doibase 10.1103/PhysRevLett.118.073201} {\bibfield
   {journal} {\bibinfo  {journal} {\prl}\ }\textbf {\bibinfo {volume} {118}},\
  \bibinfo {eid} {073201} (\bibinfo {year} {2017})},\ \Eprint
  {http://arxiv.org/abs/1607.06536} {arXiv:1607.06536 [cond-mat.quant-gas]}
  \BibitemShut {NoStop}%
\bibitem [{\citenamefont {{Guo}}\ \emph {et~al.}(2016)\citenamefont {{Guo}},
  \citenamefont {{Zhu}}, \citenamefont {{Lu}}, \citenamefont {{Ye}},
  \citenamefont {{Wang}}, \citenamefont {{Vexiau}}, \citenamefont
  {{Bouloufa-Maafa}}, \citenamefont {{Qu{\'e}m{\'e}ner}}, \citenamefont
  {{Dulieu}},\ and\ \citenamefont {{Wang}}}]{2016PhRvL.116t5303G}%
  \BibitemOpen
  \bibfield  {author} {\bibinfo {author} {\bibfnamefont {M.}~\bibnamefont
  {{Guo}}}, \bibinfo {author} {\bibfnamefont {B.}~\bibnamefont {{Zhu}}},
  \bibinfo {author} {\bibfnamefont {B.}~\bibnamefont {{Lu}}}, \bibinfo {author}
  {\bibfnamefont {X.}~\bibnamefont {{Ye}}}, \bibinfo {author} {\bibfnamefont
  {F.}~\bibnamefont {{Wang}}}, \bibinfo {author} {\bibfnamefont
  {R.}~\bibnamefont {{Vexiau}}}, \bibinfo {author} {\bibfnamefont
  {N.}~\bibnamefont {{Bouloufa-Maafa}}}, \bibinfo {author} {\bibfnamefont
  {G.}~\bibnamefont {{Qu{\'e}m{\'e}ner}}}, \bibinfo {author} {\bibfnamefont
  {O.}~\bibnamefont {{Dulieu}}}, \ and\ \bibinfo {author} {\bibfnamefont
  {D.}~\bibnamefont {{Wang}}},\ }\href {\doibase
  10.1103/PhysRevLett.116.205303} {\bibfield  {journal} {\bibinfo  {journal}
  {\prl}\ }\textbf {\bibinfo {volume} {116}},\ \bibinfo {eid} {205303}
  (\bibinfo {year} {2016})},\ \Eprint {http://arxiv.org/abs/1602.03947}
  {arXiv:1602.03947 [cond-mat.quant-gas]} \BibitemShut {NoStop}%
\bibitem [{\citenamefont {Heo}\ \emph {et~al.}(2012)\citenamefont {Heo},
  \citenamefont {Wang}, \citenamefont {Christensen}, \citenamefont {Rvachov},
  \citenamefont {Cotta}, \citenamefont {Choi}, \citenamefont {Lee},\ and\
  \citenamefont {Ketterle}}]{heo2012formation}%
  \BibitemOpen
  \bibfield  {author} {\bibinfo {author} {\bibfnamefont {M.-S.}\ \bibnamefont
  {Heo}}, \bibinfo {author} {\bibfnamefont {T.~T.}\ \bibnamefont {Wang}},
  \bibinfo {author} {\bibfnamefont {C.~A.}\ \bibnamefont {Christensen}},
  \bibinfo {author} {\bibfnamefont {T.~M.}\ \bibnamefont {Rvachov}}, \bibinfo
  {author} {\bibfnamefont {D.~A.}\ \bibnamefont {Cotta}}, \bibinfo {author}
  {\bibfnamefont {J.-H.}\ \bibnamefont {Choi}}, \bibinfo {author}
  {\bibfnamefont {Y.-R.}\ \bibnamefont {Lee}}, \ and\ \bibinfo {author}
  {\bibfnamefont {W.}~\bibnamefont {Ketterle}},\ }\href@noop {} {\bibfield
  {journal} {\bibinfo  {journal} {\pra}\ }\textbf {\bibinfo {volume} {86}},\
  \bibinfo {pages} {021602} (\bibinfo {year} {2012})}\BibitemShut {NoStop}%
\bibitem [{\citenamefont {Son}\ \emph {et~al.}(2020)\citenamefont {Son},
  \citenamefont {Park}, \citenamefont {Ketterle},\ and\ \citenamefont
  {Jamison}}]{son2020collisional}%
  \BibitemOpen
  \bibfield  {author} {\bibinfo {author} {\bibfnamefont {H.}~\bibnamefont
  {Son}}, \bibinfo {author} {\bibfnamefont {J.~J.}\ \bibnamefont {Park}},
  \bibinfo {author} {\bibfnamefont {W.}~\bibnamefont {Ketterle}}, \ and\
  \bibinfo {author} {\bibfnamefont {A.~O.}\ \bibnamefont {Jamison}},\
  }\href@noop {} {\bibfield  {journal} {\bibinfo  {journal} {Nature}\ }\textbf
  {\bibinfo {volume} {580}},\ \bibinfo {pages} {197} (\bibinfo {year}
  {2020})}\BibitemShut {NoStop}%
\bibitem [{\citenamefont {{Gaubatz}}\ \emph {et~al.}(1990)\citenamefont
  {{Gaubatz}}, \citenamefont {{Rudecki}}, \citenamefont {{Schiemann}},\ and\
  \citenamefont {{Bergmann}}}]{1990JChPh..92.5363G}%
  \BibitemOpen
  \bibfield  {author} {\bibinfo {author} {\bibfnamefont {U.}~\bibnamefont
  {{Gaubatz}}}, \bibinfo {author} {\bibfnamefont {P.}~\bibnamefont
  {{Rudecki}}}, \bibinfo {author} {\bibfnamefont {S.}~\bibnamefont
  {{Schiemann}}}, \ and\ \bibinfo {author} {\bibfnamefont {K.}~\bibnamefont
  {{Bergmann}}},\ }\href {\doibase 10.1063/1.458514} {\bibfield  {journal}
  {\bibinfo  {journal} {\jcp}\ }\textbf {\bibinfo {volume} {92}},\ \bibinfo
  {pages} {5363} (\bibinfo {year} {1990})}\BibitemShut {NoStop}%
\bibitem [{\citenamefont {Thorsheim}\ \emph {et~al.}(1987)\citenamefont
  {Thorsheim}, \citenamefont {Weiner},\ and\ \citenamefont
  {Julienne}}]{PhysRevLett.58.2420}%
  \BibitemOpen
  \bibfield  {author} {\bibinfo {author} {\bibfnamefont {H.~R.}\ \bibnamefont
  {Thorsheim}}, \bibinfo {author} {\bibfnamefont {J.}~\bibnamefont {Weiner}}, \
  and\ \bibinfo {author} {\bibfnamefont {P.~S.}\ \bibnamefont {Julienne}},\
  }\href {\doibase 10.1103/PhysRevLett.58.2420} {\bibfield  {journal} {\bibinfo
   {journal} {\prl}\ }\textbf {\bibinfo {volume} {58}},\ \bibinfo {pages}
  {2420} (\bibinfo {year} {1987})}\BibitemShut {NoStop}%
\bibitem [{\citenamefont {{Jones}}\ \emph {et~al.}(2006)\citenamefont
  {{Jones}}, \citenamefont {{Tiesinga}}, \citenamefont {{Lett}},\ and\
  \citenamefont {{Julienne}}}]{2006RvMP...78..483J}%
  \BibitemOpen
  \bibfield  {author} {\bibinfo {author} {\bibfnamefont {K.~M.}\ \bibnamefont
  {{Jones}}}, \bibinfo {author} {\bibfnamefont {E.}~\bibnamefont {{Tiesinga}}},
  \bibinfo {author} {\bibfnamefont {P.~D.}\ \bibnamefont {{Lett}}}, \ and\
  \bibinfo {author} {\bibfnamefont {P.~S.}\ \bibnamefont {{Julienne}}},\ }\href
  {\doibase 10.1103/RevModPhys.78.483} {\bibfield  {journal} {\bibinfo
  {journal} {Rev. Mod. Phys.}\ }\textbf {\bibinfo {volume} {78}},\ \bibinfo
  {pages} {483} (\bibinfo {year} {2006})}\BibitemShut {NoStop}%
\bibitem [{\citenamefont {{Pellegrini}}\ \emph {et~al.}(2008)\citenamefont
  {{Pellegrini}}, \citenamefont {{Gacesa}},\ and\ \citenamefont
  {{C{\^o}t{\'e}}}}]{2008PhRvL.101e3201P}%
  \BibitemOpen
  \bibfield  {author} {\bibinfo {author} {\bibfnamefont {P.}~\bibnamefont
  {{Pellegrini}}}, \bibinfo {author} {\bibfnamefont {M.}~\bibnamefont
  {{Gacesa}}}, \ and\ \bibinfo {author} {\bibfnamefont {R.}~\bibnamefont
  {{C{\^o}t{\'e}}}},\ }\href {\doibase 10.1103/PhysRevLett.101.053201}
  {\bibfield  {journal} {\bibinfo  {journal} {\prl}\ }\textbf {\bibinfo
  {volume} {101}},\ \bibinfo {eid} {053201} (\bibinfo {year} {2008})},\ \Eprint
  {http://arxiv.org/abs/0806.1295} {arXiv:0806.1295 [physics.atom-ph]}
  \BibitemShut {NoStop}%
\bibitem [{\citenamefont {Pellegrini}\ and\ \citenamefont
  {C\^{o}t\'{e}}(2009)}]{pellegrini2009probing}%
  \BibitemOpen
  \bibfield  {author} {\bibinfo {author} {\bibfnamefont {P.}~\bibnamefont
  {Pellegrini}}\ and\ \bibinfo {author} {\bibfnamefont {R.}~\bibnamefont
  {C\^{o}t\'{e}}},\ }\href@noop {} {\bibfield  {journal} {\bibinfo  {journal}
  {New Journal of Physics}\ }\textbf {\bibinfo {volume} {11}},\ \bibinfo
  {pages} {055047} (\bibinfo {year} {2009})}\BibitemShut {NoStop}%
\bibitem [{\citenamefont {Krzyzewski}\ \emph {et~al.}(2015)\citenamefont
  {Krzyzewski}, \citenamefont {Akin}, \citenamefont {Dizikes}, \citenamefont
  {Morrison},\ and\ \citenamefont {Abraham}}]{FOPA-2015-85Rb2}%
  \BibitemOpen
  \bibfield  {author} {\bibinfo {author} {\bibfnamefont {S.~P.}\ \bibnamefont
  {Krzyzewski}}, \bibinfo {author} {\bibfnamefont {T.~G.}\ \bibnamefont
  {Akin}}, \bibinfo {author} {\bibfnamefont {J.}~\bibnamefont {Dizikes}},
  \bibinfo {author} {\bibfnamefont {M.~A.}\ \bibnamefont {Morrison}}, \ and\
  \bibinfo {author} {\bibfnamefont {E.~R.~I.}\ \bibnamefont {Abraham}},\ }\href
  {\doibase 10.1103/PhysRevA.92.062714} {\bibfield  {journal} {\bibinfo
  {journal} {Phys. Rev. A}\ }\textbf {\bibinfo {volume} {92}},\ \bibinfo
  {pages} {062714} (\bibinfo {year} {2015})}\BibitemShut {NoStop}%
\bibitem [{\citenamefont {Hai}\ \emph {et~al.}(2020)\citenamefont {Hai},
  \citenamefont {Li}, \citenamefont {Li}, \citenamefont {Wang},\ and\
  \citenamefont {Cong}}]{FOPA-2020-KCs}%
  \BibitemOpen
  \bibfield  {author} {\bibinfo {author} {\bibfnamefont {Y.}~\bibnamefont
  {Hai}}, \bibinfo {author} {\bibfnamefont {L.-H.}\ \bibnamefont {Li}},
  \bibinfo {author} {\bibfnamefont {J.-L.}\ \bibnamefont {Li}}, \bibinfo
  {author} {\bibfnamefont {G.-R.}\ \bibnamefont {Wang}}, \ and\ \bibinfo
  {author} {\bibfnamefont {S.-L.}\ \bibnamefont {Cong}},\ }\href {\doibase
  10.1063/5.0001794} {\bibfield  {journal} {\bibinfo  {journal} {The Journal of
  Chemical Physics}\ }\textbf {\bibinfo {volume} {152}},\ \bibinfo {pages}
  {174307} (\bibinfo {year} {2020})},\ \Eprint
  {http://arxiv.org/abs/https://doi.org/10.1063/5.0001794}
  {https://doi.org/10.1063/5.0001794} \BibitemShut {NoStop}%
\bibitem [{\citenamefont {Dulieu}\ and\ \citenamefont
  {Gabbanini}(2009)}]{dulieu2009formation}%
  \BibitemOpen
  \bibfield  {author} {\bibinfo {author} {\bibfnamefont {O.}~\bibnamefont
  {Dulieu}}\ and\ \bibinfo {author} {\bibfnamefont {C.}~\bibnamefont
  {Gabbanini}},\ }\href@noop {} {\bibfield  {journal} {\bibinfo  {journal}
  {Rep. Prog. Phys.}\ }\textbf {\bibinfo {volume} {72}},\ \bibinfo {pages}
  {086401} (\bibinfo {year} {2009})}\BibitemShut {NoStop}%
\bibitem [{\citenamefont {Balakrishnan}(2016)}]{balakrishnan2016perspective}%
  \BibitemOpen
  \bibfield  {author} {\bibinfo {author} {\bibfnamefont {N.}~\bibnamefont
  {Balakrishnan}},\ }\href@noop {} {\bibfield  {journal} {\bibinfo  {journal}
  {\jcp}\ }\textbf {\bibinfo {volume} {145}},\ \bibinfo {pages} {150901}
  (\bibinfo {year} {2016})}\BibitemShut {NoStop}%
\bibitem [{\citenamefont {{Byrd}}\ \emph {et~al.}(2010)\citenamefont {{Byrd}},
  \citenamefont {{Montgomery}},\ and\ \citenamefont
  {{C{\^o}t{\'e}}}}]{2010PhRvA..82a0502B}%
  \BibitemOpen
  \bibfield  {author} {\bibinfo {author} {\bibfnamefont {J.~N.}\ \bibnamefont
  {{Byrd}}}, \bibinfo {author} {\bibfnamefont {J.~A.}\ \bibnamefont
  {{Montgomery}}, \bibfnamefont {Jr.}}, \ and\ \bibinfo {author} {\bibfnamefont
  {R.}~\bibnamefont {{C{\^o}t{\'e}}}},\ }\href {\doibase
  10.1103/PhysRevA.82.010502} {\bibfield  {journal} {\bibinfo  {journal}
  {\pra}\ }\textbf {\bibinfo {volume} {82}},\ \bibinfo {eid} {010502} (\bibinfo
  {year} {2010})},\ \Eprint {http://arxiv.org/abs/1003.4514} {arXiv:1003.4514
  [physics.chem-ph]} \BibitemShut {NoStop}%
\bibitem [{\citenamefont {{Buchachenko}}\ \emph {et~al.}(2012)\citenamefont
  {{Buchachenko}}, \citenamefont {{Stolyarov}}, \citenamefont {{Szcz{\c
  e}{\'s}niak}},\ and\ \citenamefont
  {{Cha{\l}asi{\'n}ski}}}]{2012JChPh.137k4305B}%
  \BibitemOpen
  \bibfield  {author} {\bibinfo {author} {\bibfnamefont {A.~A.}\ \bibnamefont
  {{Buchachenko}}}, \bibinfo {author} {\bibfnamefont {A.~V.}\ \bibnamefont
  {{Stolyarov}}}, \bibinfo {author} {\bibfnamefont {M.~M.}\ \bibnamefont
  {{Szcz{\c e}{\'s}niak}}}, \ and\ \bibinfo {author} {\bibfnamefont
  {G.}~\bibnamefont {{Cha{\l}asi{\'n}ski}}},\ }\href {\doibase
  10.1063/1.4752740} {\bibfield  {journal} {\bibinfo  {journal} {\jcp}\
  }\textbf {\bibinfo {volume} {137}},\ \bibinfo {pages} {114305} (\bibinfo
  {year} {2012})}\BibitemShut {NoStop}%
\bibitem [{\citenamefont {{Byrd}}\ \emph
  {et~al.}(2012{\natexlab{a}})\citenamefont {{Byrd}}, \citenamefont {{Harvey
  Michels}}, \citenamefont {{Montgomery}}, \citenamefont {{C{\^o}t{\'e}}},\
  and\ \citenamefont {{Stwalley}}}]{2012JChPh.136a4306B}%
  \BibitemOpen
  \bibfield  {author} {\bibinfo {author} {\bibfnamefont {J.~N.}\ \bibnamefont
  {{Byrd}}}, \bibinfo {author} {\bibfnamefont {H.}~\bibnamefont {{Harvey
  Michels}}}, \bibinfo {author} {\bibfnamefont {J.~A.}\ \bibnamefont
  {{Montgomery}}}, \bibinfo {author} {\bibfnamefont {R.}~\bibnamefont
  {{C{\^o}t{\'e}}}}, \ and\ \bibinfo {author} {\bibfnamefont {W.~C.}\
  \bibnamefont {{Stwalley}}},\ }\href {\doibase 10.1063/1.3672646} {\bibfield
  {journal} {\bibinfo  {journal} {\jcp}\ }\textbf {\bibinfo {volume} {136}},\
  \bibinfo {pages} {014306} (\bibinfo {year} {2012}{\natexlab{a}})}\BibitemShut
  {NoStop}%
\bibitem [{\citenamefont {{Byrd}}\ \emph
  {et~al.}(2012{\natexlab{b}})\citenamefont {{Byrd}}, \citenamefont
  {{Montgomery}},\ and\ \citenamefont {{C{\^o}t{\'e}}}}]{2012PhRvL.109h3003B}%
  \BibitemOpen
  \bibfield  {author} {\bibinfo {author} {\bibfnamefont {J.~N.}\ \bibnamefont
  {{Byrd}}}, \bibinfo {author} {\bibfnamefont {J.~A.}\ \bibnamefont
  {{Montgomery}}, \bibfnamefont {Jr.}}, \ and\ \bibinfo {author} {\bibfnamefont
  {R.}~\bibnamefont {{C{\^o}t{\'e}}}},\ }\href {\doibase
  10.1103/PhysRevLett.109.083003} {\bibfield  {journal} {\bibinfo  {journal}
  {\prl}\ }\textbf {\bibinfo {volume} {109}},\ \bibinfo {eid} {083003}
  (\bibinfo {year} {2012}{\natexlab{b}})},\ \Eprint
  {http://arxiv.org/abs/1207.2797} {arXiv:1207.2797 [physics.chem-ph]}
  \BibitemShut {NoStop}%
\bibitem [{\citenamefont {{Byrd}}\ \emph
  {et~al.}(2012{\natexlab{c}})\citenamefont {{Byrd}}, \citenamefont
  {{Montgomery}},\ and\ \citenamefont {{C{\^o}t{\'e}}}}]{2012PhRvA..86c2711B}%
  \BibitemOpen
  \bibfield  {author} {\bibinfo {author} {\bibfnamefont {J.~N.}\ \bibnamefont
  {{Byrd}}}, \bibinfo {author} {\bibfnamefont {J.~A.}\ \bibnamefont
  {{Montgomery}}, \bibfnamefont {Jr.}}, \ and\ \bibinfo {author} {\bibfnamefont
  {R.}~\bibnamefont {{C{\^o}t{\'e}}}},\ }\href {\doibase
  10.1103/PhysRevA.86.032711} {\bibfield  {journal} {\bibinfo  {journal}
  {\pra}\ }\textbf {\bibinfo {volume} {86}},\ \bibinfo {eid} {032711} (\bibinfo
  {year} {2012}{\natexlab{c}})},\ \Eprint {http://arxiv.org/abs/1207.3546}
  {arXiv:1207.3546 [physics.chem-ph]} \BibitemShut {NoStop}%
\bibitem [{\citenamefont {Lepers}\ \emph {et~al.}(2010)\citenamefont {Lepers},
  \citenamefont {Dulieu},\ and\ \citenamefont
  {Kokoouline}}]{PhysRevA.82.042711}%
  \BibitemOpen
  \bibfield  {author} {\bibinfo {author} {\bibfnamefont {M.}~\bibnamefont
  {Lepers}}, \bibinfo {author} {\bibfnamefont {O.}~\bibnamefont {Dulieu}}, \
  and\ \bibinfo {author} {\bibfnamefont {V.}~\bibnamefont {Kokoouline}},\
  }\href {\doibase 10.1103/PhysRevA.82.042711} {\bibfield  {journal} {\bibinfo
  {journal} {\pra}\ }\textbf {\bibinfo {volume} {82}},\ \bibinfo {pages}
  {042711} (\bibinfo {year} {2010})}\BibitemShut {NoStop}%
\bibitem [{\citenamefont {{P{\'e}rez-R{\'{\i}}os}}\ \emph
  {et~al.}(2015)\citenamefont {{P{\'e}rez-R{\'{\i}}os}}, \citenamefont
  {{Lepers}},\ and\ \citenamefont {{Dulieu}}}]{2015PhRvL.115g3201P}%
  \BibitemOpen
  \bibfield  {author} {\bibinfo {author} {\bibfnamefont {J.}~\bibnamefont
  {{P{\'e}rez-R{\'{\i}}os}}}, \bibinfo {author} {\bibfnamefont
  {M.}~\bibnamefont {{Lepers}}}, \ and\ \bibinfo {author} {\bibfnamefont
  {O.}~\bibnamefont {{Dulieu}}},\ }\href {\doibase
  10.1103/PhysRevLett.115.073201} {\bibfield  {journal} {\bibinfo  {journal}
  {\prl}\ }\textbf {\bibinfo {volume} {115}},\ \bibinfo {eid} {073201}
  (\bibinfo {year} {2015})},\ \Eprint {http://arxiv.org/abs/1505.03288}
  {arXiv:1505.03288 [physics.atom-ph]} \BibitemShut {NoStop}%
\bibitem [{\citenamefont {{Vexiau}}\ \emph {et~al.}(2015)\citenamefont
  {{Vexiau}}, \citenamefont {{Lepers}}, \citenamefont {{Aymar}}, \citenamefont
  {{Bouloufa-Maafa}},\ and\ \citenamefont {{Dulieu}}}]{2015JChPh.142u4303V}%
  \BibitemOpen
  \bibfield  {author} {\bibinfo {author} {\bibfnamefont {R.}~\bibnamefont
  {{Vexiau}}}, \bibinfo {author} {\bibfnamefont {M.}~\bibnamefont {{Lepers}}},
  \bibinfo {author} {\bibfnamefont {M.}~\bibnamefont {{Aymar}}}, \bibinfo
  {author} {\bibfnamefont {N.}~\bibnamefont {{Bouloufa-Maafa}}}, \ and\
  \bibinfo {author} {\bibfnamefont {O.}~\bibnamefont {{Dulieu}}},\ }\href
  {\doibase 10.1063/1.4921622} {\bibfield  {journal} {\bibinfo  {journal}
  {\jcp}\ }\textbf {\bibinfo {volume} {142}},\ \bibinfo {eid} {214303}
  (\bibinfo {year} {2015})},\ \Eprint {http://arxiv.org/abs/1502.05636}
  {arXiv:1502.05636 [physics.atom-ph]} \BibitemShut {NoStop}%
\bibitem [{\citenamefont {{Lepers}}\ \emph {et~al.}(2016)\citenamefont
  {{Lepers}}, \citenamefont {{Qu{\'e}m{\'e}ner}}, \citenamefont
  {{Luc-Koenig}},\ and\ \citenamefont {{Dulieu}}}]{2016JPhB...49a4004L}%
  \BibitemOpen
  \bibfield  {author} {\bibinfo {author} {\bibfnamefont {M.}~\bibnamefont
  {{Lepers}}}, \bibinfo {author} {\bibfnamefont {G.}~\bibnamefont
  {{Qu{\'e}m{\'e}ner}}}, \bibinfo {author} {\bibfnamefont {E.}~\bibnamefont
  {{Luc-Koenig}}}, \ and\ \bibinfo {author} {\bibfnamefont {O.}~\bibnamefont
  {{Dulieu}}},\ }\href {\doibase 10.1088/0953-4075/49/1/014004} {\bibfield
  {journal} {\bibinfo  {journal} {J. Phys. B}\ }\textbf {\bibinfo {volume}
  {49}},\ \bibinfo {eid} {014004} (\bibinfo {year} {2016})},\ \Eprint
  {http://arxiv.org/abs/1508.06066} {arXiv:1508.06066 [physics.atom-ph]}
  \BibitemShut {NoStop}%
\bibitem [{\citenamefont {{Qu{\'e}m{\'e}ner}}(2017)}]{2017arXiv170309174Q}%
  \BibitemOpen
  \bibfield  {author} {\bibinfo {author} {\bibfnamefont {G.}~\bibnamefont
  {{Qu{\'e}m{\'e}ner}}},\ }\href@noop {} {\bibfield  {journal} {\bibinfo
  {journal} {ArXiv e-prints}\ } (\bibinfo {year} {2017})},\ \Eprint
  {http://arxiv.org/abs/1703.09174} {arXiv:1703.09174 [physics.atom-ph]}
  \BibitemShut {NoStop}%
\bibitem [{\citenamefont {Schnabel}\ \emph {et~al.}(2021)\citenamefont
  {Schnabel}, \citenamefont {Kampschulte}, \citenamefont {Rupp}, \citenamefont
  {Hecker~Denschlag},\ and\ \citenamefont {K\"ohn}}]{PhysRevA.103.022820}%
  \BibitemOpen
  \bibfield  {author} {\bibinfo {author} {\bibfnamefont {J.}~\bibnamefont
  {Schnabel}}, \bibinfo {author} {\bibfnamefont {T.}~\bibnamefont
  {Kampschulte}}, \bibinfo {author} {\bibfnamefont {S.}~\bibnamefont {Rupp}},
  \bibinfo {author} {\bibfnamefont {J.}~\bibnamefont {Hecker~Denschlag}}, \
  and\ \bibinfo {author} {\bibfnamefont {A.}~\bibnamefont {K\"ohn}},\ }\href
  {\doibase 10.1103/PhysRevA.103.022820} {\bibfield  {journal} {\bibinfo
  {journal} {Phys. Rev. A}\ }\textbf {\bibinfo {volume} {103}},\ \bibinfo
  {pages} {022820} (\bibinfo {year} {2021})}\BibitemShut {NoStop}%
\bibitem [{\citenamefont {{Lett}}\ \emph {et~al.}(1995)\citenamefont {{Lett}},
  \citenamefont {{Julienne}},\ and\ \citenamefont
  {{Phillips}}}]{1995ARPC...46..423L}%
  \BibitemOpen
  \bibfield  {author} {\bibinfo {author} {\bibfnamefont {P.~D.}\ \bibnamefont
  {{Lett}}}, \bibinfo {author} {\bibfnamefont {P.~S.}\ \bibnamefont
  {{Julienne}}}, \ and\ \bibinfo {author} {\bibfnamefont {W.~D.}\ \bibnamefont
  {{Phillips}}},\ }\href {\doibase 10.1146/annurev.pc.46.100195.002231}
  {\bibfield  {journal} {\bibinfo  {journal} {Annu. Rev. Phys. Chem.}\ }\textbf
  {\bibinfo {volume} {46}},\ \bibinfo {pages} {423} (\bibinfo {year}
  {1995})}\BibitemShut {NoStop}%
\bibitem [{\citenamefont {{Stwalley}}\ and\ \citenamefont
  {{Wang}}(1999)}]{1999JMoSp.195..194S}%
  \BibitemOpen
  \bibfield  {author} {\bibinfo {author} {\bibfnamefont {W.~C.}\ \bibnamefont
  {{Stwalley}}}\ and\ \bibinfo {author} {\bibfnamefont {H.}~\bibnamefont
  {{Wang}}},\ }\href {\doibase 10.1006/jmsp.1999.7838} {\bibfield  {journal}
  {\bibinfo  {journal} {J. Mol. Spectr.}\ }\textbf {\bibinfo {volume} {195}},\
  \bibinfo {pages} {194} (\bibinfo {year} {1999})}\BibitemShut {NoStop}%
\bibitem [{\citenamefont {C{\^o}t{\'e}}(2010)}]{cote2010forming}%
  \BibitemOpen
  \bibfield  {author} {\bibinfo {author} {\bibfnamefont {R.}~\bibnamefont
  {C{\^o}t{\'e}}},\ }in\ \href@noop {} {\emph {\bibinfo {booktitle}
  {Proceedings Of The Dalgarno Celebratory Symposium}}}\ (\bibinfo
  {organization} {World Scientific},\ \bibinfo {year} {2010})\ pp.\ \bibinfo
  {pages} {262--280}\BibitemShut {NoStop}%
\bibitem [{\citenamefont {Fioretti}\ \emph {et~al.}(1998)\citenamefont
  {Fioretti}, \citenamefont {Comparat}, \citenamefont {Crubellier},
  \citenamefont {Dulieu}, \citenamefont {Masnou-Seeuws},\ and\ \citenamefont
  {Pillet}}]{PhysRevLett.80.4402}%
  \BibitemOpen
  \bibfield  {author} {\bibinfo {author} {\bibfnamefont {A.}~\bibnamefont
  {Fioretti}}, \bibinfo {author} {\bibfnamefont {D.}~\bibnamefont {Comparat}},
  \bibinfo {author} {\bibfnamefont {A.}~\bibnamefont {Crubellier}}, \bibinfo
  {author} {\bibfnamefont {O.}~\bibnamefont {Dulieu}}, \bibinfo {author}
  {\bibfnamefont {F.}~\bibnamefont {Masnou-Seeuws}}, \ and\ \bibinfo {author}
  {\bibfnamefont {P.}~\bibnamefont {Pillet}},\ }\href {\doibase
  10.1103/PhysRevLett.80.4402} {\bibfield  {journal} {\bibinfo  {journal}
  {\prl}\ }\textbf {\bibinfo {volume} {80}},\ \bibinfo {pages} {4402} (\bibinfo
  {year} {1998})}\BibitemShut {NoStop}%
\bibitem [{\citenamefont {{Vanhaecke}}\ \emph {et~al.}(2002)\citenamefont
  {{Vanhaecke}}, \citenamefont {{de Souza Melo}}, \citenamefont {{Laburthe
  Tolra}}, \citenamefont {{Comparat}},\ and\ \citenamefont
  {{Pillet}}}]{2002PhRvL..89f3001V}%
  \BibitemOpen
  \bibfield  {author} {\bibinfo {author} {\bibfnamefont {N.}~\bibnamefont
  {{Vanhaecke}}}, \bibinfo {author} {\bibfnamefont {W.}~\bibnamefont {{de Souza
  Melo}}}, \bibinfo {author} {\bibfnamefont {B.}~\bibnamefont {{Laburthe
  Tolra}}}, \bibinfo {author} {\bibfnamefont {D.}~\bibnamefont {{Comparat}}}, \
  and\ \bibinfo {author} {\bibfnamefont {P.}~\bibnamefont {{Pillet}}},\ }\href
  {\doibase 10.1103/PhysRevLett.89.063001} {\bibfield  {journal} {\bibinfo
  {journal} {\prl}\ }\textbf {\bibinfo {volume} {89}},\ \bibinfo {eid} {063001}
  (\bibinfo {year} {2002})}\BibitemShut {NoStop}%
\bibitem [{\citenamefont {{Vanhaecke}}\ \emph {et~al.}(2004)\citenamefont
  {{Vanhaecke}}, \citenamefont {{Lisdat}}, \citenamefont {{T'jampens}},
  \citenamefont {{Comparat}}, \citenamefont {{Crubellier}},\ and\ \citenamefont
  {{Pillet}}}]{2004EPJD...28..351V}%
  \BibitemOpen
  \bibfield  {author} {\bibinfo {author} {\bibfnamefont {N.}~\bibnamefont
  {{Vanhaecke}}}, \bibinfo {author} {\bibfnamefont {C.}~\bibnamefont
  {{Lisdat}}}, \bibinfo {author} {\bibfnamefont {B.}~\bibnamefont
  {{T'jampens}}}, \bibinfo {author} {\bibfnamefont {D.}~\bibnamefont
  {{Comparat}}}, \bibinfo {author} {\bibfnamefont {A.}~\bibnamefont
  {{Crubellier}}}, \ and\ \bibinfo {author} {\bibfnamefont {P.}~\bibnamefont
  {{Pillet}}},\ }\href {\doibase 10.1140/epjd/e2004-00001-y} {\bibfield
  {journal} {\bibinfo  {journal} {\epjd}\ }\textbf {\bibinfo {volume} {28}},\
  \bibinfo {pages} {351} (\bibinfo {year} {2004})}\BibitemShut {NoStop}%
\bibitem [{\citenamefont {{C{\^o}t{\'e}}}\ and\ \citenamefont
  {{Dalgarno}}(1997)}]{1997CPL...279...50C}%
  \BibitemOpen
  \bibfield  {author} {\bibinfo {author} {\bibfnamefont {R.}~\bibnamefont
  {{C{\^o}t{\'e}}}}\ and\ \bibinfo {author} {\bibfnamefont {A.}~\bibnamefont
  {{Dalgarno}}},\ }\href {\doibase 10.1016/S0009-2614(97)00937-8} {\bibfield
  {journal} {\bibinfo  {journal} {Chem. Phys. Lett.}\ }\textbf {\bibinfo
  {volume} {279}},\ \bibinfo {pages} {50} (\bibinfo {year} {1997})}\BibitemShut
  {NoStop}%
\bibitem [{\citenamefont {{C{\^o}t{\'e}}}\ and\ \citenamefont
  {{Dalgarno}}(1999)}]{1999JMoSp.195..236C}%
  \BibitemOpen
  \bibfield  {author} {\bibinfo {author} {\bibfnamefont {R.}~\bibnamefont
  {{C{\^o}t{\'e}}}}\ and\ \bibinfo {author} {\bibfnamefont {A.}~\bibnamefont
  {{Dalgarno}}},\ }\href {\doibase 10.1006/jmsp.1999.7837} {\bibfield
  {journal} {\bibinfo  {journal} {\jms}\ }\textbf {\bibinfo {volume} {195}},\
  \bibinfo {pages} {236} (\bibinfo {year} {1999})}\BibitemShut {NoStop}%
\bibitem [{\citenamefont {{Hu}}\ \emph {et~al.}(2014)\citenamefont {{Hu}},
  \citenamefont {{Xie}}, \citenamefont {{Huang}},\ and\ \citenamefont
  {{Cong}}}]{2014PhRvA..89e2712H}%
  \BibitemOpen
  \bibfield  {author} {\bibinfo {author} {\bibfnamefont {X.-J.}\ \bibnamefont
  {{Hu}}}, \bibinfo {author} {\bibfnamefont {T.}~\bibnamefont {{Xie}}},
  \bibinfo {author} {\bibfnamefont {Y.}~\bibnamefont {{Huang}}}, \ and\
  \bibinfo {author} {\bibfnamefont {S.-L.}\ \bibnamefont {{Cong}}},\ }\href
  {\doibase 10.1103/PhysRevA.89.052712} {\bibfield  {journal} {\bibinfo
  {journal} {\pra}\ }\textbf {\bibinfo {volume} {89}},\ \bibinfo {eid} {052712}
  (\bibinfo {year} {2014})}\BibitemShut {NoStop}%
\bibitem [{\citenamefont {{Gacesa}}\ and\ \citenamefont
  {{C{\^o}t{\'e}}}(2014)}]{2014JMoSp.300..124G}%
  \BibitemOpen
  \bibfield  {author} {\bibinfo {author} {\bibfnamefont {M.}~\bibnamefont
  {{Gacesa}}}\ and\ \bibinfo {author} {\bibfnamefont {R.}~\bibnamefont
  {{C{\^o}t{\'e}}}},\ }\href {\doibase 10.1016/j.jms.2014.03.005} {\bibfield
  {journal} {\bibinfo  {journal} {\jms}\ }\textbf {\bibinfo {volume} {300}},\
  \bibinfo {pages} {124} (\bibinfo {year} {2014})},\ \Eprint
  {http://arxiv.org/abs/1402.0494} {arXiv:1402.0494 [physics.atom-ph]}
  \BibitemShut {NoStop}%
\bibitem [{\citenamefont {{Hai}}\ \emph {et~al.}(2020)\citenamefont {{Hai}},
  \citenamefont {{Li}}, \citenamefont {{Li}}, \citenamefont {{Wang}},\ and\
  \citenamefont {{Cong}}}]{2020JChPh.152q4307H}%
  \BibitemOpen
  \bibfield  {author} {\bibinfo {author} {\bibfnamefont {Y.}~\bibnamefont
  {{Hai}}}, \bibinfo {author} {\bibfnamefont {L.-H.}\ \bibnamefont {{Li}}},
  \bibinfo {author} {\bibfnamefont {J.-L.}\ \bibnamefont {{Li}}}, \bibinfo
  {author} {\bibfnamefont {G.-R.}\ \bibnamefont {{Wang}}}, \ and\ \bibinfo
  {author} {\bibfnamefont {S.-L.}\ \bibnamefont {{Cong}}},\ }\href {\doibase
  10.1063/5.0001794} {\bibfield  {journal} {\bibinfo  {journal} {\jcp}\
  }\textbf {\bibinfo {volume} {152}},\ \bibinfo {eid} {174307} (\bibinfo {year}
  {2020})}\BibitemShut {NoStop}%
\bibitem [{\citenamefont {Sun}\ \emph {et~al.}(2020)\citenamefont {Sun},
  \citenamefont {Hai}, \citenamefont {Lyu}, \citenamefont {Wang},\ and\
  \citenamefont {Cong}}]{sun2020formation}%
  \BibitemOpen
  \bibfield  {author} {\bibinfo {author} {\bibfnamefont {Z.-X.}\ \bibnamefont
  {Sun}}, \bibinfo {author} {\bibfnamefont {Y.}~\bibnamefont {Hai}}, \bibinfo
  {author} {\bibfnamefont {B.-K.}\ \bibnamefont {Lyu}}, \bibinfo {author}
  {\bibfnamefont {G.-R.}\ \bibnamefont {Wang}}, \ and\ \bibinfo {author}
  {\bibfnamefont {S.-L.}\ \bibnamefont {Cong}},\ }\href@noop {} {\bibfield
  {journal} {\bibinfo  {journal} {Journal of Physics B: Atomic, Molecular and
  Optical Physics}\ }\textbf {\bibinfo {volume} {53}},\ \bibinfo {pages}
  {205204} (\bibinfo {year} {2020})}\BibitemShut {NoStop}%
\bibitem [{\citenamefont {{Koch}}\ \emph {et~al.}(2006)\citenamefont {{Koch}},
  \citenamefont {{Luc-Koenig}},\ and\ \citenamefont
  {{Masnou-Seeuws}}}]{2006PhRvA..73c3408K}%
  \BibitemOpen
  \bibfield  {author} {\bibinfo {author} {\bibfnamefont {C.~P.}\ \bibnamefont
  {{Koch}}}, \bibinfo {author} {\bibfnamefont {E.}~\bibnamefont
  {{Luc-Koenig}}}, \ and\ \bibinfo {author} {\bibfnamefont {F.}~\bibnamefont
  {{Masnou-Seeuws}}},\ }\href {\doibase 10.1103/PhysRevA.73.033408} {\bibfield
  {journal} {\bibinfo  {journal} {\pra}\ }\textbf {\bibinfo {volume} {73}},\
  \bibinfo {eid} {033408} (\bibinfo {year} {2006})},\ \Eprint
  {http://arxiv.org/abs/physics/0508090} {arXiv:physics/0508090
  [physics.atom-ph]} \BibitemShut {NoStop}%
\bibitem [{\citenamefont {{Ghosal}}\ \emph {et~al.}(2009)\citenamefont
  {{Ghosal}}, \citenamefont {{Doyle}}, \citenamefont {{Koch}},\ and\
  \citenamefont {{Hutson}}}]{2009NJPh...11e5011G}%
  \BibitemOpen
  \bibfield  {author} {\bibinfo {author} {\bibfnamefont {S.}~\bibnamefont
  {{Ghosal}}}, \bibinfo {author} {\bibfnamefont {R.~J.}\ \bibnamefont
  {{Doyle}}}, \bibinfo {author} {\bibfnamefont {C.~P.}\ \bibnamefont {{Koch}}},
  \ and\ \bibinfo {author} {\bibfnamefont {J.~M.}\ \bibnamefont {{Hutson}}},\
  }\href {\doibase 10.1088/1367-2630/11/5/055011} {\bibfield  {journal}
  {\bibinfo  {journal} {\njp}\ }\textbf {\bibinfo {volume} {11}},\ \bibinfo
  {eid} {055011} (\bibinfo {year} {2009})},\ \Eprint
  {http://arxiv.org/abs/0810.5703} {arXiv:0810.5703 [physics.atom-ph]}
  \BibitemShut {NoStop}%
\bibitem [{\citenamefont {{Carini}}\ \emph {et~al.}(2015)\citenamefont
  {{Carini}}, \citenamefont {{Kallush}}, \citenamefont {{Kosloff}},\ and\
  \citenamefont {{Gould}}}]{2015PhRvL.115q3003C}%
  \BibitemOpen
  \bibfield  {author} {\bibinfo {author} {\bibfnamefont {J.~L.}\ \bibnamefont
  {{Carini}}}, \bibinfo {author} {\bibfnamefont {S.}~\bibnamefont {{Kallush}}},
  \bibinfo {author} {\bibfnamefont {R.}~\bibnamefont {{Kosloff}}}, \ and\
  \bibinfo {author} {\bibfnamefont {P.~L.}\ \bibnamefont {{Gould}}},\ }\href
  {\doibase 10.1103/PhysRevLett.115.173003} {\bibfield  {journal} {\bibinfo
  {journal} {\prl}\ }\textbf {\bibinfo {volume} {115}},\ \bibinfo {eid}
  {173003} (\bibinfo {year} {2015})},\ \Eprint
  {http://arxiv.org/abs/1602.08026} {arXiv:1602.08026 [physics.atom-ph]}
  \BibitemShut {NoStop}%
\bibitem [{\citenamefont {{Wang}}\ \emph {et~al.}(2017)\citenamefont {{Wang}},
  \citenamefont {{Li}}, \citenamefont {{Hu}}, \citenamefont {{Chen}},\ and\
  \citenamefont {{Cong}}}]{2017PhRvA..96d3417W}%
  \BibitemOpen
  \bibfield  {author} {\bibinfo {author} {\bibfnamefont {M.}~\bibnamefont
  {{Wang}}}, \bibinfo {author} {\bibfnamefont {J.-L.}\ \bibnamefont {{Li}}},
  \bibinfo {author} {\bibfnamefont {X.-J.}\ \bibnamefont {{Hu}}}, \bibinfo
  {author} {\bibfnamefont {M.-D.}\ \bibnamefont {{Chen}}}, \ and\ \bibinfo
  {author} {\bibfnamefont {S.-L.}\ \bibnamefont {{Cong}}},\ }\href {\doibase
  10.1103/PhysRevA.96.043417} {\bibfield  {journal} {\bibinfo  {journal}
  {\pra}\ }\textbf {\bibinfo {volume} {96}},\ \bibinfo {eid} {043417} (\bibinfo
  {year} {2017})}\BibitemShut {NoStop}%
\bibitem [{\citenamefont {{Gacesa}}\ \emph {et~al.}(2013)\citenamefont
  {{Gacesa}}, \citenamefont {{Ghosal}}, \citenamefont {{Byrd}},\ and\
  \citenamefont {{C{\^o}t{\'e}}}}]{2013PhRvA..88f3418G}%
  \BibitemOpen
  \bibfield  {author} {\bibinfo {author} {\bibfnamefont {M.}~\bibnamefont
  {{Gacesa}}}, \bibinfo {author} {\bibfnamefont {S.}~\bibnamefont {{Ghosal}}},
  \bibinfo {author} {\bibfnamefont {J.~N.}\ \bibnamefont {{Byrd}}}, \ and\
  \bibinfo {author} {\bibfnamefont {R.}~\bibnamefont {{C{\^o}t{\'e}}}},\ }\href
  {\doibase 10.1103/PhysRevA.88.063418} {\bibfield  {journal} {\bibinfo
  {journal} {\pra}\ }\textbf {\bibinfo {volume} {88}},\ \bibinfo {eid} {063418}
  (\bibinfo {year} {2013})},\ \Eprint {http://arxiv.org/abs/1310.7140}
  {arXiv:1310.7140 [physics.atom-ph]} \BibitemShut {NoStop}%
\bibitem [{\citenamefont {{Hu}}\ \emph {et~al.}(2015)\citenamefont {{Hu}},
  \citenamefont {{Li}}, \citenamefont {{Xie}}, \citenamefont {{Han}},\ and\
  \citenamefont {{Cong}}}]{2015PhRvA..92c2709H}%
  \BibitemOpen
  \bibfield  {author} {\bibinfo {author} {\bibfnamefont {X.-J.}\ \bibnamefont
  {{Hu}}}, \bibinfo {author} {\bibfnamefont {J.-L.}\ \bibnamefont {{Li}}},
  \bibinfo {author} {\bibfnamefont {T.}~\bibnamefont {{Xie}}}, \bibinfo
  {author} {\bibfnamefont {Y.-C.}\ \bibnamefont {{Han}}}, \ and\ \bibinfo
  {author} {\bibfnamefont {S.-L.}\ \bibnamefont {{Cong}}},\ }\href {\doibase
  10.1103/PhysRevA.92.032709} {\bibfield  {journal} {\bibinfo  {journal}
  {\pra}\ }\textbf {\bibinfo {volume} {92}},\ \bibinfo {eid} {032709} (\bibinfo
  {year} {2015})}\BibitemShut {NoStop}%
\bibitem [{\citenamefont {Mulder}\ \emph {et~al.}(1979)\citenamefont {Mulder},
  \citenamefont {van~der Avoird},\ and\ \citenamefont
  {Wormer}}]{mulder1979anisotropy}%
  \BibitemOpen
  \bibfield  {author} {\bibinfo {author} {\bibfnamefont {F.}~\bibnamefont
  {Mulder}}, \bibinfo {author} {\bibfnamefont {A.}~\bibnamefont {van~der
  Avoird}}, \ and\ \bibinfo {author} {\bibfnamefont {P.~E.}\ \bibnamefont
  {Wormer}},\ }\href@noop {} {\bibfield  {journal} {\bibinfo  {journal} {Mol.
  Phys.}\ }\textbf {\bibinfo {volume} {37}},\ \bibinfo {pages} {159} (\bibinfo
  {year} {1979})}\BibitemShut {NoStop}%
\bibitem [{\citenamefont {van~der Avoird}\ \emph {et~al.}(1980)\citenamefont
  {van~der Avoird}, \citenamefont {Wormer}, \citenamefont {Mulder},\ and\
  \citenamefont {Berns}}]{van1980ab}%
  \BibitemOpen
  \bibfield  {author} {\bibinfo {author} {\bibfnamefont {A.}~\bibnamefont
  {van~der Avoird}}, \bibinfo {author} {\bibfnamefont {P.~E.}\ \bibnamefont
  {Wormer}}, \bibinfo {author} {\bibfnamefont {F.}~\bibnamefont {Mulder}}, \
  and\ \bibinfo {author} {\bibfnamefont {R.~M.}\ \bibnamefont {Berns}},\ }in\
  \href@noop {} {\emph {\bibinfo {booktitle} {Van der Waals Systems}}}\
  (\bibinfo  {publisher} {Springer},\ \bibinfo {year} {1980})\ pp.\ \bibinfo
  {pages} {1--51}\BibitemShut {NoStop}%
\bibitem [{\citenamefont {{Byrd}}\ \emph {et~al.}(2011)\citenamefont {{Byrd}},
  \citenamefont {{C{\^o}t{\'e}}},\ and\ \citenamefont
  {{Montgomery}}}]{2011JChPh.135x4307B}%
  \BibitemOpen
  \bibfield  {author} {\bibinfo {author} {\bibfnamefont {J.~N.}\ \bibnamefont
  {{Byrd}}}, \bibinfo {author} {\bibfnamefont {R.}~\bibnamefont
  {{C{\^o}t{\'e}}}}, \ and\ \bibinfo {author} {\bibfnamefont {J.~A.}\
  \bibnamefont {{Montgomery}}},\ }\href {\doibase 10.1063/1.3671371} {\bibfield
   {journal} {\bibinfo  {journal} {\jcp}\ }\textbf {\bibinfo {volume} {135}},\
  \bibinfo {pages} {244307} (\bibinfo {year} {2011})},\ \Eprint
  {http://arxiv.org/abs/1109.4410} {arXiv:1109.4410 [physics.chem-ph]}
  \BibitemShut {NoStop}%
\bibitem [{\citenamefont {De~Miranda}\ \emph {et~al.}(2011)\citenamefont
  {De~Miranda}, \citenamefont {Chotia}, \citenamefont {Neyenhuis},
  \citenamefont {Wang}, \citenamefont {Qu{\'e}m{\'e}ner}, \citenamefont
  {Ospelkaus}, \citenamefont {Bohn}, \citenamefont {Ye},\ and\ \citenamefont
  {Jin}}]{de2011controlling}%
  \BibitemOpen
  \bibfield  {author} {\bibinfo {author} {\bibfnamefont {M.}~\bibnamefont
  {De~Miranda}}, \bibinfo {author} {\bibfnamefont {A.}~\bibnamefont {Chotia}},
  \bibinfo {author} {\bibfnamefont {B.}~\bibnamefont {Neyenhuis}}, \bibinfo
  {author} {\bibfnamefont {D.}~\bibnamefont {Wang}}, \bibinfo {author}
  {\bibfnamefont {G.}~\bibnamefont {Qu{\'e}m{\'e}ner}}, \bibinfo {author}
  {\bibfnamefont {S.}~\bibnamefont {Ospelkaus}}, \bibinfo {author}
  {\bibfnamefont {J.}~\bibnamefont {Bohn}}, \bibinfo {author} {\bibfnamefont
  {J.}~\bibnamefont {Ye}}, \ and\ \bibinfo {author} {\bibfnamefont
  {D.}~\bibnamefont {Jin}},\ }\href@noop {} {\bibfield  {journal} {\bibinfo
  {journal} {Nat. Phys.}\ }\textbf {\bibinfo {volume} {7}},\ \bibinfo {pages}
  {502} (\bibinfo {year} {2011})}\BibitemShut {NoStop}%
\bibitem [{\citenamefont {{Bohn}}\ \emph {et~al.}(2017)\citenamefont {{Bohn}},
  \citenamefont {{Rey}},\ and\ \citenamefont {{Ye}}}]{2017Sci...357.1002B}%
  \BibitemOpen
  \bibfield  {author} {\bibinfo {author} {\bibfnamefont {J.~L.}\ \bibnamefont
  {{Bohn}}}, \bibinfo {author} {\bibfnamefont {A.~M.}\ \bibnamefont {{Rey}}}, \
  and\ \bibinfo {author} {\bibfnamefont {J.}~\bibnamefont {{Ye}}},\ }\href
  {\doibase 10.1126/science.aam6299} {\bibfield  {journal} {\bibinfo  {journal}
  {Science}\ }\textbf {\bibinfo {volume} {357}},\ \bibinfo {pages} {1002}
  (\bibinfo {year} {2017})},\ \Eprint {http://arxiv.org/abs/1708.02806}
  {arXiv:1708.02806 [physics.atom-ph]} \BibitemShut {NoStop}%
\bibitem [{\citenamefont {De~Marco}\ \emph {et~al.}(2019)\citenamefont
  {De~Marco}, \citenamefont {Valtolina}, \citenamefont {Matsuda}, \citenamefont
  {Tobias}, \citenamefont {Covey},\ and\ \citenamefont {Ye}}]{De_Marco853}%
  \BibitemOpen
  \bibfield  {author} {\bibinfo {author} {\bibfnamefont {L.}~\bibnamefont
  {De~Marco}}, \bibinfo {author} {\bibfnamefont {G.}~\bibnamefont {Valtolina}},
  \bibinfo {author} {\bibfnamefont {K.}~\bibnamefont {Matsuda}}, \bibinfo
  {author} {\bibfnamefont {W.~G.}\ \bibnamefont {Tobias}}, \bibinfo {author}
  {\bibfnamefont {J.~P.}\ \bibnamefont {Covey}}, \ and\ \bibinfo {author}
  {\bibfnamefont {J.}~\bibnamefont {Ye}},\ }\href {\doibase
  10.1126/science.aau7230} {\bibfield  {journal} {\bibinfo  {journal}
  {Science}\ }\textbf {\bibinfo {volume} {363}},\ \bibinfo {pages} {853}
  (\bibinfo {year} {2019})}\BibitemShut {NoStop}%
\bibitem [{\citenamefont {{Aymar}}\ and\ \citenamefont
  {{Dulieu}}(2005)}]{2005JChPh.122t4302A}%
  \BibitemOpen
  \bibfield  {author} {\bibinfo {author} {\bibfnamefont {M.}~\bibnamefont
  {{Aymar}}}\ and\ \bibinfo {author} {\bibfnamefont {O.}~\bibnamefont
  {{Dulieu}}},\ }\href {\doibase 10.1063/1.1903944} {\bibfield  {journal}
  {\bibinfo  {journal} {J. Chem. Phys.}\ }\textbf {\bibinfo {volume} {122}},\
  \bibinfo {pages} {204302} (\bibinfo {year} {2005})},\ \Eprint
  {http://arxiv.org/abs/quant-ph/0502059} {quant-ph/0502059} \BibitemShut
  {NoStop}%
\bibitem [{\citenamefont {{Purvis III}}\ and\ \citenamefont
  {Bartlett}(1982)}]{purvis1982}%
  \BibitemOpen
  \bibfield  {author} {\bibinfo {author} {\bibfnamefont {G.~D.}\ \bibnamefont
  {{Purvis III}}}\ and\ \bibinfo {author} {\bibfnamefont {R.~J.}\ \bibnamefont
  {Bartlett}},\ }\href@noop {} {\bibfield  {journal} {\bibinfo  {journal} {J.
  Chem. Phys.}\ }\textbf {\bibinfo {volume} {76}},\ \bibinfo {pages} {1910}
  (\bibinfo {year} {1982})}\BibitemShut {NoStop}%
\bibitem [{\citenamefont {Raghavachari}\ \emph {et~al.}(1989)\citenamefont
  {Raghavachari}, \citenamefont {Trucks}, \citenamefont {Pople},\ and\
  \citenamefont {Head-Gordon}}]{raghavachari1989}%
  \BibitemOpen
  \bibfield  {author} {\bibinfo {author} {\bibfnamefont {K.}~\bibnamefont
  {Raghavachari}}, \bibinfo {author} {\bibfnamefont {G.~W.}\ \bibnamefont
  {Trucks}}, \bibinfo {author} {\bibfnamefont {J.~A.}\ \bibnamefont {Pople}}, \
  and\ \bibinfo {author} {\bibfnamefont {M.}~\bibnamefont {Head-Gordon}},\
  }\href@noop {} {\bibfield  {journal} {\bibinfo  {journal} {Chem. Phys.
  Lett.}\ }\textbf {\bibinfo {volume} {157}},\ \bibinfo {pages} {479} (\bibinfo
  {year} {1989})}\BibitemShut {NoStop}%
\bibitem [{\citenamefont {Bartlett}\ and\ \citenamefont
  {Musia\l}(2007)}]{bartlett2007}%
  \BibitemOpen
  \bibfield  {author} {\bibinfo {author} {\bibfnamefont {R.~J.}\ \bibnamefont
  {Bartlett}}\ and\ \bibinfo {author} {\bibfnamefont {M.}~\bibnamefont
  {Musia\l}},\ }\href@noop {} {\bibfield  {journal} {\bibinfo  {journal} {Rev.
  Mod. Phys.}\ }\textbf {\bibinfo {volume} {79}},\ \bibinfo {pages} {291}
  (\bibinfo {year} {2007})}\BibitemShut {NoStop}%
\bibitem [{\citenamefont {Weigend}\ and\ \citenamefont
  {Ahlrichs}(2005)}]{weigend2005}%
  \BibitemOpen
  \bibfield  {author} {\bibinfo {author} {\bibfnamefont {F.}~\bibnamefont
  {Weigend}}\ and\ \bibinfo {author} {\bibfnamefont {R.}~\bibnamefont
  {Ahlrichs}},\ }\href@noop {} {\bibfield  {journal} {\bibinfo  {journal}
  {Phys. Chem. Chem. Phys.}\ }\textbf {\bibinfo {volume} {7}},\ \bibinfo
  {pages} {3297} (\bibinfo {year} {2005})}\BibitemShut {NoStop}%
\bibitem [{\citenamefont {Leininger}\ \emph {et~al.}(1996)\citenamefont
  {Leininger}, \citenamefont {Nicklass}, \citenamefont {K{\"{u}}chle},
  \citenamefont {Stoll}, \citenamefont {Dolg},\ and\ \citenamefont
  {Bergner}}]{leininger1996}%
  \BibitemOpen
  \bibfield  {author} {\bibinfo {author} {\bibfnamefont {T.}~\bibnamefont
  {Leininger}}, \bibinfo {author} {\bibfnamefont {A.}~\bibnamefont {Nicklass}},
  \bibinfo {author} {\bibfnamefont {W.}~\bibnamefont {K{\"{u}}chle}}, \bibinfo
  {author} {\bibfnamefont {H.}~\bibnamefont {Stoll}}, \bibinfo {author}
  {\bibfnamefont {M.}~\bibnamefont {Dolg}}, \ and\ \bibinfo {author}
  {\bibfnamefont {A.}~\bibnamefont {Bergner}},\ }\href@noop {} {\bibfield
  {journal} {\bibinfo  {journal} {Chem. Phys. Lett.}\ }\textbf {\bibinfo
  {volume} {255}},\ \bibinfo {pages} {274} (\bibinfo {year}
  {1996})}\BibitemShut {NoStop}%
\bibitem [{\citenamefont {Stanton}\ and\ \citenamefont
  {Bartlett}(1993)}]{stanton1993}%
  \BibitemOpen
  \bibfield  {author} {\bibinfo {author} {\bibfnamefont {J.~F.}\ \bibnamefont
  {Stanton}}\ and\ \bibinfo {author} {\bibfnamefont {R.~J.}\ \bibnamefont
  {Bartlett}},\ }\href@noop {} {\bibfield  {journal} {\bibinfo  {journal} {J.
  Chem. Phys.}\ }\textbf {\bibinfo {volume} {98}},\ \bibinfo {pages} {7029}
  (\bibinfo {year} {1993})}\BibitemShut {NoStop}%
\bibitem [{\citenamefont {Werner}\ \emph {et~al.}(2010)\citenamefont {Werner},
  \citenamefont {Knowles}, \citenamefont {Manby}, \citenamefont {{Sch\"{u}tz}}
  \emph {et~al.}}]{molpro10_short}%
  \BibitemOpen
  \bibfield  {author} {\bibinfo {author} {\bibfnamefont {H.-J.}\ \bibnamefont
  {Werner}}, \bibinfo {author} {\bibfnamefont {P.~J.}\ \bibnamefont {Knowles}},
  \bibinfo {author} {\bibfnamefont {F.~R.}\ \bibnamefont {Manby}}, \bibinfo
  {author} {\bibfnamefont {M.}~\bibnamefont {{Sch\"{u}tz}}},  \emph {et~al.},\
  }\href@noop {} {\enquote {\bibinfo {title} {{MOLPRO, version 2010.1, a
  package of ab initio programs}},}\ } (\bibinfo {year} {2010}),\ \bibinfo
  {note} {{see http://www.molpro.net}}\BibitemShut {NoStop}%
\bibitem [{cod()}]{codelongrange}%
  \BibitemOpen
  \href@noop {} {\enquote {\bibinfo {title} {The sum in eq. (9) was evaluated
  up to n=8. the code to evaluate long-range potentials is available on
  request},}\ }\BibitemShut {NoStop}%
\bibitem [{\citenamefont {Werner}\ and\ \citenamefont
  {Knowles}(1985)}]{werner1985}%
  \BibitemOpen
  \bibfield  {author} {\bibinfo {author} {\bibfnamefont {H.-J.}\ \bibnamefont
  {Werner}}\ and\ \bibinfo {author} {\bibfnamefont {P.~J.}\ \bibnamefont
  {Knowles}},\ }\href@noop {} {\bibfield  {journal} {\bibinfo  {journal} {J.
  Chem. Phys.}\ }\textbf {\bibinfo {volume} {82}},\ \bibinfo {pages} {5053}
  (\bibinfo {year} {1985})}\BibitemShut {NoStop}%
\bibitem [{\citenamefont {Knowles}\ and\ \citenamefont
  {Werner}(1985)}]{knowles1985}%
  \BibitemOpen
  \bibfield  {author} {\bibinfo {author} {\bibfnamefont {P.~J.}\ \bibnamefont
  {Knowles}}\ and\ \bibinfo {author} {\bibfnamefont {H.~J.}\ \bibnamefont
  {Werner}},\ }\href@noop {} {\bibfield  {journal} {\bibinfo  {journal} {Chem.
  Phys. Lett.}\ }\textbf {\bibinfo {volume} {115}},\ \bibinfo {pages} {259}
  (\bibinfo {year} {1985})}\BibitemShut {NoStop}%
\bibitem [{\citenamefont {{Napolitano}}\ \emph {et~al.}(1994)\citenamefont
  {{Napolitano}}, \citenamefont {{Weiner}}, \citenamefont {{Williams}},\ and\
  \citenamefont {{Julienne}}}]{1994PhRvL..73.1352N}%
  \BibitemOpen
  \bibfield  {author} {\bibinfo {author} {\bibfnamefont {R.}~\bibnamefont
  {{Napolitano}}}, \bibinfo {author} {\bibfnamefont {J.}~\bibnamefont
  {{Weiner}}}, \bibinfo {author} {\bibfnamefont {C.~J.}\ \bibnamefont
  {{Williams}}}, \ and\ \bibinfo {author} {\bibfnamefont {P.~S.}\ \bibnamefont
  {{Julienne}}},\ }\href {\doibase 10.1103/PhysRevLett.73.1352} {\bibfield
  {journal} {\bibinfo  {journal} {\prl}\ }\textbf {\bibinfo {volume} {73}},\
  \bibinfo {pages} {1352} (\bibinfo {year} {1994})}\BibitemShut {NoStop}%
\bibitem [{\citenamefont {{Juarros}}\ \emph {et~al.}(2006)\citenamefont
  {{Juarros}}, \citenamefont {{Kirby}},\ and\ \citenamefont
  {{C{\^o}t{\'e}}}}]{2006JPhB...39S.965J}%
  \BibitemOpen
  \bibfield  {author} {\bibinfo {author} {\bibfnamefont {E.}~\bibnamefont
  {{Juarros}}}, \bibinfo {author} {\bibfnamefont {K.}~\bibnamefont {{Kirby}}},
  \ and\ \bibinfo {author} {\bibfnamefont {R.}~\bibnamefont {{C{\^o}t{\'e}}}},\
  }\href {\doibase 10.1088/0953-4075/39/19/S11} {\bibfield  {journal} {\bibinfo
   {journal} {J. Phys. B}\ }\textbf {\bibinfo {volume} {39}},\ \bibinfo {pages}
  {965} (\bibinfo {year} {2006})}\BibitemShut {NoStop}%
\bibitem [{\citenamefont {{Gacesa}}\ \emph {et~al.}(2016)\citenamefont
  {{Gacesa}}, \citenamefont {{Montgomery}}, \citenamefont {{Michels}},\ and\
  \citenamefont {{C{\^o}t{\'e}}}}]{2016PhRvA..94a3407G}%
  \BibitemOpen
  \bibfield  {author} {\bibinfo {author} {\bibfnamefont {M.}~\bibnamefont
  {{Gacesa}}}, \bibinfo {author} {\bibfnamefont {J.~A.}\ \bibnamefont
  {{Montgomery}}}, \bibinfo {author} {\bibfnamefont {H.~H.}\ \bibnamefont
  {{Michels}}}, \ and\ \bibinfo {author} {\bibfnamefont {R.}~\bibnamefont
  {{C{\^o}t{\'e}}}},\ }\href {\doibase 10.1103/PhysRevA.94.013407} {\bibfield
  {journal} {\bibinfo  {journal} {\pra}\ }\textbf {\bibinfo {volume} {94}},\
  \bibinfo {eid} {013407} (\bibinfo {year} {2016})},\ \Eprint
  {http://arxiv.org/abs/1603.08032} {arXiv:1603.08032 [physics.atom-ph]}
  \BibitemShut {NoStop}%
\bibitem [{\citenamefont {Juarros}\ \emph {et~al.}(2006)\citenamefont
  {Juarros}, \citenamefont {Pellegrini}, \citenamefont {Kirby},\ and\
  \citenamefont {C{\^o}t{\'e}}}]{juarros2006one}%
  \BibitemOpen
  \bibfield  {author} {\bibinfo {author} {\bibfnamefont {E.}~\bibnamefont
  {Juarros}}, \bibinfo {author} {\bibfnamefont {P.}~\bibnamefont {Pellegrini}},
  \bibinfo {author} {\bibfnamefont {K.}~\bibnamefont {Kirby}}, \ and\ \bibinfo
  {author} {\bibfnamefont {R.}~\bibnamefont {C{\^o}t{\'e}}},\ }\href@noop {}
  {\bibfield  {journal} {\bibinfo  {journal} {Physical Review A}\ }\textbf
  {\bibinfo {volume} {73}},\ \bibinfo {pages} {041403} (\bibinfo {year}
  {2006})}\BibitemShut {NoStop}%
\bibitem [{\citenamefont {{Herzberg}}(1950)}]{Herzberg}%
  \BibitemOpen
  \bibfield  {author} {\bibinfo {author} {\bibfnamefont {G.}~\bibnamefont
  {{Herzberg}}},\ }\href@noop {} {\emph {\bibinfo {title} {New York: Van
  Nostrand Reinhold, 1950, 2nd ed.}}}\ (\bibinfo {year} {1950})\BibitemShut
  {NoStop}%
\bibitem [{\citenamefont {{Kokoouline}}\ \emph {et~al.}(1999)\citenamefont
  {{Kokoouline}}, \citenamefont {{Dulieu}}, \citenamefont {{Kosloff}},\ and\
  \citenamefont {{Masnou-Seeuws}}}]{1999JChPh.110.9865K}%
  \BibitemOpen
  \bibfield  {author} {\bibinfo {author} {\bibfnamefont {V.}~\bibnamefont
  {{Kokoouline}}}, \bibinfo {author} {\bibfnamefont {O.}~\bibnamefont
  {{Dulieu}}}, \bibinfo {author} {\bibfnamefont {R.}~\bibnamefont {{Kosloff}}},
  \ and\ \bibinfo {author} {\bibfnamefont {F.}~\bibnamefont
  {{Masnou-Seeuws}}},\ }\href {\doibase 10.1063/1.478860} {\bibfield  {journal}
  {\bibinfo  {journal} {J. Chem. Phys.}\ }\textbf {\bibinfo {volume} {110}},\
  \bibinfo {pages} {9865} (\bibinfo {year} {1999})}\BibitemShut {NoStop}%
\bibitem [{\citenamefont {Johnson}(1978)}]{ren_numerov}%
  \BibitemOpen
  \bibfield  {author} {\bibinfo {author} {\bibfnamefont {B.~R.}\ \bibnamefont
  {Johnson}},\ }\href@noop {} {\bibfield  {journal} {\bibinfo  {journal} {J.
  Chem. Phys.}\ }\textbf {\bibinfo {volume} {69}},\ \bibinfo {pages} {4678}
  (\bibinfo {year} {1978})}\BibitemShut {NoStop}%
\bibitem [{\citenamefont {{Zemke}}\ \emph {et~al.}(2010)\citenamefont
  {{Zemke}}, \citenamefont {{Byrd}}, \citenamefont {{Michels}}, \citenamefont
  {{Montgomery}},\ and\ \citenamefont {{Stwalley}}}]{2010JChPh.132x4305Z}%
  \BibitemOpen
  \bibfield  {author} {\bibinfo {author} {\bibfnamefont {W.~T.}\ \bibnamefont
  {{Zemke}}}, \bibinfo {author} {\bibfnamefont {J.~N.}\ \bibnamefont {{Byrd}}},
  \bibinfo {author} {\bibfnamefont {H.~H.}\ \bibnamefont {{Michels}}}, \bibinfo
  {author} {\bibfnamefont {J.~A.}\ \bibnamefont {{Montgomery}}}, \ and\
  \bibinfo {author} {\bibfnamefont {W.~C.}\ \bibnamefont {{Stwalley}}},\ }\href
  {\doibase 10.1063/1.3454656} {\bibfield  {journal} {\bibinfo  {journal}
  {\jcp}\ }\textbf {\bibinfo {volume} {132}},\ \bibinfo {pages} {244305}
  (\bibinfo {year} {2010})}\BibitemShut {NoStop}%
\bibitem [{\citenamefont {{Moses}}\ \emph {et~al.}(2017)\citenamefont
  {{Moses}}, \citenamefont {{Covey}}, \citenamefont {{Miecnikowski}},
  \citenamefont {{Jin}},\ and\ \citenamefont {{Ye}}}]{2017NatPh..13...13M}%
  \BibitemOpen
  \bibfield  {author} {\bibinfo {author} {\bibfnamefont {S.~A.}\ \bibnamefont
  {{Moses}}}, \bibinfo {author} {\bibfnamefont {J.~P.}\ \bibnamefont
  {{Covey}}}, \bibinfo {author} {\bibfnamefont {M.~T.}\ \bibnamefont
  {{Miecnikowski}}}, \bibinfo {author} {\bibfnamefont {D.~S.}\ \bibnamefont
  {{Jin}}}, \ and\ \bibinfo {author} {\bibfnamefont {J.}~\bibnamefont {{Ye}}},\
  }\href {\doibase 10.1038/nphys3985} {\bibfield  {journal} {\bibinfo
  {journal} {Nat. Phys.}\ }\textbf {\bibinfo {volume} {13}},\ \bibinfo {pages}
  {13} (\bibinfo {year} {2017})},\ \Eprint {http://arxiv.org/abs/1610.07711}
  {arXiv:1610.07711 [cond-mat.quant-gas]} \BibitemShut {NoStop}%
\bibitem [{\citenamefont {Bowman}(2014)}]{bowman2014roaming}%
  \BibitemOpen
  \bibfield  {author} {\bibinfo {author} {\bibfnamefont {J.~M.}\ \bibnamefont
  {Bowman}},\ }\href@noop {} {\bibfield  {journal} {\bibinfo  {journal}
  {Molecular Physics}\ }\textbf {\bibinfo {volume} {112}},\ \bibinfo {pages}
  {2516} (\bibinfo {year} {2014})}\BibitemShut {NoStop}%
\end{thebibliography}%

\end{document}